\documentclass{article}

\usepackage{arxiv}

\usepackage[utf8]{inputenc} 
\usepackage[T1]{fontenc}    
\usepackage{stix}
\usepackage{hyperref}       
\usepackage{url}            
\usepackage{booktabs}       
\usepackage{amsfonts}       
\usepackage{nicefrac}       
\usepackage{microtype}      
\usepackage{graphicx}
\usepackage[super,sort&compress]{natbib}
\setcitestyle{comma,numbers,super,open={},close={}}
\usepackage{doi}
\usepackage{ntheorem}

\title{A comparative study on different neural network architectures to model inelasticity}

\author{ Max Rosenkranz \\
	Institute of Solid Mechancis\\
	TU Dresden\\
	01062 Dresden, Germany \\
	\And
	Karl A. Kalina \\
	Institute of Solid Mechancis\\
	TU Dresden\\
	01062 Dresden, Germany \\
	\And
	J\"{o}rg Brummund \\
	Institute of Solid Mechancis\\
	TU Dresden\\
	01062 Dresden, Germany \\
	\And
	Markus K\"{a}stner\thanks{Corresponding author, email: \texttt{markus.kaestner@tu-dresden.de}.} \\
	Institute of Solid Mechancis\\
	TU Dresden\\
	01062 Dresden, Germany \\
}

\renewcommand{\shorttitle}{A comparative study on different neural network architectures to model inelasticity}

\theorembodyfont{\rm}
\newtheorem{remark}{Remark}

\hypersetup{
pdftitle={Preprint_RosenkranzEtAl_2023},
pdfsubject={},
pdfauthor={Rosenkranz, Kalina},
pdfkeywords={First keyword, Second keyword, More},
}

\usepackage{mathtools}
\usepackage{subfig}
\usepackage{pgfplots}
\usepgfplotslibrary[groupplots]
\usepackage{transparent}
\usepackage{bm}
\usepackage{siunitx}
\usepackage{lipsum}
\usepackage{tcolorbox}
\usepackage{verbatim}
\usepackage{makecell}
\usepackage{cleveref}       

\makeatletter
\def\smallunderbrace#1{\mathop{\vtop{\m@th\ialign{##\crcr
   $\hfil\displaystyle{#1}\hfil$\crcr
   \noalign{\kern3\p@\nointerlineskip}%
   \small\upbracefill\crcr\noalign{\kern3\p@}}}}\limits}
\makeatother

\DeclareMathOperator{\sign}{sign}
\DeclareMathOperator{\relu}{ReLU}
\DeclareMathOperator{\softplus}{SP}
\DeclareMathOperator{\mae}{MAE}

\newcommand{\Ve}[1]{\bm{#1}} 
\newcommand{\VeGr}[1]{\bm{#1}} 

\newcommand{\reff}[1]{Fig.~\ref{#1}} 
\newcommand{\reft}[1]{Tab.~\ref{#1}} 
\newcommand{\refe}[1]{Eq.~\eqref{#1}} 
\newcommand{\refs}[1]{Sect.~\ref{#1}} 

\newcolumntype{C}[1]{>{\centering\let\newline\\\arraybackslash\hspace{0pt}}m{#1}}

\begin{document}
\maketitle

\begin{abstract}
The mathematical formulation of constitutive models to describe the path-dependent, i.\,e., inelastic, behavior of materials is a challenging task and has been a focus in mechanics research for several decades. There have been increased efforts to facilitate or automate this task through data-driven techniques, impelled in particular by the recent revival of neural networks (NNs) in computational mechanics. However, it seems questionable to simply not consider fundamental findings of constitutive modeling originating from the last decades research within NN-based approaches. Herein, we propose a comparative study on different feedforward and recurrent neural network architectures to model inelasticity. Within this study, we divide the models into three basic classes: black box NNs, NNs enforcing physics in a weak form, and NNs enforcing physics in a strong form. Thereby, the first class of networks can learn constitutive relations from data while the underlying physics are completely ignored, whereas the latter two are constructed such that they can account for fundamental physics, where special attention is paid to the second law of thermodynamics in this work. Conventional linear and nonlinear viscoelastic as well as elastoplastic models are used for training data generation and, later on, as reference. After training with random walk time sequences containing information on stress, strain, and -- for some models -- internal variables, the NN-based models are compared to the reference solution, whereby interpolation and extrapolation are considered. Besides the quality of the stress prediction, the related free energy and dissipation rate are analyzed to evaluate the models. Overall, the presented study enables a clear recording of the advantages and disadvantages of different NN architectures to model inelasticity and gives guidance on how to train and apply these models.
\end{abstract}

\keywords{neural networks \and recurrent neural networks \and enforcing physics \and constitutive modeling \and thermodynamic consistency \and viscoelasticity \and plasticity}

\section{Introduction}\label{sec:Intro}

Accurately describing the behavior of materials under mechanical loading by constitutive models has been a focus in mechanics research for several decades now.
The formulation and parametrization of constitutive models is however still a challenging task especially for materials showing path-dependent, i.\,e., inelastic, behavior. As an alternative to traditional models, \emph{data-based} or \emph{data-driven} techniques are very promising and have the potential to improve or replace conventional models. These techniques have become increasingly popular in the computational mechanics community during the last years \cite{Bock2019,Montans2019}, where the application of neural networks (NNs) is probably the most common technique. In the following, a brief overview on NNs in constitutive modeling is given.  

The concept of using NNs in constitutive modeling was initially put out by Ghaboussi~et~al.\cite{Ghaboussi1991} in the early 1990s. However, in this early stage, generally pure black-box techniques were employed, i.e., networks that do not account for any physical principles and can therefore only accurately recreate the training data, in this case composed of stress-strain pairs, but perform badly when extrapolating. 
To address this issue, a relatively new approach in \emph{NN-based constitutive modeling}, and scientific \emph{machine learning} (ML) in general, is to integrate crucial underlying physics in either a strong or weak form. These methods, known as \emph{physics-informed}\cite{Raissi2019,Henkes2022}, \emph{mechanics-informed}\cite{Asad2022,Asad2022a}, \emph{physics-augmented}\cite{Klein2022,Linden2023}, \emph{physics-constrained}\cite{Kalina2023}, or \emph{thermodynamics-based}\cite{Masi2021}, improve extrapolation capability and allow for the use of sparse training data.

The easiest material behavior to model is elasticity, since here a suitable model only needs to predict the stresses for specific deformation states.
In the context of ML, the works\cite{Shen2004,Liang2008} seek to approximate the elastic potential by using a \emph{feedforward neural network} (FNN) with three deformation-type invariants as input. Thus, a number of constitutive requirements is fulfilled by construction, e.g., \emph{thermodynamic consistency}, objectivity, or material symmetry. However, training of these models directly requires the elastic potential. 
Meanwhile, FNNs using invariants as input and the hyperelastic potential as output are a very well established approach \cite{Linka2021,Linden2021,Klein2021,Klein2022,Kalina2023,Fuhg2022b,Tac2022a,Linden2023}. Thereby, an improved training is applied that allows calibration of the network directly by tuples of stress and strain, i.e., the derivative of energy with respect to the deformation is included into the loss, which is also called Sobolev training \cite{Vlassis2020,Vlassis2022a}. 
Alternatively, a network previously trained to predict stress coefficients can be used to construct a pseudopotential, thus ensuring thermodynamic consistency of NN-based elastic models a posteriori\cite{Kalina2021}.

Compared to elasticity, the modeling of path-dependent, i.e., inelastic, constitutive behavior by  NN-based approaches is more complex. Some early proposals\cite{Furukawa1998,Ghaboussi1998a,Hashash2004,Al-Haik2006,Jung2006} can already achieve quite good predictions by, for example, adding stress and strain states from previous time steps into the input layer of an FNN\cite{Ghaboussi1998a}. This allows the network to indirectly learn a kind of evolution equation. The model\cite{Furukawa1998} uses internal variables of the material to reliably reproduce stress-strain curves of a viscoplastic material.
Alternatively, load history dependent behavior can also be represented without the availability of the internal variables by so-called \emph{recurrent neural networks} (RNNs), which have been shown to be universal and accurate, particularly for more sophisticated recurrent cells, e.g., according to Hochreiter~and~Schmidhuber\cite{Hochreiter1997}. 
RNNs, especially \emph{long short-term memory} (LSTM) cells, have been intensively used to model inelasticity, e.g., in the works\cite{Ghavamian2019,Wu2020,Li2020a,Fuchs2021}, and are very promising regarding their prediction quality. Very recently, spiking LSTMs, which enable a massive reduction in memory and energy consumption over conventional neural networks, have been applied to model isotropic hardening plasticity\cite{Henkes2022a}.
In addition, a new type of RNN named linearized minimal state cell\cite{Bonatti2022} prevents that its response depends on path-sampling and is therefore advantageous in modeling elastoplasticity. This approach has been used for both 2D and 3D datasets matching to the real mechanical behavior of an aluminum alloy as determined by simulations of crystal plasticity\cite{Bonatti2022a}. 
A further promising approach to represent anisotropic elastoplasticity combines Lie algebra with RNNs\cite{Heider2020}. Finally, although being trained entirely on monotonic data, a hybrid model\cite{Rocha2023} combining a data-driven encoder and a physics-based decoder allows for accurate predictions of elastoplastic unloading/reloading paths.

Despite the great progress in ML-based constitutive modeling, the NN approaches to describing inelasticity mentioned so far are united by their lack of knowledge of the \emph{second law of thermodynamics}. However, following Masi~et~al.\cite{Masi2021}, the incorporation of such fundamental physical principles offers decisive advantages as a more targeted and therefore faster training, that also requires only a small amount of data, and a significantly improved extrapolation capability.
A data-driven framework called \emph{deep material network} is shown in the works \cite{Liu2019a,Liu2019,Gajek2020,Gajek2022}. Thereby, the response of a representative volume element is reproduced by a network including a collection of connected mechanistic building blocks with analytical homogenization solutions, which enables to describe a complex effective response without the loss of essential physics. 
Within the works\cite{Settgast2020,Malik2021,Vlassis2021,Vlassis2021b,Fuhg2022d} the idea to replace parts of classical models with NNs is pursued to achieve this. E.g., the yield
function or the evolution equations are described by FNNs instead of using a particular model. A coupling of NNs to the so-called micro-sphere approach is shown in the work\cite{Zopf2017}.
A more freely formulated approach for the consideration of rate-independent inelasticity based on an adapted network architecture consisting of two FNNs is presented in the works\cite{Masi2021,Masi2022}. Thereby, the first network is used to learn the internal variables' evolution and the second for the approximation of the free energy, where the training procedure requires internal state variables. To account for the thermodynamic consistency, i.e., that the rate of dissipation $\mathcal D$ is always greater equal to zero, this term, which follows from the free energy, is added into the loss function.  
A similar approach tailored for the modeling of inelasticity is shown in the work\cite{He2022}. In contrast to the former model, internal state variables capturing the path-dependency are inferred automatically from the hidden state of an RNN. Thus, this method has the  advantage of requiring only stresses and strains for training. 
The two mentioned models\cite{Masi2021,He2022} nevertheless have the weakness that the requirement $\mathcal D\ge 0$ is not satisfied by design for arbitrary load cases, but is merely enforced by adding a penalty term to the loss function.
Within the works\cite{Huang2022,Asad2022a}, on the other hand, thermodynamic consistency is fulfilled by design of the network architecture. This is achieved by combining the concept of \emph{generalized standard materials}\cite{Miehe2002} with \emph{input convex neural networks} (ICNNs)\cite{Amos2017}. Within the mentioned works, an application to viscoelasticity is shown.  
In a similar approach, finite viscoelasticity is modeled by replacing the Helmholtz free energy function and dissipation potential with data-driven functions that a priori satisfy the second law of thermodynamics, using neural ordinary differential equations (NODEs)\cite{Tac2023}.

After the brief overview given above, it can be summarized that there are a variety of NN-based approaches to modeling inelasticity, with very different levels of incorporated physics. Most approaches were applied exclusively for describing one specific material class, elastoplasticity or viscoelasticity. Thus, this work aims on bringing the different approaches into a uniform framework and comparing them by applying to both elastoplastic as well as viscoelastic data in the 1D case. Thereby, special attention is paid to the fulfillment of the second law of thermodynamics.
With regard to this, a division of the models into three basic classes is done: \emph{black box NNs, NNs enforcing physics in a weak form}, and \emph{NNs enforcing physics in a strong form}. Networks belonging to the first class learn constitutive relations from data while the underlying physics are completely ignored, whereas the latter two are constructed such that they can account for fundamental physical principles. 
However, NNs enforcing physics in a weak form do not necessarily satisfy the second law for arbitrary load cases, which is due to the fact that $\mathcal D\ge 0$ is only integrated into the loss function by a penalty term. In contrast, the network architecture of NNs enforcing physics in a strong form is designed in such a way that this condition is fulfilled in every case, i.e., by construction.
In this paper, conventional linear and nonlinear viscoelastic as well as elastoplastic models are used for training data generation and, later on, as reference. After training with random walk time sequences
containing information on stress, strain, and – for some models – internal variables,
the NN-based models are compared to the reference solution. Besides the quality of
the stress prediction, the predicted free energy and dissipation rate are analyzed to
evaluate the models for both interpolation and extrapolation. In addition to the provided comparison, some of the NN-based models are extended and/or modified at several points, in particular the approaches belonging to NNs enforcing physics in a strong form.

The organization of the paper is as follows: In Sects.~\ref{sec:Classical} and \ref{sec:ANNs}, the basics of constitutive modeling in continuum solid mechanics as well as artificial neural networks are given, respectively. After this, the considered NN-based constitutive models are introduced in Sect.~\ref{sec:Models}. The generation of the database for training is given in Sect.~\ref{sec:Database}. This is followed by a study of the prediction quality of the various NN-based models in Sect.~\ref{sec:App}. After a discussion of the results, the paper is closed by concluding remarks and an outlook to necessary future work in Sect.~\ref{sec:Concl}.

\section{Classical constitutive models}\label{sec:Classical}
    The description of the behavior of materials requires constitutive equations. A framework for the formulation of these equations for different types of material behavior is shown in this section.
    According to Haupt\cite{Haupt2000}, in terms of their behavior, materials can be categorized into the four classes (i) \emph{elasticity}, (ii) \emph{elastoplasticity}, (iii) \emph{viscoelasticity}, and (iv) \emph{viscoelastoplasticity}. Classes (ii) to (iv) are referred to as \emph{inelastic} and exhibit dissipative behavior. In order to ensure the irreversibility of such processes, \emph{thermodynamic consistency} must be taken into account during the formulation of the material laws.
    
    \subsection{General framework}\label{subsec:GenFra}
    \subsubsection{Dissipation inequality}
        Using entropy balance, energy balance and the second law of thermodynamics, an expression known as Clausius-Duhem inequality can be formulated, which, assuming isothermal 1D processes, takes the form
        \begin{equation}
            \label{eq:CDI}
            \sigma\dot{\varepsilon} - \dot{\psi} \geq 0 \quad ,
        \end{equation}
        with $\sigma$ being the stress, $\varepsilon$ the strain and $\psi$ the Helmholtz free energy density, which for the sake of brevity will simply be referred to as free energy in the following. Starting from \refe{eq:CDI}, depending on the particular choice
        of $\psi$, different constitutive models which strictly satisfy the second law can be derived. Generally, the free energy \mbox{$\psi \coloneqq \psi\left( \varepsilon, \xi^1, \xi^2, \dots, \xi^N \right)$} is a function of $\varepsilon$ and the internal variables \mbox{$\xi^{\alpha}, \alpha \in \left\{ 1, \dots, N \right\}$}. This set of internal variables is required to describe the load history dependent internal state of a material point and does not necessarily represent measurable physical quantities. To shorten notation, the internal variables are summarized in the generalized vector $\Ve \xi \in \mathbb R^N$ where appropriate in the following. Applying the principle of equipresence \cite{Haupt2000}, the stress \mbox{$\sigma \coloneqq \sigma\left( \varepsilon, \Ve \xi \right)$} is assumed to be a function of the same set of variables. Evaluating \refe{eq:CDI} yields
        \begin{equation}
            \label{eq:CDI2}
            \mathcal{D}\coloneqq\left(\sigma - \frac{\partial\psi}{\partial\varepsilon} \right) \dot{\varepsilon} - \smallunderbrace{\frac{\partial\psi}{\partial\Ve\xi}}_{\eqqcolon -\Ve\tau}\cdot\dot{\Ve\xi} \geq 0 \quad ,
        \end{equation}
        where $\Ve a \cdot \Ve b$ is the scalar product of two vectors $\Ve a, \Ve b \in \mathbb R^N$.
        The quantities $\mathcal{D}$ and $\Ve\tau \in \mathbb R^N$ denote the dissipation rate and the vector of \emph{thermodynamic conjugate forces}, also called \emph{internal forces}\cite{Miehe2002}, with respect to the internal variables $\Ve\xi$. In order to comply with inequality \eqref{eq:CDI2}, the necessary and sufficient conditions
        \begin{equation}
            \label{eq:Cond}
            \mathcal{D}\geq 0 \quad\Longleftrightarrow\quad \sigma = \frac{\partial\psi}{\partial\varepsilon}\quad \land \quad \Ve\tau \cdot \dot{\Ve\xi} = \mathcal{D} \geq 0
        \end{equation}
        arise.
        Thus, the evolution equations for the internal variable are yet to be defined such that $\mathcal{D} \geq 0$ is ensured at all times. 
        
        \subsubsection{Generalized standard materials}\label{subsubsec:Biot}
        A common way to formulate the necessary evolution equations is to use the concept of \emph{generalized standard materials}\cite{Miehe2002,Miehe2002a,Miehe2011a} which is briefly explained in the following. Within this concept, in addition to $\psi$, a \emph{dissipation potential} $\phi:=\phi(\dot{\Ve \xi},\Ve \xi,\varepsilon)$ which is defined to be (i) \emph{convex} with respect to its first argument $\dot{\Ve \xi}$ and is additionally normalized with respect to $\dot{\Ve{\xi}}$, i.e., it fulfills the conditions (ii) \mbox{$\phi(\Ve 0,\Ve\xi,\varepsilon)=0$} and (iii) \mbox{$\phi(\Ve{0},\Ve\xi,\varepsilon)\ge 0$} is a minimum in $\dot{\Ve{\xi}}$, is introduced. The dissipation potential may be non-smooth for rate independent, i.\,e. elastoplastic, materials. The internal forces can now be determined from $\phi$ according to $\Ve\tau \in \partial_{\dot{\Ve\xi}} \phi$, where the operator $\partial_{\dot{\Ve \xi}}(\cdot)$ denotes the subdifferential of a non smooth convex function. On the other hand, if $\phi$ is smooth, the relation changes to $\Ve\tau = \partial_{\dot{\Ve\xi}} \phi$, where the introduced operator now represents the standard partial derivative. Thus, using Eq.~\eqref{eq:CDI2}, it follows the Biot equation which describes the internal variables' evolution and is given by 
        \begin{align}
            \Ve 0 \in \partial_{\Ve\xi} \psi + \partial_{\dot{\Ve \xi}} \phi 
            \quad \text{or} \quad 
            \Ve 0 = \partial_{\Ve\xi} \psi + \partial_{\dot{\Ve \xi}} \phi 
            \quad \text{with} \quad \Ve\xi(t=0) = \Ve\xi_0
            \label{eq:Biot}
        \end{align}
        for rate independent and rate dependent constitutive behavior, respectively. Note that the inequality  given in Eq.~\eqref{eq:Cond}, i.\,e., $\mathcal D \ge 0$, is automatically fulfilled by Eq.~\eqref{eq:Biot} due to the stated requirements on $\phi$, i.\,e., convexity and normalization. An alternative formulation follows with the \emph{dual dissipation potential} $\phi^*$ obtained by the Legendre-Fenchel transformation
        \begin{align}
            \phi^*(\Ve\tau,\Ve \xi,\varepsilon) := \underset{\dot{\Ve\xi}}{\sup} \left[\Ve\tau \cdot \dot{\Ve \xi} - \phi(\dot{\Ve \xi},\Ve \xi,\varepsilon)\right] \; .
            \label{eq:Legendre}
        \end{align}
        By using the dual dissipation one gets 
        \begin{align}
            \dot{\Ve \xi} \in \partial_{\Ve \tau} \phi^* 
            \quad \text{or} \quad 
            \dot{\Ve \xi} = \partial_{\Ve\tau} \phi^* 
            \quad \text{with} \quad \Ve \xi(t=0) = \Ve\xi_0 \quad ,
            \label{eq:BiotDual}
        \end{align}
        respectively, instead of Eq.~\eqref{eq:Biot}.
        
		\begin{remark}\label{remark:constraints}
         It should be noted, that Eqs. \eqref{eq:Biot} and \eqref{eq:BiotDual} in this stated form only hold if the rates of the internal variables or the internal forces, respectively, are independent from each other. If there are constraints between the individual quantities, these must be explicitly taken into account during the evaluation of Eqs. \eqref{eq:Biot} and \eqref{eq:BiotDual}. Otherwise, the dissipation potential might for example be expressed in terms of a reduced set of internal variable rates and the partial derivatives with respect to the omitted rates yield zero. This is the case, e.g., for the elastoplastic model as formulated in \reft{tab:analyt_models}.
         \end{remark}

        Within the special case of \emph{rate-independent} constitutive behavior, the dissipation function $\phi$ is now obtained by using the \emph{concept of maximum dissipation}\cite{Miehe2002,Miehe2011a}. Thus, it is defined by the constrained maximization problem 
        \begin{align}
            \phi(\dot{\Ve \xi},\Ve \xi,\varepsilon) := \underset{\Ve\tau\in\mathcal E}{\sup}\left(\Ve\tau\cdot\dot{\Ve\xi}\right)
            \quad \text{with} \quad \mathcal E := \left\{ \Ve \tau \in \mathbb R^N \; | \; f(\Ve \tau, \Ve \xi) \le 0 \right\} \; .
            \label{eq:maxDissipation}
        \end{align}
        In the equation above, $\mathcal E$ denotes the admissible domain of internal forces with the yield function $f(\Ve \tau,\Ve \xi)$ which is assumed to be convex with respect to $\Ve \tau$, normalized and homogeneous of degree one.
        The solution of Eq.~\eqref{eq:maxDissipation} yields the evolution equations of internal variables together with the Karush-Kuhn-Tucker conditions:
        \begin{align}
            \dot{\Ve\xi} = \lambda \partial_{\Ve \tau} f \; \wedge \; \lambda \ge 0 \; \wedge \; f \le 0 \; \wedge \; \lambda f = 0 \; .
        \end{align}
        Therein, the scalar $\lambda \in \mathbb R_+$ denotes the plastic multiplier.

    \subsection{Specific constitutive models}
    The outlined framework given in Sect.~\ref{subsec:GenFra} can be applied to describe, e.g., viscoelasticity or elastoplasticity, for which the corresponding models are briefly summarized in the following.
        \subsubsection{Viscoelasticity}
        \label{subsubsec:Viscoelasticity}
            \begin{figure}[t]
    			\centering
    			\graphicspath{{images/Classical/}}
    			\newlength\figw
		        \newlength\figh 
    			\setlength{\figw}{4cm}
    			\setlength{\figh}{4cm}
                \subfloat[\label{subfig:ViscoLinRheo}]{\includegraphics{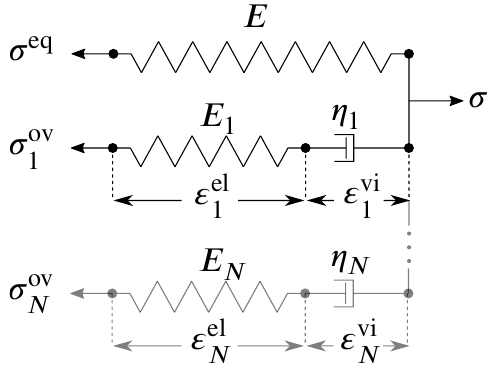}}
    			\hfil
    			\graphicspath{{images/Classical/}}
    			\setlength{\figw}{4cm}
    			\setlength{\figh}{4cm}
    			\subfloat[\label{subfig:ViscoNonLinRheo}]{\includegraphics{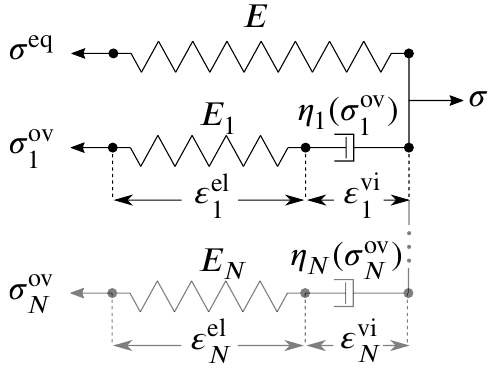}}
    			\hfil
    			\subfloat[\label{subfig:ViscoLinResp}]{\includegraphics{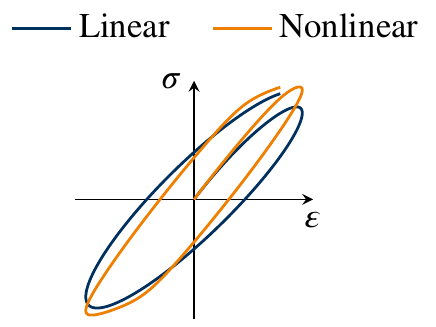}}
    			\caption{Viscoelastic constitutive model: Rheological models of \textbf{(a)} a linear generalized Maxwell model and \textbf{(b)} a nonlinear generalized Maxwell model and \textbf{(c)} typical hystereses for (a) and (b).}
                \label{fig:ViscoBoth}
            \end{figure}
        
            Viscoelastic behavior of solids is characterized by a strain rate dependent stress response with elastic equilibrium curve, see \reff{subfig:ViscoLinResp}. These properties can be described by a \textit{generalized Maxwell model}\cite{Haupt2000} as shown in \reff{subfig:ViscoLinRheo}, where $E$ and $E_\alpha$ are the Young's moduli of the springs within the rheological model, $\eta_\alpha$ the respective viscosities, $\varepsilon_\alpha^{\text{el}}$ and $\varepsilon_\alpha^{\text{vi}}$ the elastic and viscous strains, and $\sigma_\alpha^{\text{ov}}$ the non-equilibrium stresses also denoted as overstresses. The viscosity $\eta_\alpha$ may be a function of the overstress $\sigma_\alpha^{\text{ov}}$, see \reff{subfig:ViscoNonLinRheo}. A model with overstress dependent viscosities is referred to as nonlinear in the following and linear otherwise. The behaviour of such a generalized Maxwell model is described by the governing equations in Tab.~\ref{tab:analyt_models}.\\

            \begin{table}
                \caption{Governing equations of the considered viscoelastic and elastoplastic models.}
                \centering
                \begin{small}
                \begin{tcolorbox}[width=1.0\linewidth, arc=1.5em] 
                    \renewcommand{\arraystretch}{1.5}
                    \begin{tabular}{rll}
                    & \textbf{Viscoelasticity} & \textbf{Elastoplasticity}\\
                    \rule{0pt}{4ex}\textbf{Free energy}: & $\psi \coloneqq \frac{1}{2}E\varepsilon^2 + \sum_{\alpha=1}^{N}\frac{1}{2}E_\alpha\left(\varepsilon - \varepsilon_\alpha^{\text{vi}}\right)^2$ & $\psi \coloneqq \frac{1}{2}E\left( \varepsilon - \varepsilon^{\text{pl}} \right)^2 +  \frac{1}{2}H\alpha^2 +  \frac{1}{2}\hat{H}\hat{\alpha}^2$\\
                    \textbf{Dissipation potential}: & $\phi = \sum_{\alpha=1}^N \frac{\eta_\alpha}{2}\left( \dot\varepsilon^{\text{vi}}_\alpha \right)^2$ & $\phi = \sigma_{\text{y}_0} |\dot{\varepsilon}^{\text{pl}}|$  with $\dot{\alpha}=\dot{\varepsilon}^{\text{pl}}$ and $\dot{\hat{\alpha}}=|\dot{\varepsilon}^{\text{pl}}|$\\
                    \textbf{Dual dissipation potential}: & $\phi^* = \sum_{\alpha=1}^N \frac{1}{2\eta_\alpha}\left( \sigma_\alpha^{\text{ov}}\right)^2$ & 
                    $\phi^* = \begin{cases}
                        0, & \text{if} \phantom{m} (\sigma^\text{pl},p,\hat p)\in \mathcal{E}\\[-5pt]
                        \infty, & \text{else}
                        \end{cases}\, $\\
                    \textbf{Stress}: & $\sigma = E\varepsilon + \sum_{\alpha=1}^N E_\alpha\left(\varepsilon-\varepsilon_\alpha^{\text{vi}}\right)$ &  $\sigma = \partial_{\varepsilon}\psi = E\left( \varepsilon - \varepsilon^{\text{pl}} \right)$\\
                    \textbf{Internal forces}: & $\sigma^{\text{ov}}_\alpha = -\partial_{\varepsilon^{\text{vi}}_\alpha}\psi = E_\alpha\left(\varepsilon-\varepsilon_\alpha^{\text{vi}}\right)$ & \makecell[l]{ $\sigma^{\text{pl}}=-\partial_{\varepsilon^{\text{pl}}}\psi=\sigma \phantom{m} , \phantom{m} p=-\partial_{\alpha}\psi=-H\alpha \phantom{m} ,$ \\ $ \hat{p}=-\partial_{\hat{\alpha}}\psi=-\hat{H}\hat{\alpha}$}\\
                    \textbf{Evolution equations}: & $\dot{\varepsilon}_\alpha^{\text{vi}} \coloneqq \frac{1}{\eta_\alpha}\sigma_\alpha^{\text{ov}}$ & $\dot{\varepsilon}^{\text{pl}} = \dot{\alpha} = \lambda\sign\left(\sigma+p\right) \phantom{m} , \phantom{m} \dot{\hat{\alpha}} = \lambda$\\
                    \textbf{Dissipation rate}: & $\mathcal{D} = \sum_{\alpha=1}^N \sigma_\alpha^{\text{ov}}\dot{\varepsilon}_\alpha^{\text{vi}} \geq 0$ & $\mathcal{D} = \sigma\dot{\varepsilon}^{\text{pl}}+ p\dot{\alpha} + \hat{p}\dot{\hat{\alpha}} \geq 0$ \\
                    \textbf{Viscosity}: & $\eta_\alpha\left(\sigma^{\text{ov}}_\alpha\right) \coloneqq \hat{\eta}_\alpha \exp{\left(a_\alpha\left|\sigma^{\text{ov}}_\alpha\right|^{b_\alpha}\right)}$ & \hspace{2.5cm}---\\
                    \textbf{Yield function}: & \hspace{2cm}--- & $f \coloneqq \left| \sigma+p \right| - \sigma_\text{y}(\hat{p}) \phantom{m} , \phantom{m}  \sigma_\text{y}(\hat{p}) = \sigma_{\text{y}_0} - \hat{p}$\\
                    \textbf{Karush-Kuhn-Tucker conditions}: & \hspace{2cm}--- & $\lambda \ge 0 \; \wedge \; f \le 0 \; \wedge \; \lambda f = 0 \;$\\
                    \end{tabular}
                    \renewcommand{\arraystretch}{1}
                \end{tcolorbox}
                \end{small}                
            \label{tab:analyt_models}
            \end{table}

        \subsubsection{Elastoplasticity}
        \label{subsubsec:Elastoplasticity}
            In contrast to viscoelasticity, elastoplastic material behavior is characterized by strain rate independence but the presence of an equilibrium hysteresis\cite{Haupt2000}. Within an initial region \mbox{$\sigma\in\left(-\sigma_{\text{y}_0},\,\sigma_{\text{y}_0}\right)$}, the material acts purely elastic. If the limits of this initial region are exceeded, i.e., \mbox{$|\sigma | > \sigma_{\text{y}_0}$}, elastic-plastic deformation occurs and the elastic region can be both, shifted in the respective direction (\emph{kinematic hardening}) as well as expanded (\emph{isotropic hardening})\cite{Simo2000}. In order to model these properties, a so-called yield function \mbox{$f\coloneqq f\left(\Ve \tau, \Ve{\xi} \right) \leq 0$} is defined. It determines the boundaries of the elastic region such that the deformation is purely elastic if \mbox{$f < 0$} and elastic-plastic if \mbox{$f = 0$}. The case \mbox{$f > 0$} is not admissible.
            The evolution equations can be obtained from the maximum dissipation principle, see Sect.~\ref{subsec:GenFra}. Note, that the rates of the internal variables are not independent from each other, which has to be taken into account when evaluating \refe{eq:Biot}, as explained in Remark \ref{remark:constraints}.
            With that, the governing equations of elastoplastic behavior with mixed kinematic and isotropic hardening are derived, cf. Tab.~\ref{tab:analyt_models}. The corresponding rheological model is shown in \ref{subfig:RheoPlast}. Therein, $E$ is the Young's modulus, $H$ and $\hat{H}$ are the kinematic and isotropic hardening modulus, respectively, and $\sigma_{\text{y}_0}$ is the initial yield stress. Moreover, $p$ denotes the back stress and $\varepsilon^{\text{el}}$ and $\varepsilon^{\text{pl}}$ the elastic and plastic strain. 
            
            Setting the hardening modules $H$ or $\hat{H}$ to zero results in the special cases with ideal plasticity (\mbox{$H=\hat{H}=0$}), only kinematic hardening (\mbox{$\hat{H}=0$}) or only isotropic hardening (\mbox{$H=0$}) see Figs.~\ref{subfig:idePl}--\ref{subfig:kiniso}. The corresponding internal variables of the hardening modules set to zero then still take on values unequal to zero, but no longer have any influence on $\sigma$, $\psi$ or $\mathcal{D}$ and can therefore be neglected. For instance, in the case of ideal plasticity \mbox{$H=\hat{H}=0$}, the set of internal variables can be reduced to the  plastic strain $\varepsilon^{\text{pl}}$.
            
            {
    		\begin{figure}[t]
    			\graphicspath{{images/Classical/}}
                \centering
    			\setlength{\figw}{3cm}
    			\setlength{\figh}{3cm}
    			\subfloat[\label{subfig:RheoPlast}]{\includegraphics{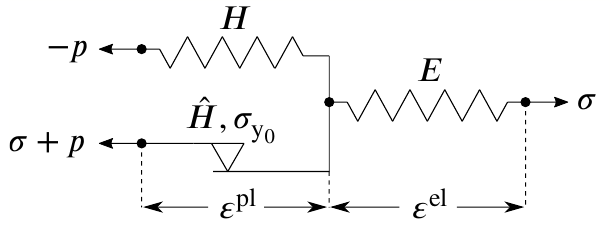}}
    			\hfil
    			\subfloat[\label{subfig:idePl}]{\includegraphics{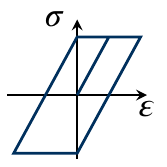}}
                \hfil
    			\subfloat[\label{subfig:kin}]{\includegraphics{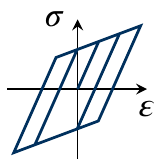}}
                \hfil
    			\subfloat[\label{subfig:iso}]{\includegraphics{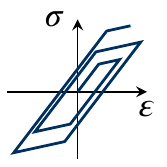}}
                \hfil
    			\subfloat[\label{subfig:kiniso}]{\includegraphics{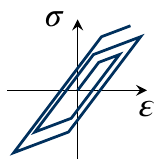}}

    			\caption{Elastoplastic constitutive model: \textbf{(a)} Rheological model with isotropic and kinematic hardening and typical hystereses for \textbf{(b)} ideal plasticity ($H=\hat{H}=0$), \textbf{(c)} kinematic hardening ($\hat{H}=0$), \textbf{(d)} isotropic hardening ($H=0$) and \textbf{(e)} kinematic and isotropic hardening.}
    			\label{fig:plast}
    		\end{figure}
    	    }

\section{Basics of artificial neural networks}\label{sec:ANNs}
NNs can be divided into different classes, each adapted to different types of tasks\cite{Kruse2016,Kollmannsberger2021}. For constitutive modeling, the use of FNNs and RNNs is particularly suitable.

\subsection{Feedforward neural networks}
{
	\begin{figure}[b]
		\centering
		\graphicspath{{images/NN/}}
		\subfloat[\label{subfig:netw}]{\includegraphics{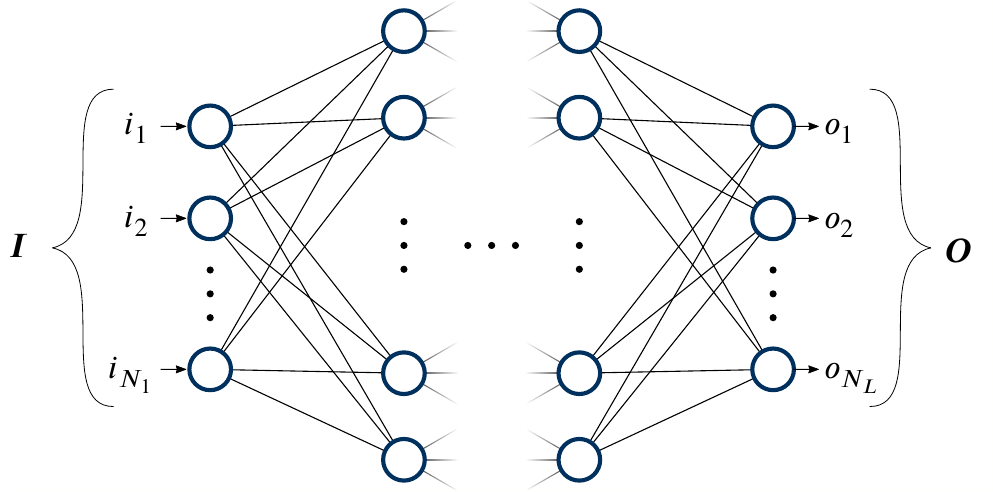}} 
		\hfil
		\subfloat[\label{subfig:neuron}]{\raisebox{1cm}{\includegraphics{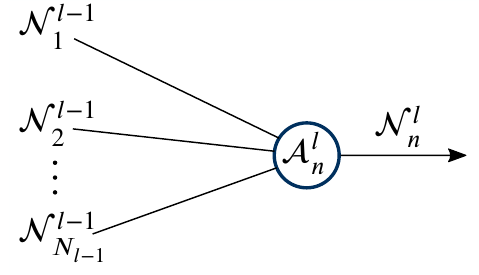}}}
		\caption{Structure of an FNN: \textbf{(a)} General representation of a full network with arbitrary number of layers and neurons and \textbf{(b)} functionality of the $n$th neuron of layer $l$.}
		\label{fig:neur}
	\end{figure}
}
FNNs are composed of artificial neurons arranged in several layers as shown in \reff{subfig:netw}. The information of the input variables (features) $i_n$, summarised in the input \mbox{$\Ve{I} \coloneqq {(\,i_1,\ldots,i_{N_1}\,)}$}, is passed on to each neuron of the subsequent layer 2 and is used in these to calculate the outputs of this layer. The entire output of layer 2 in turn serves as input for each neuron in the next layer and so forth. The forward flow of information from the input layer to the final output \mbox{$\Ve{O} \coloneqq {(\,o_1,\ldots,o_{N_L}\,)}$} of layer $L$ without feedback of intermediate results is called \textit{feedforward}. The $n$th neuron of layer $l$ is shown in \reff{subfig:neuron}. The output $\mathcal{N}^l_n$ of this neuron is determined via
\begin{equation}
	\label{eq:Output}
	\mathcal{N}^l_n =  \mathcal{A}^l_n\left( x^l_n \right) = \mathcal{A}^l_n\left( \sum_{m=0}^{N_{l-1}} \mathcal{N}^{l-1}_m w^l_{nm} \right) \quad .
\end{equation}
Therein, $\mathcal{A}^l_n$ denotes the activation function of this neuron, $N_{l-1}$ the number of neurons in layer $l-1$ and $w^l_{nm}$ the weight of the output $\mathcal{N}_m^{l-1}$ in the argument 
of the activation function $\mathcal{A}^l_n$. The weight $w^{l}_{n0}$ of the output of the imaginary neuron \mbox{$\mathcal{N}^{l-1}_{0} \equiv 1$} is referred to as the bias of the neuron. Common activation functions are the hyperbolic tangent $\tanh(x)$, the rectifier \mbox{$\relu(x)\coloneqq \max\left(0, x \right)$} or the softplus activation $\softplus(x) \coloneqq \ln(1+\exp(x))$\cite{Kruse2016,Kollmannsberger2021,Masi2021}.

Particularly, the outputs $\mathcal{N}^L_n$ of the output layer $L$ can also be expressed with \eqref{eq:Output}. Recursively, the $\mathcal{N}^l_n$ can be replaced until layer 1 with its inputs \mbox{$\mathcal{N}^1_n=i_n$} is reached. This clarifies that such a network may be arbitrarily nested, but for fixed $w^l_{nm}$ is  a well-defined function of the input $\Ve{I}$. In this sense, the network as a function
\begin{equation}
	\label{eq:funktionNN}
	\mathcal{F} : \mathbb{R}^{N_1} \rightarrow \mathbb{R}^{N_L} \quad , \quad \Ve{I}\mapsto \mathcal{F}(\Ve{I}) = \Ve{O}
\end{equation}
maps the input $\Ve{I}$ to the output $\Ve{O}$. In order to adapt the weights $w^l_{nm}$ such that the predictions of the NN match the expected values, a training data set \mbox{$\mathcal{S}\coloneqq\left\{ \mathcal{T}_1,\ldots\,,\mathcal{T}_{N^{\text{ds}}} \right\}$} consisting of $N^{\text{ds}}\in \mathbb{N}$ data tuples $\mathcal{T}_\alpha$ is required. In each of these data tuples \mbox{$\mathcal{T}_\alpha\coloneqq (\,\Ve{I}_\alpha\,,\,\bar{\Ve{O}}_\alpha\,)$}, an Input $\Ve{I}_\alpha$ is assigned its expected output $\bar{\Ve{O}}_\alpha$.
The expected output usually contains the desired values of the neurons in the last layer, but may also contain additional information, such as desired derivatives with respect to a certain input. 
The error of the predictions of an NN with respect to this set of data tuples is summarized in the loss function $\mathcal{L}\left( \Ve{w}, \mathcal{S} \right)$,  where $\Ve{w}$ denotes the vector of all weights $w^l_{nm}$ of the NN. For the applications shown here, the loss function is mostly composed of mean absolute error terms, which for FNN read 
\begin{equation}
	\mae\left(q\right) \coloneqq \frac{1}{N^{\text{ds}}} \sum_{i=1}^{N^{\text{ds}}}\left| q(\Ve I_i,\Ve{w}) - \bar{q}_i \right| \quad .
\end{equation}
Therein, $q$ is the quantity, whose prediction is evaluated.
The minimization of this error function with respect to the weights $\Ve{w}$
\begin{equation}
	\label{eq:Opt}
	\Ve{w} = \arg \min_{\VeGr{\kappa}} \mathcal{L}\left( \VeGr{\kappa}, \mathcal{S} \right)
\end{equation}
is called the training process of the NN. Various methods can be used to solve this minimization problem, e.g. \textit{Stochastic Gradient Descent} (SGD) or \textit{Adam}
, to name two of the most common optimizers. Within this work, \textit{Sequential Least Squares Programming} (SLSQP) is used for the optimization.

A special class of FNNs are ICNNs initially proposed by Amos~et~al.\cite{Amos2017}, i.e., networks that have the property to be convex with respect to their input arguments. This is achieved by using a convex and non-decreasing activation function and non-negative weights.

\subsection{Recurrent neural networks}
{
	\begin{figure}[b]
		\centering
		\graphicspath{{images/NN/}}
		\subfloat[\label{subfig:RNN}]{\includegraphics{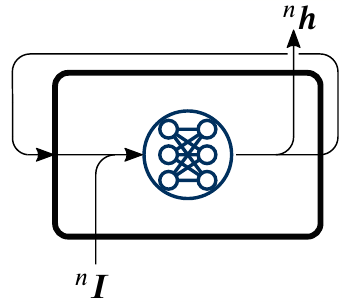}}
		\hfil
		\subfloat[\label{subfig:RNNtime}]{\raisebox{0.25cm}{\includegraphics{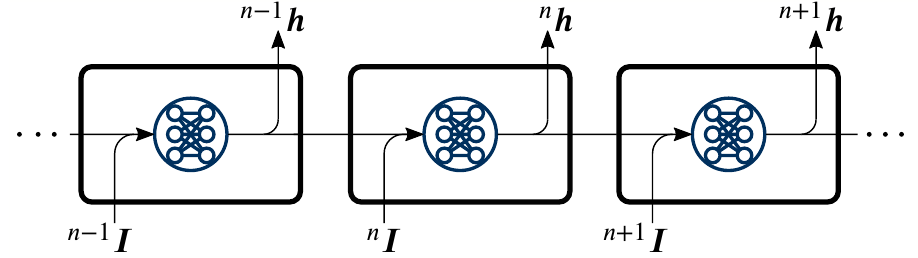}}}
		\caption{Functionality and internal structure of a standard RNN cell in two equivalent illustrations: the output is \textbf{(a)} fed back into the RNN cell and \textbf{(b)} passed from one time step to the next. I.e., (b) can be understood as (a) unfolded in time.}
		\label{fig:RNN1}
	\end{figure}
}

In contrast to FNNs, RNNs allow intermediate results to be fed back, see \reff{subfig:RNN}. This feature enables the network to take into account not only one, but several time steps for the output and thus to learn to interpret the input variables in their (pseudo) temporal context. This can be particularly useful, for example, in text translation, speech recognition or prediction of the evolution of physical quantities. Thus, in RNNs, entire sequences of successive data are evaluated, which in the applications shown within this work are sets of time steps. RNNs are thus very different from FNNs in the way they operate. Nevertheless, their internal structure is also composed of a single or multiple FNNs.

These FNNs receive the current input vector ${}^{n}\Ve{I}$ and the so called hidden state ${}^{n-1}\Ve{h}$ of the previous time step and output the hidden state ${}^{n}\Ve{h}$ of the current time step. Outside of the recurrent cell, the new hidden state is subsequently processed further to obtain the output ${}^{n}\Ve{O}$. \reff{subfig:RNNtime} illustrates this process for a simple RNN cell unfolded in time.

The depicted cell consists of only a single FNN and is not capable of incorporating many preceding time steps into the output of the current hidden state. Therefore, more complex internal structures have been developed using multiple FNNs in order to enable the cell to detect long term dependencies. The RNNs presented in this work use the so-called \emph{long short-term memory} (LSTM)\cite{Hochreiter1997} cell to overcome this problem. Besides the hidden state, this LSTM cell passes another set of information, the so-called cell state ${}^{n}\Ve{c}$, from time step to time step. The number $N^\text{c}$ of entries in ${}^{n}\Ve{c}$ can be interpreted as the memory capacity of the cell and is a crucial parameter for the network performance. 
In contrast to FNNs, the training data set \mbox{$\mathcal{S}\coloneqq\left\{ \mathcal{T}_1,\ldots\,,\mathcal{T}_{N^{\text{seq}}} \right\}$} contains \mbox{$N^{\text{seq}}\in\mathbb{N}$} tuples \mbox{$\mathcal{T}_s\coloneqq(\mathcal{T}_s^{\text{I}}, \bar{\mathcal{T}}_s^{\text{O}})$} of input sequences $\mathcal{T}_s^{\text{I}}$ and assigned expected output sequences $\bar{\mathcal{T}}_s^{\text{O}}$. Each input sequence \mbox{$\mathcal{T}_s^{\text{I}}\coloneqq({}^{1}\Ve{I}_s,\ldots,{}^{N_s^{\text{ts}}}\Ve{I}_s)$} is an ordered series of \mbox{$N_s^{\text{ts}}\in\mathbb{N}$} input vectors, 
each representing one of the $N_s^{\text{ts}}$ time steps of the sequence. The expected output sequence \mbox{$\bar{\mathcal{T}}_s^{\text{O}}\coloneqq({}^{1}\bar{\Ve{O}}_s,\ldots,{}^{N_s^{\text{ts}}}\bar{\Ve{O}}_s)$} contains the same number of expected output vectors.\\ 
The error of the RNNs predictions is measured in the loss function $\mathcal{L}\left( \Ve{w}, \mathcal{S} \right)$, where $\Ve{w}$ now contains the weights of all FNNs inside and outside the recurrent cell. For RNN, the mean absolute error measure of a quantity $q$ takes the form
\begin{equation}
	\mae\left(q\right) \coloneqq \frac{1}{N^{\text{seq}}} \sum_{s=1}^{N^{\text{seq}}} \left( \frac{1}{N_s^{\text{ts}}} \sum_{n=1}^{N_s^{\text{ts}}} \left| q(\mathcal{T}^\text{I}_s,\Ve{w}) - {}^{n}\bar{q}_s \right| \right) \quad .
\end{equation}
The training process, similar to the FNNs, is performed as minimization of the loss function with respect to all weights present in the network.

\section{NN-based constitutive models}\label{sec:Models}
Although the models presented in \refs{sec:Classical} are able to reproduce essential aspects of real material behavior in a physically consistent manner, the applied approach to find a set of governing equations has some disadvantages. On the one hand, this set of equations has to be formulated manually, which becomes increasingly difficult for more complex materials in three dimensions and, on the other hand, it can only reproduce the phenomena taken into account during modeling. In this section, some possible methods are presented that, based on NNs, are intended to find arbitrary correlations automatically, i.e., without any additional manual modeling effort. These methods are divided into three categories, which describe the amount of incorporated physics: (i) the \emph{black box models} with no physical information, (ii) \emph{NNs enforcing physics in a weak form}, which are trained to satisfy the second law and (iii) \emph{NN-based models enforcing physics in a strong form}, in which the second law is satisfied a priori.
Each method requires a training data set, whose generation is outlined in \refs{sec:Database}. The implementation is done in Python using the Tensorflow library, with the results being presented in \refs{sec:App}.

\subsection{Black box models}
Black box models are characterized by their lack of incorporated physical knowledge, i.e., they take a set of input values (depending on the concrete model) and output the desired quantity (the stress) without respecting relationships known from continuum mechanics. Two simple architectures are examined, one using a feedforward network, the other using a recurrent cell.
\subsubsection{Feedforward architecture with stress as output (FNN$^\sigma$)}\label{subsubsec:FFsig}

This architecture is the simplest of the six architectures presented in this work and consists of only a single FNN, which outputs the stress directly. It receives the new strain ${}^{n+1}\varepsilon$ as well as strains ${}^{m}\varepsilon$, stresses ${}^{m}\sigma$ and time increments \mbox{${}^{m}\Delta t \coloneqq {}^{m+1}t - {}^{m}t$} of an arbitrary number of \mbox{$N^{\text{pt}}\geq 1$} preceding time steps and outputs the new stress, that is
\begin{equation}
	\label{eq:IFFsig}
	\mbox{$\Ve{I}\coloneqq {(\,{}^{n+1}\varepsilon, {}^{n}\varepsilon,{}^{n}\sigma,{}^{n}\Delta t,\ldots,{}^{n+1-N^{\text{pt}}}\varepsilon,{}^{n+1-N^{\text{pt}}}\sigma,{}^{n+1-N^{\text{pt}}}\Delta t\,)} \mapsto O \coloneqq {}^{n+1}\sigma \quad .$}
\end{equation}
The time steps \mbox{${}^{m}\Delta t$} in the input are only necessary for rate-dependent behavior, i.e., for viscoelasticity.
The loss function contains only the stress prediction,
\begin{equation}
	\label{eq:LossFFsigma}
	\mathcal{L} \coloneqq \mathcal{L}^{\sigma} \coloneqq \mae(\sigma) \quad .
\end{equation}
Similar approaches, e.g., the nested adaptive neural networks for constitutive modeling, have been described earlier\cite{Ghaboussi1998a,Hashash2004,Jung2006}. However, as the results in \refs{sec:App} show, this method can only be applied to specific classes of material behavior. An improved architecture based on a recurrent cell can be used to overcome this restriction.

\subsubsection{Recurrent architecture with stress as output (RNN$^\sigma$)}
\begin{figure}[h]
	\centering
	\begin{minipage}{0.45\linewidth}
		\centering
		\graphicspath{{images/Architectures/}}
		\includegraphics{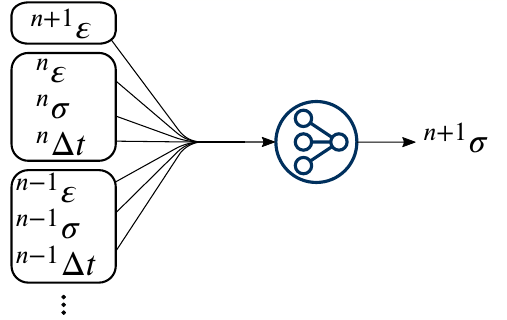}
		\caption{Functionality of the FNN$^\sigma$ architecture: A single FNN maps information from multiple time steps to the new stress.}
		\label{fig:my_label}
	\end{minipage}%
	\hfil
	\begin{minipage}{0.5\linewidth}
		\centering
		\graphicspath{{images/Architectures/}}
		\includegraphics{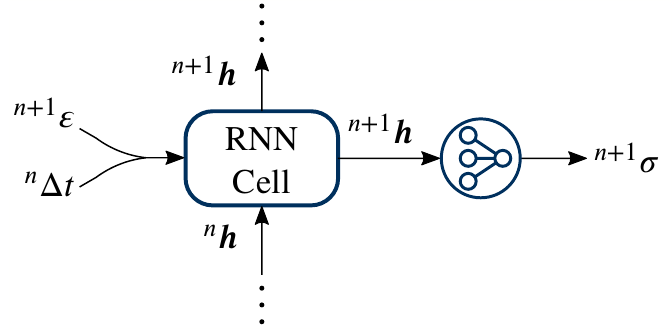}
		\caption{Functionality of the RNN$^\sigma$ architecture: Based on new strain and time increment, the RNN cell outputs a new hidden state, which is fed into a FNN to predict the new stress.}
		\label{fig:RNNsig}
	\end{minipage}
\end{figure}

The second black box architecture consists of a \emph{recurrent cell} followed by an FNN as shown in \reff{fig:RNNsig}. Such models are used in the works\cite{Wu2020}.
For each time step in every sequence, the RNN cell takes as input the new strain and if necessary the time increment and outputs the new hidden state, which is subsequently fed into the FNN to receive the new stress,
\begin{equation}
	{}^{n}\Ve{I}^{\text{RNN}}\coloneqq {(\,{}^{n}\varepsilon, {}^{n-1}\Delta t\,)} \mapsto {}^{n}\Ve{h} \eqqcolon {}^{n}\Ve{I}^{\text{FNN}} \mapsto {}^{n}O^{\text{FNN}} \coloneqq {}^{n}\sigma \quad .
\end{equation}
The initial state before the first time step of hidden state as well as cell state are set to \mbox{${}^{0}\Ve{h}=\Ve{0}$} and \mbox{${}^{0}\Ve{c}=\Ve{0}$}. The loss function
\begin{equation}
	\mathcal{L} \coloneqq \mathcal{L}^{\sigma} \coloneqq \mae(\sigma)
\end{equation}
averages the error in the stress prediction over all time steps and sequences.

This architecture is capable of modeling a broad variety of materials. Nevertheless, neither is any physical knowledge provided to the network nor can any additional physical information be obtained besides the stress response. Therefore, two physically informed models are presented below.

\subsection{Neural networks enforcing physics in a weak form}
In contrast to black box models, \emph{NN-based models enforcing physics in a weak form} take into account known relations, namely \refe{eq:Cond}$_2$ and \refe{eq:Cond}$_3$. This aims at making predictions that are actually physically consistent and gaining additional information, particularly the free energy and the dissipation rate. The networks are trained to respect these relations, but the architecture itself cannot guarantee $\mathcal D\ge 0$ a priori.
To incorporate the relations into the training process, the network must be differentiated with respect to its inputs. The differentiation of a network results in restrictions regarding the choice of activation functions: since gradients of the loss function are required during the optimization, the activation functions are differentiated a second time. Consequently, the activations must be differentiable twice to enable an optimization\cite{Masi2021,Masi2022,Kalina2023}. Therefore, the hyperbolic tangent and the softplus activation are used for the corresponding networks in the following.

\subsubsection{Feedforward architecture with internal variables and free energy as output (FNN$^{\xi+\psi}$)}

\begin{figure}[h]
	\centering
	\graphicspath{{images/Architectures/}}
	\includegraphics{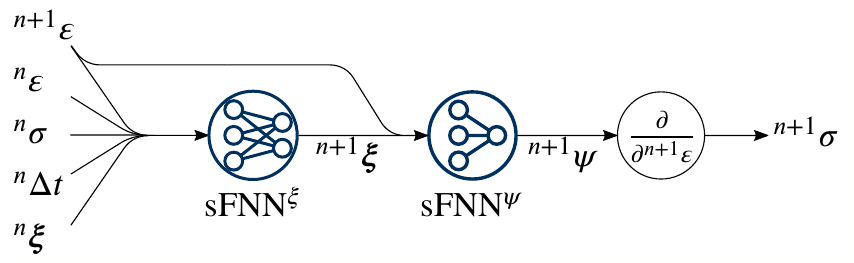}
	\caption{Functionality of the FNN$^{\xi+\psi}$ architecture: Based on the input values, the first subnetwork sFNN$^\xi$ predicts a new set of internal variables, which alongside the new strain is passed on to the second subnetwork sFNN$^\psi$ to predict the new free energy. Differentiating with respect to the new strain yields the new stress.}
	\label{fig:FF2}
\end{figure}

The architecture described below is based on Masi~et~al.\cite{Masi2021} and is studied in detail therein. It uses two FNNs with different tasks, as shown in \reff{fig:FF2}. The first subnetwork sFNN$^\xi$ receives the new and old strain, old stress as well as the time increment and internal variables of the previous time step and is trained to predict new internal variables:
\begin{equation}
	\label{eq:InputFF21}
	\Ve{I}^{\text{sFNN}^\xi}\coloneqq {(\,{}^{n+1}\varepsilon, {}^{n}\varepsilon,{}^{n}\sigma,{}^{n}\Delta t, {}^{n}\xi^1,\ldots,{}^{n}\xi^N \,)} \mapsto \Ve{O}^{\text{sFNN}^\xi} \coloneqq {}^{n+1}\VeGr{\xi} \quad .
\end{equation}
The following second subnetwork sFNN$^\psi$ uses the output of the first network alongside the new strain to determine the free energy, i.e.,
\begin{equation}
	\label{eq:InputFF22}
	\Ve{I}^{\text{sFNN}^\psi}\coloneqq {(\,{}^{n+1}\varepsilon, {}^{n+1}\xi^1,\ldots,{}^{n+1}\xi^N \,)} \mapsto O^{\text{sFNN}^\psi} \coloneqq {}^{n+1}\psi \quad .
\end{equation}
Since sFNN$^{\xi}$ takes ${}^{n}\Ve{\xi}$ as input, the availability of internal variables in the training data set is essential. 
The free energy is not required but improves the network performance. Within this study, the goal is to use as few information as possible, thus considering the free energy as not available during training. This applies to all of the following architectures as well.

In contrast to the original work\cite{Masi2021}, the training process is not performed as a single optimization considering both networks, but is split into two smaller steps, speeding up the optimization: First, sFNN$^\xi$ is trained independently, i.e., detached from sFNN$^\psi$, using
\begin{equation}
	\label{eq:LossFF21}
	\mathcal{L}^{\text{sFNN}^\xi} \coloneqq \mathcal{L}^{\xi} \quad \text{with} \quad \mathcal{L}^{\xi}\coloneqq \frac{1}{N} \sum_{\alpha=1}^{N} \mae\left( \xi^\alpha \right)
\end{equation}
to learn to predict the internal variables. Subsequently, both networks are used together in order to train the second network via the loss function
\begin{equation}
	\mathcal{L}^{\text{sFNN}^\psi} \coloneqq w^{\sigma} \mathcal{L}^{\sigma} + w^{\psi} \mathcal{L}^{\psi} + w^{\mathcal{D}\geq 0} \mathcal{L}^{\mathcal{D}\geq 0}  \quad \text{with} \quad \mathcal{L}^{\sigma} \coloneqq \mae\left(\sigma\right) \quad ,
\end{equation}
\begin{align}
	\mathcal{L}^{\psi} \coloneqq \mae\left(\psi\right) \quad \text{and} \quad \mathcal{L}^{\mathcal{D}\geq 0} \coloneqq \sum_{k=1}^{N^{\text{ds}}} \relu \left( -{}^{n+1}\mathcal{D}_k \right) \quad \text{where} \quad {}^{n+1}\mathcal{D} = - \sum_{\alpha=1}^{N} \frac{\partial{}^{n+1}\psi}{\partial{}^{n+1}\xi^\alpha}\frac{{}^{n+1}\xi^\alpha - {}^{n}\xi^\alpha}{{}^{n}\Delta t} \quad .
\end{align}
The non-trainable parameters $w^{\sigma}$, $w^{\psi}$ and $w^{\mathcal{D}\geq 0}$ regulate the influence of each term on the value of the loss function. During this optimization, the loss is only differentiated with respect to the weights of sFNN$^\psi$, so that only these weights are adjusted and the weights of sFNN$^\xi$ retain their values as obtained from the first training step.\\
However, the provision of internal variables for the training data for more complex material behavior is not a trivial task, which is addressed in Masi~and~Stefanou\cite{Masi2022}. In order to avoid providing internal variables, a similar architecture seems appropriate, in which sFNN$^\xi$ is replaced by a recurrent cell.

\subsubsection{Recurrent architecture with internal variables and free energy as output (RNN$^{\xi+\psi}$)}

\begin{figure}[b]
	\centering
	\graphicspath{{images/Architectures/}}
	\includegraphics{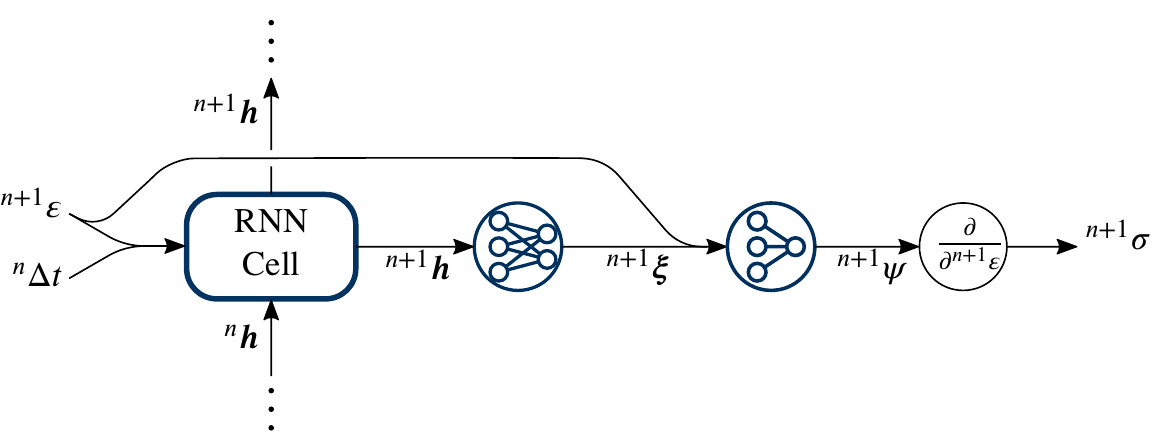}
	\caption{Functionality of the RNN$^{\xi+\psi}$ architecture: For each time step in a sequence, new strain and time increment are fed into the RNN cell, whose output is passed on to the feedforward network sFNN$^\xi$ to obtain the new set of internal variables. Together with the new strain, these serve as input in a final feedforward network sFNN$^\psi$ which predicts the new free energy. Differentiating sFNN$^\psi$ yields the new stress.}
	\label{fig:RNNpsi}
\end{figure}

Since RNNs are capable of storing information from time steps far in the past, it seems reasonable to exploit this property to mimic internal variables. \reff{fig:RNNpsi} shows a possible implementation of this idea, which is similar to He~and~Chen\cite{He2022}. Therein, at each time step, the RNN cell takes the new strain as argument as well as the time increment in case of viscous behavior and produces a new vector-valued output ${}^{n}\Ve{h}$, which carries the history information and is initialized with \mbox{${}^{0}\Ve{h}={}^{0}\Ve{c}=\Ve{0}$}. Subsequently, these values are fed into an FNN and are reduced to freely selectable number of internal variables $N^\xi$:
\begin{equation}
	\label{eq:InputRNNpsiRNN}
	\mbox{${}^{n}\Ve{I}^{\text{RNN}}\coloneqq {(\,{}^{n}\varepsilon, {}^{n-1}\Delta t \,)} \mapsto {}^{n}\Ve{h} \eqqcolon {}^{n}\Ve{I}^{\text{FF}^\xi} \mapsto {}^{n}\Ve{O}^{\text{FF}^\xi} \coloneqq {}^{n}\VeGr{\xi} \in \mathbb{R}^{N^\xi} \quad .$}
\end{equation}
Within another FNN, these values along with the new strain are projected onto the free energy, i.e., 
\begin{equation}
	\label{eq:InputRNNpsiFFNN2}
	\Ve{I}^{\text{FF}^\psi}\coloneqq {(\,{}^{n}\varepsilon, {}^{n}\xi^1,\ldots,{}^{n}\xi^{N^\xi} \,)} \mapsto O^{\text{FF}^\psi} \coloneqq {}^{n}\psi \quad .
\end{equation}
The loss function
\begin{align}
	\mathcal{L}^{\text{FF}^\psi} \coloneqq w^{\sigma} \mathcal{L}^{\sigma} + w^{\xi} \mathcal{L}^{\xi} + w^{\psi} &\mathcal{L}^{\psi} + w^{\mathcal{D}\geq 0} \mathcal{L}^{\mathcal{D}\geq 0}  \quad \text{with} \\
	\mathcal{L}^{\sigma} \coloneqq \mae \left( \sigma \right) \quad , \quad \mathcal{L}^{\xi}\coloneqq \frac{1}{N_{\text{av}}} \sum_{\alpha=1}^{N_{\text{av}}} &\mae\left( \xi_{\text{av}}^\alpha \right) \quad , \quad
	\mathcal{L}^{\psi} \coloneqq \mae\left( \psi \right) \quad \text{and} \label{eq:LossRNNPsiZ2}\\
	\mathcal{L}^{\mathcal{D}\geq 0} \coloneqq \frac{1}{N^{\text{seq}}} \sum_{s=1}^{N^{\text{seq}}} \left( \frac{1}{N_s^{\text{ts}}} \sum_{n=1}^{N_s^{\text{ts}}} \relu \left( -{}^{n}\mathcal{D}_s \right) \right) &\quad \text{where} \quad
	{}^{n}\mathcal{D} = - \sum_{\alpha=1}^{N} \frac{\partial{}^{n}\psi}{\partial{}^{n}\xi^\alpha}\frac{{}^{n}\xi^\alpha - {}^{n-1}\xi^\alpha}{{}^{n}\Delta t} \quad ,
\end{align}
is composed of four terms that control the prediction of stress, internal variables and free energy as well as the compliance with thermodynamic consistency, i.e., $\mathcal D\ge 0$. In the error term for the internal variables $\mathcal{L}^{\xi}$, only those internal variables can be considered that are actually available in the training data set, in \refe{eq:LossRNNPsiZ2} denoted as $\VeGr{\xi}_{\text{av}}$ for $N_{\text{av}}$ available internal variables. This might be the measurable plastic strain, for example. Consequently, if no internal variables are known, this term is omitted and the network has to find reasonable representations of internal variables on its own. The free energy term \mbox{$w^{\psi}\mathcal{L}^{\psi}$} is also optional and can be omitted if the free energy is unknown.

\subsection{Neural networks enforcing physics in a strong form}
\label{subsec:Augm}
The last category of models, denoted as \emph{NNs enforcing physics in a strong form} here, in contrast to the previous model class, satisfies $\mathcal D\ge 0$ a priori by construction of the model.
To achieve this, the concept of generalized standard materials summarized in Sect.~\ref{subsubsec:Biot} is adopted into the data-driven paradigm by applying ICNNs\cite{Amos2017}. This has been proposed for viscoelasticity by Huang~et~al.~\cite{Huang2022} as well as As'ad~and~Farhat~\cite{Asad2022a}.
In the following, three models, FNN$^{\psi+\phi}$, FNN$^{\psi+\phi^*}$, and FNN$^{\psi+\phi+\xi}$, are introduced in detail.

	\begin{remark}\label{remark:regularization}
	It should be noted that an application of the introduced approaches FNN$^{\psi+\phi}$ and FNN$^{\psi+\phi^*}$ to data belonging to a rate-independent material automatically leads to a regularization\cite{Nagler2022}, i.e., an approximation of the data by a rate-dependent model. This is due to the fact that the chosen activation function cannot represent the non-smooth dissipation potentials typical for plasticity, cf. Tab.~\ref{tab:analyt_models}.
	\end{remark}

\subsubsection{FNN-architecture with free energy and dissipation potential as output (FNN$^{\psi+\phi}$)}

The first architecture of this category, FNN$^{\psi+\phi}$, is composed of two FNNs, modeling the free energy $\psi$ and the dissipation potential $\phi$. With the requirements on $\phi$ from \refs{subsec:GenFra}, i.e., (i) convexity in $\dot{\Ve{\xi}}$, (ii) \mbox{$\phi(\dot{\Ve{\xi}}=\Ve{0}, \Ve{\xi}, \varepsilon)=0$}, and (iii) \mbox{$\phi(\dot{\Ve{\xi}}=\Ve{0}, \Ve{\xi}, \varepsilon)$} is a minimum in $\dot{\Ve{\xi}}$, this model can be used to make predictions that are \emph{a priori thermodynamically consistent}.
Convexity of the dissipation potential with respect to only $\dot{\Ve{\xi}}$, but not $\Ve{\xi}$ and $\varepsilon$, is achieved through the multiplicative split
\begin{align}
	\phi^\text{NN}(\dot{\Ve{\xi}}, \Ve{\xi}, \varepsilon) := \phi^{\text{con}}(\dot{\Ve{\xi}}) \, \phi^{+}(\varepsilon, \Ve{\xi})
	\label{eq:mulitplicative}
\end{align}
into a convex part $\phi^{\text{con}}(\dot{\Ve{\xi}})$ depending on only the rate of the internal variables and a positive part $\phi^{+}(\varepsilon, \Ve{\xi})$ depending on the strain and the internal variables themselves. Each part is modeled by a single FNN.
For $\phi^{\text{con}}$ to be convex, it requires convex and non-decreasing activation functions, here the Softplus activation \mbox{$\softplus(x) \coloneqq \ln(1+\exp(x))$}, and non-negative weights across the whole network\cite{Amos2017,Klein2021,Klein2022}. Positivity of $\phi^{+}$ in turn requires only the weights and bias of the output layer to be non-negative and positive activation functions, here also the Softplus activation. As another, more general method to construct a network, that is convex in only some of its inputs, a \emph{partially input convex neural network} (PICNN) as presented by Amos~et~al.\cite{Amos2017} could be used as well.
To meet normalization conditions (ii) and (iii), the output $\phi^{\text{NN}}$ of the combined network given in Eq.~\eqref{eq:mulitplicative} is modified via
\begin{equation}
	\label{eq:phiCorr}
	\phi(\dot{\Ve{\xi}}, \Ve{\xi}, \varepsilon) \coloneqq \phi^{\text{NN}}(\dot{\Ve{\xi}}, \Ve{\xi}, \varepsilon) - \phi^{\text{NN}}(\dot{\Ve{\xi}}=\Ve{0}, \Ve{\xi}, \varepsilon) - \frac{\partial \phi^{\text{NN}}(\dot{\Ve{\xi}}=\Ve{0}, \Ve{\xi}, \varepsilon)}{\partial \dot{\Ve{\xi}}} \cdot \dot{\Ve{\xi}} \ge 0 \; \forall \dot{\Ve \xi}, \Ve \xi, \varepsilon \quad .
\end{equation}
Likewise, the ansatz
\begin{equation}
	\label{eq:psiCorr}
	\psi(\varepsilon,\Ve{\xi}) \coloneqq \psi^{\text{NN}}(\varepsilon,\Ve{\xi}) - \psi^{\text{NN}}(\varepsilon=0,\Ve{\xi}=\Ve{0}) - \frac{\partial \psi^{\text{NN}}(\varepsilon=0,\Ve{\xi}=\Ve{0})}{\partial \varepsilon} \varepsilon - \frac{\partial \psi^{\text{NN}}(\varepsilon=0,\Ve{\xi}=\Ve{0})}{\partial \Ve{\xi}} \cdot \Ve{\xi}
\end{equation}
guarantees that free energy, stress and internal forces equal zero in the initial, unloaded state\cite{Huang2022}.

Now, after the model formulation given above, it is explained how the training algorithm and the prediction of new time steps with a calibrated FNN$^{\psi+\phi}$ architecture work.
Thereby, tuples of stress $\sigma$, strain $\varepsilon$, and internal variables $\Ve \xi$ are needed for the training of $\phi$ and $\psi$ with Eqs.~\eqref{eq:phiCorr} and \eqref{eq:psiCorr}.\footnote{To be able to use only stress and strain for training, it is also possible to integrate the full path of $\Ve \xi$ in each training epoch. However, since this requires backpropagation over all time steps, the training becomes much more time consuming\cite{Tac2023}.
	Alternatively, following As'ad~and~Farhat~\cite{Asad2022a}, a further network for the approximation of internal variables $\Ve \xi$ during training can be applied. This also enables to end up with a training data set containing only tuples of $\sigma$ and $\varepsilon$, whereby the training time is reduced compared to the first method.
	Another possibility is to determine the internal variables in advance within a preprocessing step. How this can be done is described in Ladev\`{e}ze~et~al.\cite{Ladeveze2019} or Gerbaud~et~al.\cite{Gerbaud2022}. Here, the second method is used, see Sect.~\ref{subsec:FNN_psi_phi_xi}.}

Training of the three networks $\phi^{\text{con}}(\dot{\Ve{\xi}})$, $\phi^{+}(\varepsilon, \Ve{\xi})$, and $\psi^{\text{NN}}(\varepsilon,\Ve{\xi})$ is performed as follows: 
The training starts by predicting ${}^{n}\psi$ with \refe{eq:psiCorr} and obtaining ${}^{n}\sigma$ and ${}^{n}\tau^\alpha$ via differentiation. Following \refe{eq:Biot}, the internal forces shall equal the dissipation potential differentiated with respect to the rate of the corresponding internal variable. This rate is approximated with ${}^{n}\dot{\xi}^\alpha \coloneqq ({}^{n}\xi^\alpha - {}^{n-1}\xi^\alpha)/{}^{n-1}\Delta t$. The difference of the  predicted internal forces ${}^{n}\tau^\alpha = -\frac{\partial\psi}{\partial{}^{n}\xi^\alpha}$ and ${}^{n}\hat{\tau}^\alpha \coloneqq \frac{\partial\phi}{\partial{}^{n}\dot{\xi}^\alpha}$ as well as the stress prediction are included in the loss function
\begin{equation}
	\mathcal{L} \coloneqq w^\sigma\mathcal{L}^\sigma + w^{\text{Biot}}\mathcal{L}^{\text{Biot}} \quad \text{with} \quad \mathcal{L}^\sigma \coloneqq \mae\left(\sigma\right) \quad \text{and} \quad \mathcal{L}^{\text{Biot}}\coloneqq \frac{1}{N} \sum_{\alpha=1}^N \mae\left(\tau^\alpha\right) \quad ,
	\label{eq:loss_FNN_psi_phi}
\end{equation}
which is to minimize in the process.

\begin{figure}[t]
	\centering
	\graphicspath{{images/Architectures/}}
	\includegraphics{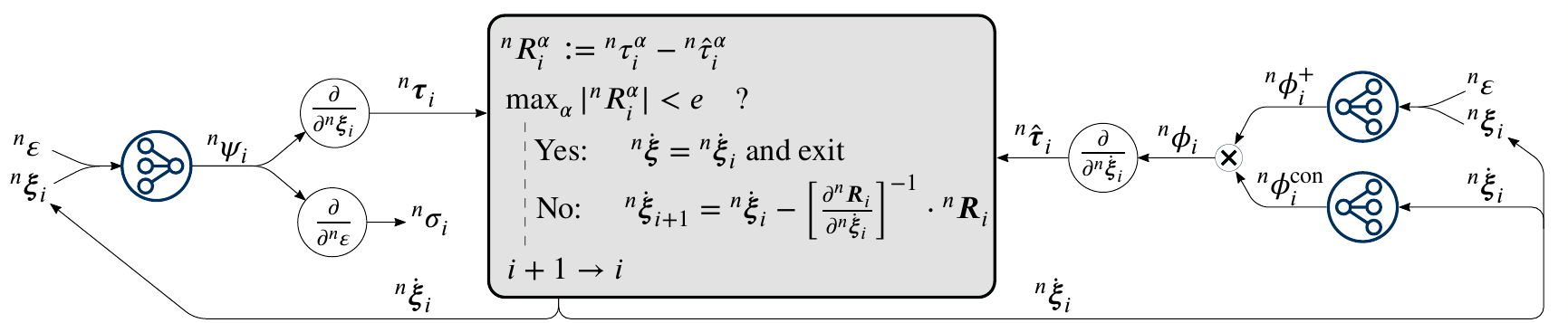}
	\caption{Functionality of the FNN$^{\psi+\phi}$ architecture: With the rate of internal variables of the current iteration, the internal forces are calculated using the network of the free energy ($\Ve{\tau}$, left) and the networks of the dissipation potential ($\hat{\Ve{\tau}}$, right). The rate of internal variables is now adapted iteratively, such that $\Ve{\tau} \approx \hat{\Ve{\tau}}$.}
	\label{fig:FFBiot}
\end{figure}

Once training is finished and good representations of $\psi$ and $\phi$ are found, these potentials can be used to predict the material response for a given strain path. Therefore, in each time step $n$, the rates of the internal variables ${}^{n}\dot{\xi}^\alpha$ are adapted iteratively using a \emph{Newton-Raphson scheme}, such that $\max_\alpha |{}^{n}\tau^\alpha - {}^{n}\hat{\tau}^\alpha | < e$ with a given tolerance $e$, where the new internal variables are obtained via ${}^{n}\xi^\alpha = {}^{n-1}\xi^\alpha + {}^{n-1}\Delta t {}^{n}\dot{\xi}^\alpha$, see the  scheme given in Fig.~\ref{fig:FFBiot}.

\subsubsection{FNN-architecture with free energy and dual dissipation potential as output (FNN$^{\psi+\phi^*}$)}

The second architecture presented here is the FNN$^{\psi+\phi^*}$ model which makes use of the \emph{dual dissipation potential} $\phi^*(\Ve{\tau}, \Ve{\xi}, \varepsilon)$ according to Eq.~\eqref{eq:Legendre}.
The model equations are thus similar to Eqs.~\eqref{eq:mulitplicative}--\eqref{eq:psiCorr} with the difference that $\phi$ has to be replaced by $\phi^*$.
The dual dissipation potential $\phi^*$ has to guarantee (i) convexity in $\Ve{\tau}$, (ii) \mbox{$\phi^*(\Ve \tau=\Ve{0}, \Ve{\xi}, \varepsilon)=0$}, and (iii) \mbox{$\phi^*(\Ve \tau=\Ve{0}, \Ve{\xi}, \varepsilon)$} is a minimum in $\Ve{\tau}$, to get an a priori thermodynamically consistent model.
Again, convexity of $\phi^*$ with respect to only $\Ve{\tau}$ is achieved through a multiplicative split similar to Eq.~\eqref{eq:mulitplicative}. 

\begin{figure}[t]
	\centering
	\graphicspath{{images/Architectures/}}
	\includegraphics{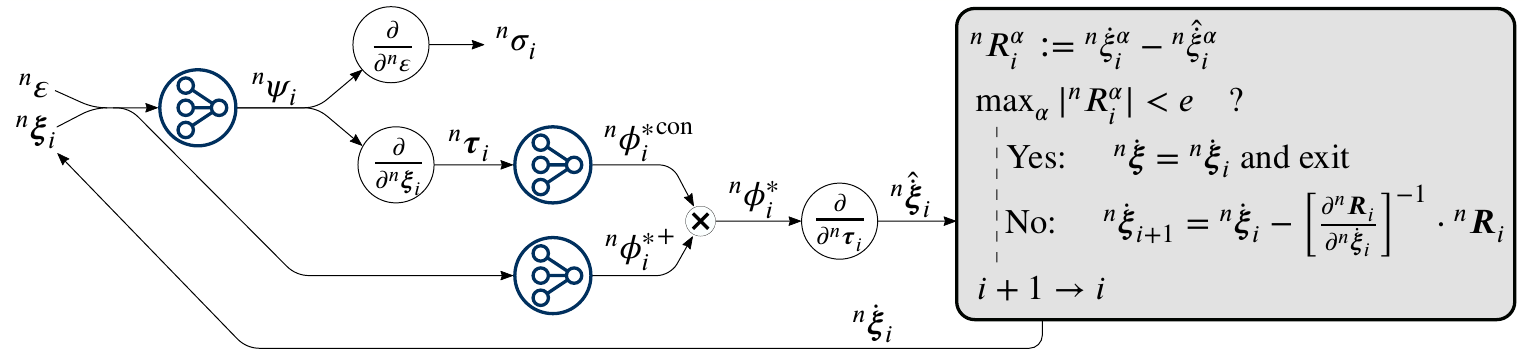}
	\caption{Functionality of the FNN$^{\psi+\phi^*}$ architecture: Using the rate of internal variables of the current iteration, the internal variables alongside the strain are fed into the network of the free energy. Stress and internal forces are obtained via differentiation. The internal forces are passed on to the convex network for ${\phi^*}^{\text{con}}$, strain and internal variables to the positive network for ${\phi^*}^{+}$. Multiplying ${\phi^*}^{\text{con}}$ and ${\phi^*}^{+}$ yields the dual dissipation potential, which is differentiated with respect to the internal forces to receive $\hat{\dot{\Ve{\xi}}}$. The rate of internal variables is now adapted iteratively, such that $\dot{\Ve{\xi}} \approx \hat{\dot{\Ve{\xi}}}$.}
	\label{fig:FFBiotStar}
\end{figure}

Besides the difference in the model equations, the FNN$^{\psi+\phi^*}$ architecture differs in its training algorithm and the method to predict the new time step.
In FNN$^{\psi+\phi^*}$, training with respect to a data set consisting of tuples of $\sigma$, $\varepsilon$, and $\Ve \xi$ is performed as follows: First, the free energy ${}^{n}\psi$ is calculated and ${}^{n}\sigma$ and ${}^{n}\tau^\alpha$ are obtained via differentiation. Subsequently, the internal forces ${}^{n}\tau^\alpha$ are fed into the combined network of the dual dissipation potential and ${}^{n}\phi^*$ is received. Differentiating ${}^{n}\phi^*$ with respect to the internal forces yields the rate of the internal variables ${}^{n}\dot{\xi}^\alpha$.
The loss function
\begin{equation}
	\mathcal{L} \coloneqq w^\sigma \mathcal{L}^\sigma + w^{\dot{\xi}} \mathcal{L}^{\dot{\xi}} \quad \text{with} \quad \mathcal{L}^\sigma \coloneqq \mae\left(\sigma\right) \quad \text{and} \quad \mathcal{L}^{\dot{\xi}} \coloneqq \frac{1}{N} \sum_{\alpha=1}^N \mae\left(\dot{\xi}^\alpha\right)
	\label{eq:loss_FNN_psi_phi*}
\end{equation}
now compares the predictions of the stress and the rates of the internal variables with their expected values ${}^{n}\bar{\sigma}$ and ${}^{n}\bar{\dot{\xi}}^\alpha$, respectively, where \mbox{${}^{n}\bar{\dot{\xi}}^\alpha = ({}^{n}\xi^\alpha - {}^{n-1}\xi^\alpha) / {}^{n-1}\Delta t$}.
Once the three networks are trained, the material state for a time step can be obtained using the pattern in \reff{fig:FFBiotStar}, where due to the implicit description of \mbox{${}^{n}\xi^\alpha={}^{n-1}\xi^\alpha + {}^{n-1}\Delta t {}^{n}\dot{\xi}^\alpha$}, the evolution of the internal variables is determined iteratively using a Newton-Raphson scheme, such that $\max_\alpha |{}^{n}\dot{\xi}^\alpha - {}^{n}\hat{\dot{\xi}}^\alpha | < e$ with a given tolerance $e$.

\subsubsection{Adapted training for FNN-architecture with free energy and dissipation potential as output (FNN$^{\psi+\phi+\xi}$)}
\label{subsec:FNN_psi_phi_xi}

The training of the former two architectures, i.e., FNN$^{\psi+\phi}$ and FNN$^{\psi+\phi^*}$, requires the internal variables to be present in the training data set. However, since these quantities are usually unknown, As'ad~and~Farhat\cite{Asad2022a} proposed an efficient training method for a model based on the dual dissipation potential, in which the internal variables are no longer required. This method is adopted analogously here for the architecture FNN$^{\psi+\phi}$.

To no longer have to specify the internal variables directly while keeping the training fast, another FNN, which takes the time $t$ as input and outputs the internal variables $\Ve\xi^{\text{NN}}(t)$, is integrated only for the training process. The values for the internal variables and their rates entering the networks for the free energy and the dissipation potential are taken from this auxiliary network, where the rate is approximated consistently to the prediction process by \mbox{${}^{n}\dot{\Ve\xi}^{\text{NN}}=\frac{1}{\Delta t}\left(\Ve\xi^{\text{NN}}({}^{n}t) - \Ve\xi^{\text{NN}}({}^{n-1}t)\right)$}. The condition $\Ve\xi^{\text{NN}}(0)=\Ve 0$ is enforced by subtracting $\Ve \xi^{\text{NN}}(0)$ from the actual output. During the training process, the weights of this network are adapted such that the predicted temporal courses of $\Ve \xi^{\text{NN}}(t)$ allow the errors for stress ($\mathcal{L}^{\sigma}$) and Biot term ($\mathcal{L}^{\text{Biot}}$) to become as small as possible. This leads to a reasonable representation of the internal variables without explicitly specifying their values. Once the training is finished, the auxiliary network is no longer necessary and the prediction process can be carried out in the same manner as described in \reff{fig:FFBiot}.
To make the training process easier for the optimizer, the courses of $\Ve \xi^{\text{NN}}(t)$ that are to be represented with the auxiliary network should to be smooth and rather simple. This is done by choosing a smooth strain path as explained in \refs{subsec:GenTrData}. As'ad~and~Farhat\cite{Asad2022a} further propose an additive split of the free energy into an equilibrium and a dissipative part. In order to retain comparability and consistency with FNN$^{\psi+\phi}$, this split is not performed herein. However, it should be noted that this additional assumption significantly facilitates training.

In addition to the loss function \refe{eq:loss_FNN_psi_phi}, another term $w^{|\xi|}\mathcal{L}^{|\xi|}$ is added. This term has the purpose to keep the internal variables small. Otherwise, the chosen internal variables become unboundedly large. To illustrate this, consider a constitutive model with free energy function \mbox{$\psi=\psi(\varepsilon, \xi)=\frac{1}{2}E(\varepsilon-\xi)^2$} with corresponding stress and internal force \mbox{$\sigma=\tau=E(\varepsilon-\xi)$}. An equivalent expression with another internal variable $\tilde{\xi}$ is \mbox{$\tilde{\psi}(\varepsilon, \tilde{\xi})=\frac{1}{2}E(\varepsilon-\frac{1}{A}\tilde{\xi})$} where \mbox{$\tilde{\xi}=A\xi$} and \mbox{$\tilde{\sigma}=E(\varepsilon-\frac{1}{A}\tilde{\xi})=\sigma$} \mbox{and $\tilde{\tau}=\frac{1}{A}E(\varepsilon-\frac{1}{A}\tilde{\xi})$}. The loss term $\mathcal{L}^{\text{Biot}}$ rewards small absolute values of the internal forces, i.e., a representation with large scaling factor $|A|$, resulting in \mbox{$|\tilde{\xi}|\gg|\xi|$}. This leads to difficulties in the prediction process later on. The new loss thus reads
\begin{equation}
	\label{eq:LossFNN3}
	\mathcal{L}\coloneqq w^{\sigma}\mathcal{L}^\sigma + w^{\text{Biot}}\mathcal{L}^{\text{Biot}} + w^{|\xi|}\mathcal{L}^{|\xi|} \quad \text{with} \quad \mathcal{L}^{|\xi|} \coloneqq \frac{1}{N^{\text{ds}}}\sum_{n=1}^{N^{\text{ds}}} \sum_{\alpha=1}^N |\xi^\text{NN}_\alpha({}^{n}t)| \quad ,
\end{equation}
where $\mathcal{L}^{|\xi^{\text{NN}}|}$ penalizes large absolute values of the chosen internal variable. Note that this does not imply, that $\sign(\tilde{\xi})=\sign(\xi)$, since $A<0$ is still a valid option. Within the scope of this study, only a single internal variable is predicted. Another network for another internal variable or a network with two outputs could be used as well, but this is not shown herein.

\section{Generation of the database for training and validation}\label{sec:Database}
Prior to studying the usability of the presented material models based on NNs, a database must be generated for the training process and, later on, as a reference for validation. Thereby, four different models according to Sects.~\ref{subsubsec:Viscoelasticity} and \ref{subsubsec:Elastoplasticity} are used to generate material states belonging to a prescribed strain path:
a \emph{nonlinear viscoelastic model with one Maxwell element} (V1), a \emph{linear viscoelastic model with two Maxwell elements} (V2), an \emph{elastoplastic model with kinematic hardening} (P1), as well as an \emph{elastoplastic model with mixed kinematic-isotropic hardening} (P2).
The models' governing equations given in Tab.~\ref{tab:analyt_models} are solved by applying an implicit Euler scheme. The chosen parameters of these reference constitutive models can be found in \reft{tab:TestMatModels}.

\begin{table}[b]
	\centering
	\caption{Chosen parameters of the four testing material models used as for training data generation and validation.}
	\label{tab:TestMatModels}
	\centering
	\begin{small}
	\begin{tabular}{C{4.7cm} C{1.8cm} l}
		\toprule
		Description & Label & Parameters\\
		\midrule
		Nonlinear viscoelastic model with one Maxwell element & V1 &  \mbox{$E=\SI{1}{\giga\pascal}$}, \mbox{$E_1=\SI{10}{\giga\pascal}$}, \mbox{$\hat{\eta}_1=\SI{200}{\giga\pascal\second}$}, \mbox{$a_1=-0.7$}, \mbox{$b_1=0.1$} \\
		Linear viscoelastic model with two Maxwell elements & V2 &  \mbox{$E=\SI{1}{\giga\pascal}$}, \mbox{$E_1=\SI{10}{\giga\pascal}$},
		\mbox{$\eta_1=\SI{10}{\giga\pascal\second}$},
		\mbox{$E_2=\SI{20}{\giga\pascal}$},  \mbox{$\eta_2=\SI{5}{\giga\pascal\second}$} \\
		\midrule
		Elastoplastic model with kinematic hardening & P1 & \mbox{$E=\SI{20}{\giga\pascal}$}, \mbox{$\sigma_\text{y}=\SI{100}{\mega\pascal}$},  \mbox{$H=\SI{10}{\giga\pascal}$} \\
		Elastoplastic model with mixed kinematic-isotropic hardening & P2 & \mbox{$E=\SI{20}{\giga\pascal}$}, \mbox{$\sigma_\text{y}=\SI{100}{\mega\pascal}$},  \mbox{$H=\SI{3}{\giga\pascal}$}, \mbox{$\hat{H}=\SI{3}{\giga\pascal}$} \\
		\bottomrule
	\end{tabular}
	\end{small}
\end{table}

\subsection{Generation of training data}
\label{subsec:GenTrData}
Two different methods to generate training data are used. The first method can be applied to all architectures except for FNN$^{\psi+\phi+\xi}$ and uses random walk sequences. The FNN$^{\psi+\phi+\xi}$ requires a smooth strain path, which is given by a cubic spline combining a chosen set of knots. Both methods are described in detail below.

\subsubsection{Random walk strain paths}
Four different data bases are generated, one for each examined material V1, V2, P1 and P2. Since the recurrent architectures require several independent sequences in their training data, the data points are generated in multiple sequences. For the feedforward architectures, these sequences are decomposed back into their separate time steps to obtain the required input-output pairs or data tuples.
Each sequence is created from a random walk regarding the strain path, starting from the initial material state with  \mbox{${}^{0}\sigma={}^{0}\varepsilon={}^{0}\xi^\alpha=0$}. A strain increment ${}^{0}\Delta\varepsilon$ as a sample of a normal distribution with standard deviation $s^{\Delta\varepsilon}$ around mean $0$ is applied to the initial state in a time increment ${}^{0}\Delta t$ sampled from a uniform distribution \mbox{${}^{0}\Delta t \in ({\Delta t}_{\text{min}}\, , \, {\Delta t}_{\text{max}})$}. The strain rate is constant within this interval. The next material state is obtained by applying another strain increment ${}^{1}\Delta\varepsilon$ within another time increment ${}^{1}\Delta t$ and so on, until the sequence contains $N^{\text{ts}}$ timesteps. Here, sequences of length $N^{\text{ts}}=100$ have shown to perform well for the recurrent architecures, independent from the examined material. In order to limit the strain to a reasonable range, the absolute value of the strain may not exceed $|\varepsilon|_{\text{max}}$. That is, if \mbox{$|{}^{n}\varepsilon + {}^{n}\Delta\varepsilon| > |\varepsilon|_{\text{max}}$}, the strain increment ${}^{n}\Delta\varepsilon$ is sampled again until \mbox{$|{}^{n}\varepsilon + {}^{n}\Delta\varepsilon| \leq |\varepsilon|_{\text{max}}$}. The parameters of the random walk are chosen to be $s^{\Delta\varepsilon}=\SI{0.25}{\percent}$, $|\varepsilon|_{\text{max}}=\SI{2}{\percent}$, ${\Delta t}_{\text{min}}=\SI{0.02}{\second}$ and ${\Delta t}_{\text{max}}=\SI{0.1}{\second}$. Exemplarily, the strain path and respective stress response for the first sequence of the data base for V1 is shown in \reff{subfig:TDRandomWalk}.

The actual training data sets, i.e., the data that is effectively used during training, comprise a number of $N^{\text{seq}}$ sequences or $N^{\text{ds}}$ data tuples taken from this database. 
It was found that $N^{\text{seq}}=100$ sequences for the recurrent or $N^{\text{ds}}=1000$ data tuples for the feedforward architectures are sufficient without a loss of prediction quality, independent from the material behavior to be modeled. The selection of data from each database starts with the first sequence or the first time step of the first sequence and is continued chronologically. This means that the training data set of all feedforward architectures consists of information from the identical material states taken from the first 10 sequences and the recurrent architectures receive the same set of 100 sequences.

\subsubsection{Smooth strain path}
In order to enable the NN representation of the internal variable with the auxiliary network, the FNN$^{\psi+\phi+\xi}$ architecture requires the training data to be generated from a simpler, more smooth strain path. This smooth strain path is described by a cubic spline connecting a set of manually chosen knots in the $\varepsilon$-$t$ space\cite{Asad2022a}. Once the spline is defined, the data points are generated by applying this strain path to the analytical models with uniformly distributed time increments $\Delta t\in (\SI{0.01}{\second}, \SI{0.02}{\second})$ until the data set contains 900 time steps. The same strain path is used for all materials V1, V2, P1 and P2 and is shown in \reff{subfig:TDSpline} with the corresponding stress response for V1. The knots were chosen such that the data generated with this path are comparable to the data from the random walk sequences.

{\begin{figure}[t]
		\centering
		\includegraphics{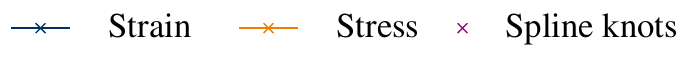}\\
		\subfloat[\label{subfig:TDRandomWalk}]{\includegraphics[width=0.49\textwidth]{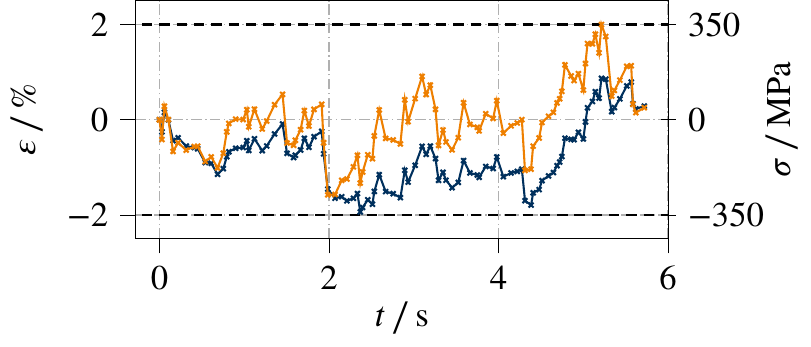}}
		\hfill
		\subfloat[\label{subfig:TDSpline}]{\includegraphics[width=0.49\textwidth]{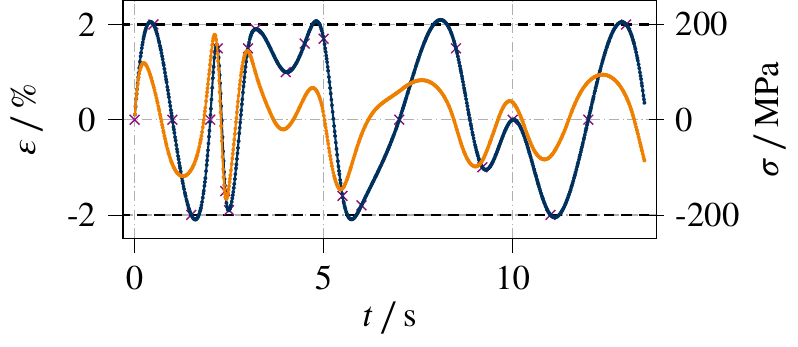}}
		\caption{Stresses and strains over time in the training data for the nonlinear viscoelastic material (V1): \textbf{(a)} the first sequence taken from the non-smooth random walk data base and \textbf{(b)} the smooth path generated with cubic splines used for FNN$^{\psi+\phi+\xi}$.}
		\label{fig:TDvisco}
\end{figure}}

\subsection{Generation of validation data}
In order to evaluate and compare the network performances for unseen data, the networks have to predict the constitutive response for a set of consecutive time steps taken from the strain path shown in \reff{subfig:Interpol}, where a constant time increment of $\SI{0.05}{\second}$ is used. The path is chosen such that strain, strain rate, and time increments do not exceed the training range. The networks thus do not have to extrapolate for this benchmark test. Reference data points are, analogously to the training data, obtained using an implicit Euler scheme.

Secondly, data for a further validation path with strain, strain rate and time increments exceeding the training range are generated, see \reff{fig:Extrapol}. This will be used to evaluate the extrapolation capabilities of the different models later on. This path includes a small hysteresis beyond the training limits of the strain (\mbox{$\SI{2}{\second} < t \leq \SI{4}{\second}$}) and a short interval (\mbox{$\SI{5}{\second} < t \leq \SI{5.08}{\second}$}) of 16 time steps with \mbox{$\Delta t = \SI{0.005}{\second}$} and \mbox{$\dot{\varepsilon}=\SI{62.5}{\percent\per\second}$}, undercutting the limits of $\Delta t$ and exceeding the limits of $\dot{\varepsilon}$ in another hysteresis. Afterwards, a short relaxation with constant strain \mbox{$\varepsilon=\SI{0}{\percent}$} is performed until \mbox{$t=\SI{6}{\second}$}. Subsequently, large time increments outside of the training range are applied (\mbox{$\Delta t=\SI{0.125}{\second}$} and \mbox{$\Delta t=\SI{0.2}{\second}$} in \mbox{$\SI{6}{\second} < t \leq \SI{7}{\second}$} and \mbox{$\SI{7}{\second} < t \leq \SI{8}{\second}$}, respectively). For all other time steps, the time increment equals \mbox{$\Delta t = \SI{0.05}{\second}$}. Finally, the strain is increased linearly until \mbox{$t=\SI{10}{\second}$} and \mbox{$\varepsilon = \SI{6}{\percent}$}.

{\begin{figure}[b]
		\centering
		\subfloat[\label{subfig:Interpol}]{\includegraphics[width=0.49\textwidth]{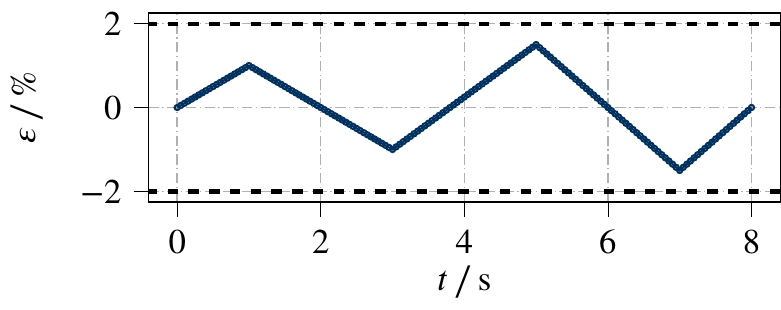}}
		\hfill
		\subfloat[\label{fig:Extrapol}]{\includegraphics[width=0.49\textwidth]{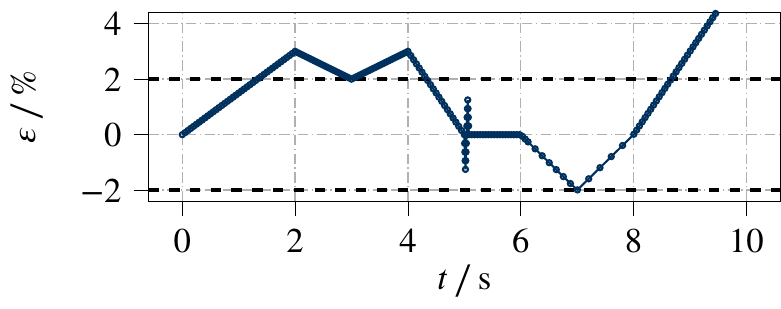}}
		\caption{Validation paths to evaluate the network performances for \textbf{(a)} interpolation and  \textbf{(b)} extrapolation. The path in (a) is chosen such that neither strain nor time increments exceed the training range. Path (b) in turn exceeds those limits separately and includes sections of large strain rates. Black dashed lines indicate the training data limits.}
		\label{fig:StrainOverTime}
\end{figure}}

\subsection{Normalization}
The first and second derivatives of most common activation functions vanish for large absolute values of their arguments, rendering training impossible. In order to ensure nonzero gradients, the training data set, which for example contains stresses in the order of several $\si{\mega\pascal}$, has to be scaled down to values of magnitude $1$. Since in general not all relevant features of the training data set are independent from each other, the scaling factors must be chosen appropriately, which is shown in the following.\\
Generally, an independent quantity $f$ can be normalized to the range $(-1,1) \ni \tilde{f}$ via
\begin{equation}
	\label{eq:norm1}
	\tilde{f} = \frac{f - m_f}{s_f} \quad \Longleftrightarrow \quad f = s_f \tilde{f} + m_f \quad \text{with} \quad m_f = \frac{1}{2}(f_{\text{max}}+f_{\text{min}}) \quad \text{and} \quad s_f = \frac{1}{2}(f_{\text{max}}-f_{\text{min}}) \quad ,
\end{equation}
wherein $\tilde{f}$ is the scaled value of $f$ and $f_{\text{max}}$ and $f_{\text{min}}$ denote the maximum and minimum value of $f$ across the whole data set\cite{Kalina2021}.

For the blackbox models FNN$^{\sigma}$ and RNN$^{\sigma}$, the necessary quantities $\varepsilon$, $\sigma$ and $\Delta t$ can be treated as independent and the scaling factors are thus obtained using \refe{eq:norm1}.
In all other architectures, not every feature is independent from the others. This is due to the occurring differential operators. Furthermore, it is reasonable to choose the scaling factors in such a way, that all relations, that hold for the unscaled quantities also hold for their scaled values to avoid back transformations during training. This simplifies and speeds up the calculation of the loss function. For the sake of simplicity, all $m_f$ are set to zero in the following. Considering all remaining architectures, the relevant quantities are $\varepsilon$, $\sigma$, $\Delta t$, $\Ve{\xi}$, $\Ve{\tau}$, $\psi$, $\phi$ and $\phi^*$.
As chosen independent quantities, the scaling factors of $\varepsilon$, $\Delta t$ and $\psi$ are obtained using \refe{eq:norm1}$_4$. All other scaling factors follow from the evaluation of the relevant equations under the stated condition, resulting in the following relations:
\begin{equation}
	\label{eq:norm2}
	s_\sigma = \frac{s_\psi}{s_\varepsilon} \quad , \quad s_\xi = s_\varepsilon \quad , \quad s_\tau = \frac{s_\psi}{s_\varepsilon} \quad , \quad s_\phi = \frac{s_\psi}{s_{\Delta t}} \quad \text{and} \quad s_{\phi^*} = \frac{s_\psi}{s_{\Delta t}} \quad .
\end{equation}
The scaling is applied to all features in the respective data set and the training is performed using only these scaled values. The network itself consequently predicts only scaled values. The back transformation \refe{eq:norm1}$_2$ yields the values in familiar physical units. The trained network can now be applied to predict the material response for a given strain path.

\section{Applications}\label{sec:App}
Within this section, building on the data generated according to Sect.~\ref{sec:Database}, each of the seven architectures presented in \refs{sec:Models}, i.e., FNN$^\sigma$, RNN$^\sigma$, FNN$^{\xi+\psi}$, RNN$^{\xi+\psi}$, FNN$^{\psi+\phi}$, FNN$^{\psi+\phi^*}$ and FNN$^{\psi+\phi+\xi}$, is tested on its capability to rebuild the constitutive response of the materials V1, V2, P1, and P2 after training. This is evaluated by the NN-based models' predicted stress responses for an unknown strain path given in \reff{subfig:Interpol}, where the path is chosen such that strain, strain rate, and time increments do not exceed the training range.
In addition, to analyze the extrapolation capability, the models' predicted responses are considered for a second validation path with strain, strain rate, and time increments exceeding the training range, cf. \reff{fig:Extrapol}. In order to reduce the scope of the presented study to a reasonable level, the second validation path is investigated for V1 only.
All of the used FNNs consist of an input layer, a single hidden layer and an output layer. The number of neurons in the hidden layer is denoted as $N_2$.

\subsection{Training results and validation path without extrapolation}\label{subsec:ResInterpolation}
\subsubsection{Black box models}\label{subsec:ResBlackbox}
\begin{figure}[h]
	\graphicspath{{images/Results/}}
	\centering
	\includegraphics{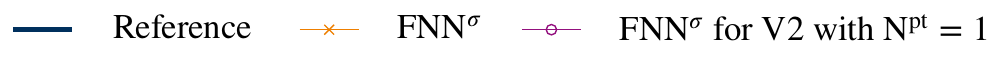}\\
	\setlength{\figw}{4cm}
	\setlength{\figh}{4cm}
	\subfloat[\label{fig:FFsig01}]{\includegraphics{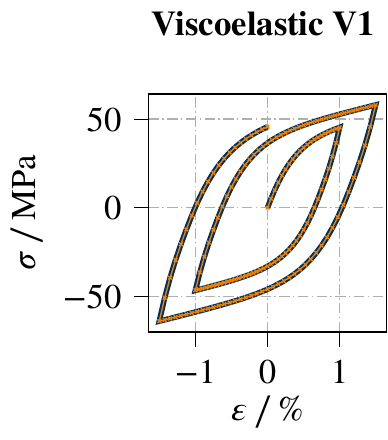}}
	\hfil
	\subfloat[\label{fig:FFsig02B}]{\includegraphics{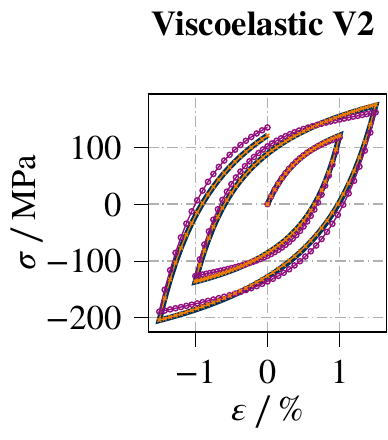}}
	\hfil
	\subfloat[\label{fig:FFsig03}]{\includegraphics{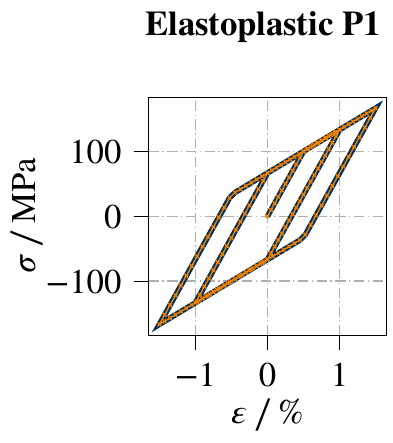}}
	\hfil
	\subfloat[\label{fig:FFsig04}]{\includegraphics{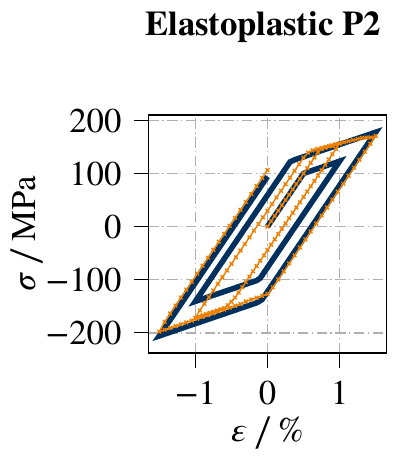}}
	\caption{Stress prediction of FNN$^\sigma$ for the four underlying test materials. The chosen hyperparameters are given in Tab.~\ref{tab:ResFFsigParams}.}
	\label{fig:ResFFsig}
\end{figure}

\paragraph{FNN$^\sigma$}
For each test material model, an FNN$^\sigma$ as described in \refs{subsubsec:FFsig} is created with the hyperparameters and training results given in \reft{tab:ResFFsigParams}. The model predictions are presented in \reff{fig:ResFFsig}. 

The results for viscoelasticity show that the prediction of the stress behavior with FNN$^\sigma$ for the materials V1 is possible without further difficulties. For V2, in contrast, a precise prediction is only possible with two preceding time steps in the input, i.e., $N^\text{pt}=2$.
These limitations of the architecture can be traced back to the ambiguity of the expected output given a particular input. That is, given the material parameters for V1, the overstress $\sigma^{\text{ov}}_1$ can be determined solely from the values of ${}^{n}\varepsilon$ and ${}^{n}\sigma$ given in the input. The material state is therefore determined unambiguously. For material V2, on the other hand, only the sum $\sigma^{\text{ov}}_1 + \sigma^{\text{ov}}_2$ of the overstresses can be determined from these quantities, but not their distinct values. Thus, the inner state is not clearly described and the stress prediction is not possible precisely, as \reff{fig:FFsig02B} shows. For this reason, the network is provided with further information of an additional time step, see also \reff{fig:FFsig02B}.

\begin{table}[h]
	\centering
	\caption{FNN$^\sigma$ architectures: network hyperparameters with number of preceding time steps $N^\text{pt}$, neurons in the hidden layer $N_2$, and activation function $\mathcal{A}^l_2$ as well as characteristic values from training with number of training data $N^\text{ds}$ and loss value $\mathcal L$ after the given number of iterations. The figure numbers with the corresponding validation load case are given in column Val.}
	\label{tab:ResFFsigParams}
	\centering
	\begin{small}
		\begin{tabular}{c c c c c c c c}
			\toprule
			Material & $N^{\text{pt}}$ & $N_2$ & $\mathcal{A}^l_2$ & $N^{\text{ds}}$ & Iterations & $\mathcal{L}$ & Val.\\
			\midrule
			V1 & 1 & 15 & $\tanh$ & 1000 & 3094 & $\num[exponent-product=\ensuremath{\cdot},round-mode=figures,round-precision=1]{2e-4}$ & \reff{fig:FFsig01} \\
			V2 & 1 & 25 & $\tanh$ & 1000 & 911 & $\num[exponent-product=\ensuremath{\cdot},round-mode=figures,round-precision=1]{2e-3}$ & \reff{fig:FFsig02B} \\
			V2 & 2 & 25 & $\tanh$ & 1000 & 2903 & $\num[exponent-product=\ensuremath{\cdot},round-mode=figures,round-precision=1]{1e-4}$ & \reff{fig:FFsig02B} \\
			P1 & 1 & 15 & $\relu$ & 1000 & 356 & $\num[exponent-product=\ensuremath{\cdot},round-mode=figures,round-precision=1]{3e-6}$ & \reff{fig:FFsig03} \\
			P2 & 1 & 15 & $\relu$ & 1000 & 1240 & $\num[exponent-product=\ensuremath{\cdot},round-mode=figures,round-precision=1]{8e-3}$ & \reff{fig:FFsig04} \\
			\bottomrule
		\end{tabular}
	\end{small}
\end{table}

The state of the elastoplastic  material with kinematic hardening P1 is, analogously to V1, sufficiently described by ${}^{n}\varepsilon$ and ${}^{n}\sigma$. However, in contrast to V2, elastoplasticity with mixed kinematic-isotropic hardening as present in P2 cannot be described by adding an additional time step to the input. This would require information about the current position and extent of the elastic region, which cannot be provided by a fixed amount of preceding times steps. Interestingly, the model tries to approximate the material behavior with kinematic hardening as good as possible.

Summarizing, FNN$^\sigma$ is thus only applicable for certain types of material behavior. Moreover, no further information is obtained apart from the stress prediction. On the other hand, the training process is computationally inexpensive and requires only stress-strain pairs and time steps for the data set.

\paragraph{RNN$^\sigma$}
Using the same reference data generated with V1, V2, P1, and P2, the capability of the RNN$^\sigma$ model according to Wu~et~al.\cite{Wu2020} is now investigated, where the network hyperparameters given in \reft{tab:ResRNNsigParams} have been used. With that, the results given in \reff{fig:ResRNNsig} could be achieved for the validation test. 
The RNN$^\sigma$ model is thus able to predict the stress response almost perfectly for all four test materials. Due to the comparatively large number of weights and the expensive gradient calculations for networks with recurrent cells, this broad applicability comes at the cost of the computational time needed for the training. Similar to FNN$^\sigma$, only stresses and strains are required for the training data set, but no additional physical information is obtained. However, both models, FNN$^\sigma$ and RNN$^\sigma$, do not allow statements to be made about whether the processes described by the respective model are embedded in a meaningful thermodynamic framework.
\begin{table}[h]
	\centering
	\caption{RNN$^\sigma$ architectures: network hyperparameters with number of values in the cell state $N^\text{c}$, neurons in the hidden layer of the feedforward network $N_2^{\text{FF}}$ with activations ${\mathcal{A}^l_2}^{\text{FF}}$ as well as characteristic values from training with number of sequences $N^\text{seq}$ of length $N^\text{ts}$ and loss value $\mathcal L$ after training for the given number of iterations.}
	\label{tab:ResRNNsigParams}
	\centering
	\begin{small}
	\begin{tabular}{c c c c c c c c c c}
		\toprule
		Material & $N^\text{c}$ & $N_2^{\text{FF}}$ & ${\mathcal{A}^l_2}^{\text{FF}}$ & $N^{\text{seq}}$ & $N^{\text{ts}}$ & Iterations & $\mathcal{L}$ & Val.\\
		\midrule
		V1 & 6 & 10 & $\tanh$ & 100 & 100 & 3465 & $\num[exponent-product=\ensuremath{\cdot},round-mode=figures,round-precision=1]{9e-4}$ & \reff{fig:RNNsig01} \\
		V2 & 10 & 10 & $\tanh$ & 100 & 100 & 2219 & $\num[exponent-product=\ensuremath{\cdot},round-mode=figures,round-precision=1]{8e-5}$ & \reff{fig:RNNsig02} \\
		P1 & 10 & 10 & $\relu$ & 100 & 100 & 5352 & $\num[exponent-product=\ensuremath{\cdot},round-mode=figures,round-precision=1]{7e-3}$ & \reff{fig:RNNsig03} \\
		P2 & 12 & 20 & $\relu$ & 100 & 100 & 5281 & $\num[exponent-product=\ensuremath{\cdot},round-mode=figures,round-precision=1]{1e-3}$ & \reff{fig:RNNsig04} \\
		\bottomrule
	\end{tabular}
	\end{small}
\end{table}
{
	\begin{figure}[h]
		\graphicspath{{images/Results/}}
		\centering
		\includegraphics{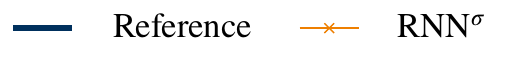}\\
		\setlength{\figw}{4cm}
		\setlength{\figh}{4cm}
		\subfloat[\label{fig:RNNsig01}]{\includegraphics{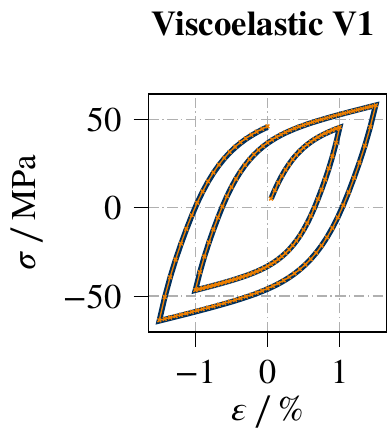}}
		\hfil
		\subfloat[\label{fig:RNNsig02}]{\includegraphics{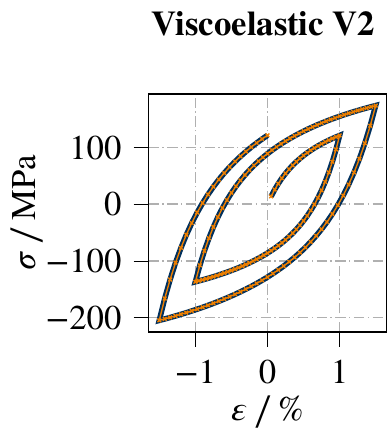}}
		\hfil
		\subfloat[\label{fig:RNNsig03}]{\includegraphics{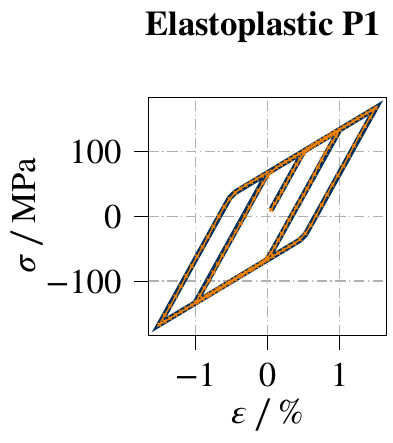}}
		\hfil
		\subfloat[\label{fig:RNNsig04}]{\includegraphics{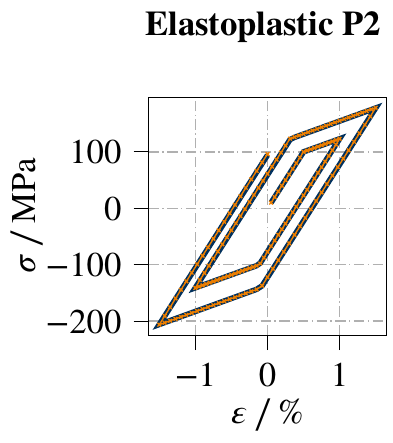}}
		\caption{Stress prediction of RNN$^\sigma$ for the four underlying test materials. The chosen hyperparameters are given in Tab.~\ref{tab:ResRNNsigParams}.}
		\label{fig:ResRNNsig}
	\end{figure}
}

\subsubsection{Neural networks enforcing physics in a weak form}\label{subsec:NN_weak}
\paragraph{FNN$^{\xi+\psi}$}
After the investigation of the pure black box models given above, the first approach belonging to the class of NNs enforcing physics in a weak form is analyzed, the FNN$^{\xi+\psi}$ model according to Masi~et~al.\cite{Masi2021}. Again, the same reference data generated with V1, V2, P1, and P2, are used here.

\begin{table}[h]
	\centering
	\caption{FNN$^{\xi+\psi}$ architectures: network hyperparameters with neurons in the hidden layer of the feedforward networks $N_2^{\text{FF}^\xi}$ and $N_2^{\text{FF}^\psi}$ with activations ${\mathcal{A}^l_2}^{\text{FF}^\xi}$ and ${\mathcal{A}^l_2}^{\text{FF}^\psi}$ as well as weighting factors of the loss function $w^\sigma$, $w^{\mathcal{D}\geq 0}$, number of training data $N^\text{ds}$ and loss values $\mathcal L^\sigma$, $\mathcal{L}^{\mathcal{D}\geq 0}$ after the given number of iterations. The figures for the validation are given in Val.}
	\label{tab:ResFF2Params}
	\centering
	\begin{small}
	\begin{tabular}{c c c c c c c c c c c c c c}
		\toprule
		Material & $N_2^{\text{FF}^{\xi}}$ & $N_2^{\text{FF}^{\psi}}$ & ${\mathcal{A}^l_2}^{\text{FF}^\xi}$ & ${\mathcal{A}^l_2}^{\text{FF}^\psi}$ & $w^\sigma$ / $w^{\mathcal{D}\geq 0}$ & $N^{\text{ds}}$ & Iter. & $\mathcal{L}^{\xi}$ & $\mathcal{L}^{\sigma}$ / $\mathcal{L}^{\mathcal{D}\geq 0}$ & Val.\\
		\midrule
		V1 & 15 & 15 & $\tanh$ & $\tanh$ & 1/1 & 1000 & 5414/504 & $\num[exponent-product=\ensuremath{\cdot},round-mode=figures,round-precision=1]{1e-4}$ & $\num[exponent-product=\ensuremath{\cdot},round-mode=figures,round-precision=1]{1e-3}$/$\num[exponent-product=\ensuremath{\cdot},round-mode=figures,round-precision=1]{4e-6}$ & \reff{fig:FF201}\\
		V2 & 15 & 15 & $\tanh$ & $\tanh$ & 1/1 & 1000 & 1410/700 & $\num[exponent-product=\ensuremath{\cdot},round-mode=figures,round-precision=1]{5e-5}$ & $\num[exponent-product=\ensuremath{\cdot},round-mode=figures,round-precision=1]{2e-4}$/$\num[exponent-product=\ensuremath{\cdot},round-mode=figures,round-precision=1]{4e-8}$ & \reff{fig:FF202}\\
		P1 & 15 & 15 & $\relu$ & $\tanh$ & 1/1 & 1000 & 973/522 & $\num[exponent-product=\ensuremath{\cdot},round-mode=figures,round-precision=1]{5e-6}$ & $\num[exponent-product=\ensuremath{\cdot},round-mode=figures,round-precision=1]{2e-4}$/$\num[exponent-product=\ensuremath{\cdot},round-mode=figures,round-precision=1]{4e-4}$ & \reff{fig:FF203}\\
		P2 & 15 & 15 & $\relu$ & $\tanh$ & 1/1 & 1000 & 1114/608 & $\num[exponent-product=\ensuremath{\cdot},round-mode=figures,round-precision=1]{1e-7}$ & $\num[exponent-product=\ensuremath{\cdot},round-mode=figures,round-precision=1]{4e-5}$/$\num[exponent-product=\ensuremath{\cdot},round-mode=figures,round-precision=1]{4e-5}$ & \reff{fig:FF204}\\
		
		\bottomrule
	\end{tabular}
	\end{small}
\end{table}

{
	\begin{figure}[h]
		\graphicspath{{images/Results/}}
		\centering
		\includegraphics{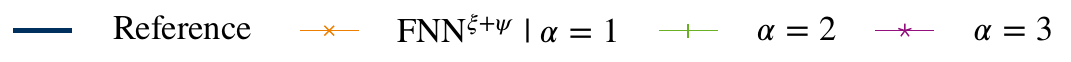}\\
		\setlength{\figw}{4cm}
		\setlength{\figh}{4cm}
		\subfloat[\label{fig:FF201}]{\includegraphics{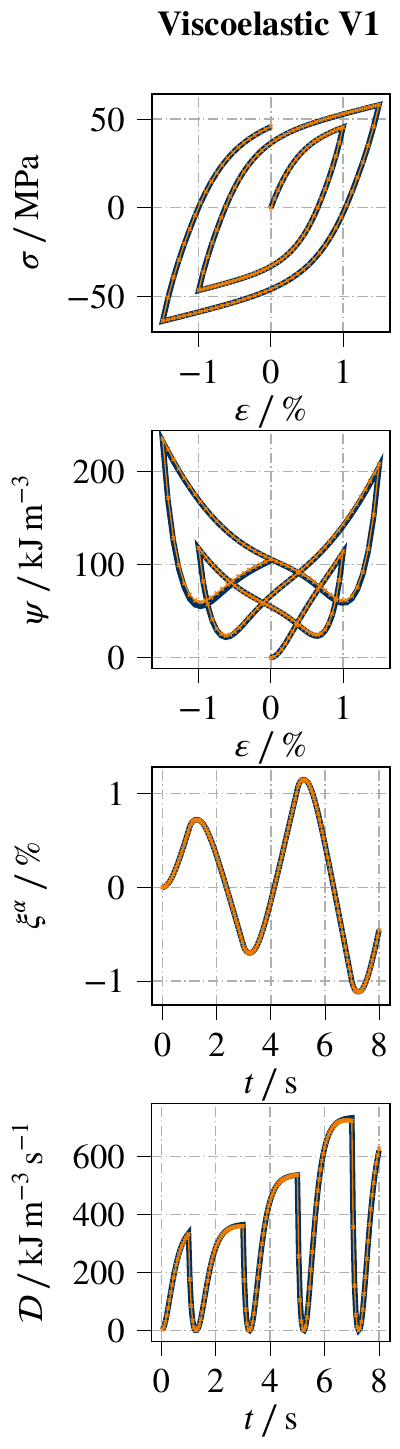}}
		\hfil
		\subfloat[\label{fig:FF202}]{\includegraphics{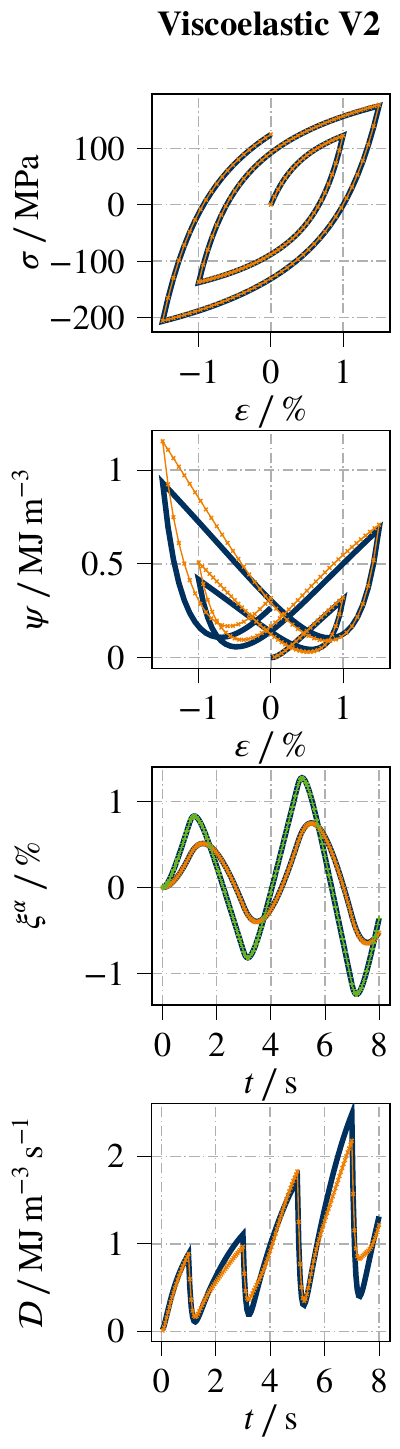}}
		\hfil
		\subfloat[\label{fig:FF203}]{\includegraphics{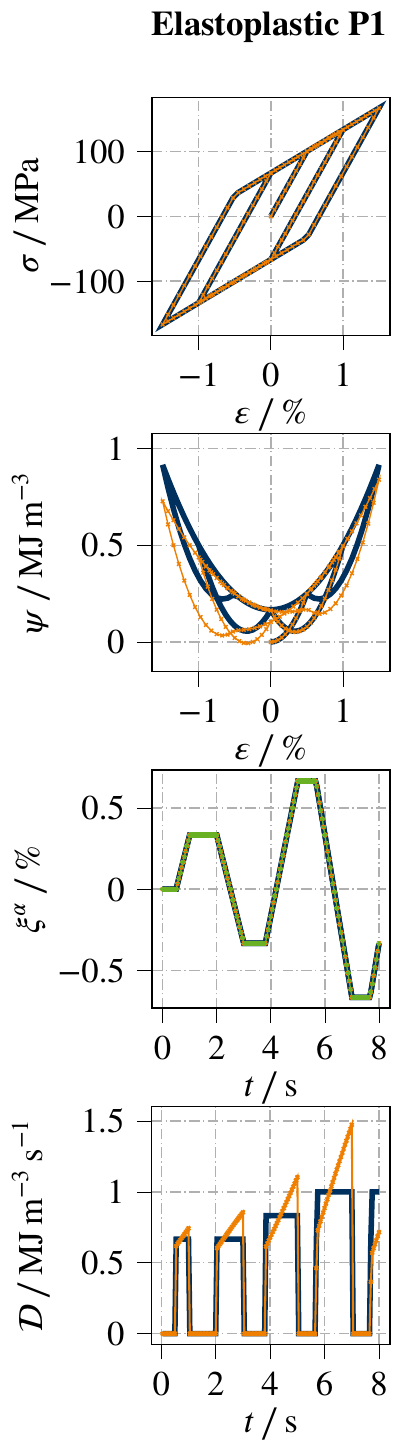}}
		\hfil
		\subfloat[\label{fig:FF204}]{\includegraphics{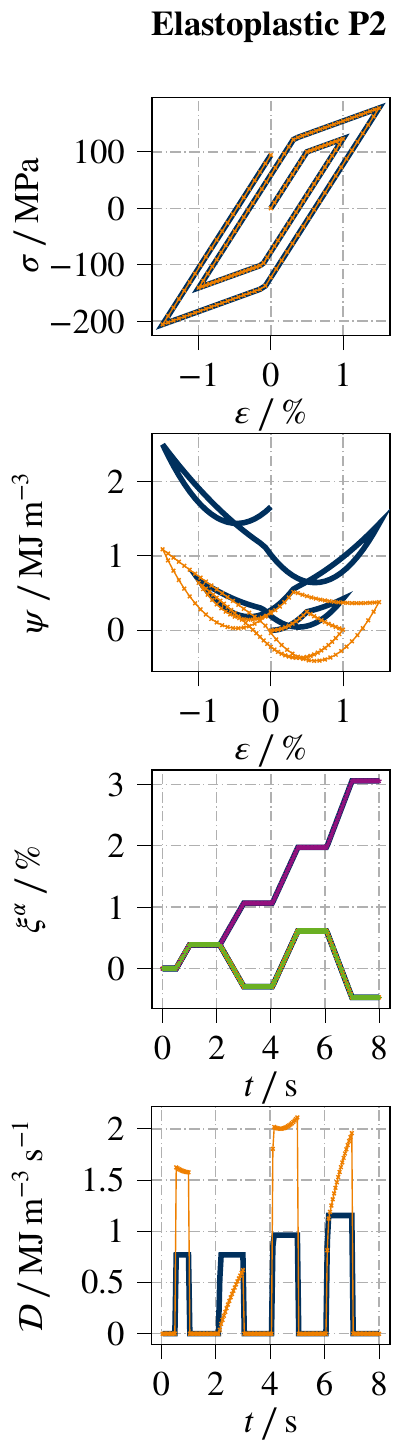}}
		\caption{Prediction of stress, free energy, internal variables and dissipation rate for the four test materials using FNN$^{\xi+\psi}$.}
		\label{fig:ResFF2}
	\end{figure}
}

Using the hyperparameters given in \reft{tab:ResFF2Params}, four FNN$^{\xi+\psi}$ networks were created. The results for the validation loading case given in \reff{fig:ResFF2} show that FNN$^{\xi+\psi}$ is able to produce precise predictions for stress and internal variables, i.e., the available quantities. In addition to the stress-strain plots, diagrams showing free energy $\psi$, internal variables $\xi^\alpha$, and dissipation rate $\mathcal D$ are added.
It can be seen that $\psi$ and $\mathcal D$ vary from the expected results, although the order of magnitude and rough course mostly coincide. The condition \mbox{$\mathcal{D}\geq 0$} is satisfied reliably. Thus, in contrast to the black box models, FNN$^{\xi+\psi}$ puts the predictions in a more physical framework, which has positive effects on the generalization capability of the network. For a detailed study on this architecture, see Masi~et~al.\cite{Masi2021}. 
However, FNN$^{\xi+\psi}$ requires the provision of information about the internal variables for the training data, which is not a trivial task. This problem is discussed in Masi~and~Stefanou\cite{Masi2022} for multiscale simulations.

\paragraph{RNN$^{\xi+\psi}$}
The second analyzed NN-based approach that enforces physics in a weak form is the RNN$^{\xi+\psi}$ approach similar to He~and~Chen\cite{He2022}. The network hyperparameters and the information on the training are given in \reft{tab:ResRNNpsiParams}. With that, the results given in \reff{fig:ResRNNpsi} could be achieved for the considered validation path.

One can see, that the stress predictions are very precise for all four materials. Furthermore, the Clausius-Duhem inequality is also complied with, with the exception of few individual time steps. The internal variables, however, vary vastly from the expected results. Since no internal variables are given in the training data set and no further restrictions on their course are made, the network is free to choose any set of internal variables that allow accurate stress predictions and compliance with \mbox{$\mathcal{D}\geq 0$}. The network thus finds another representation of the set of internal variables and the free energy function. 

Note, that for the kinematic hardening material P1, the network predicts only a single internal variable (\mbox{$N^\xi=1$}), although \mbox{$N=2$}. This is possible since plastic strain $\varepsilon^{\text{pl}}$ and kinematic hardening variable $\alpha$ coincide and the material response can be expressed in terms of only the plastic strain. The same simplification could as well be made for material P2 with \mbox{$N^\xi=2$} despite \mbox{$N=3$}, but \mbox{$N^\xi=3$} is chosen.

In contrast to FNN$^{\xi+\psi}$, the provision of internal variables is not explicitly necessary to obtain precise, physically consistent stress predictions. This advantage comes at cost of a computational expensive recurrent cell. For both FNN$^{\xi+\psi}$ and RNN$^{\xi+\psi}$, although $\mathcal D\ge 0$ is enforced by a penalty term in the loss function and is also satisfied in this validation path without extrapolation, this cannot be fully guaranteed for this class of models, see \refs{subsec:ResExtrapolation}.

\begin{table}[h]
	\centering
	\caption{RNN$^{\xi+\psi}$ architectures: network hyperparameters with number of values in the cell state $N^\text{c}$, number of predicted internal variables $N^\xi$ neurons in the hidden layer of the feedforward networks $N_2^{\text{FF}^\xi}$, $N_2^{\text{FF}^\psi}$ with activations ${\mathcal{A}^l_2}^{\text{FF}^\xi}$,  ${\mathcal{A}^l_2}^{\text{FF}^\psi}$ as well as weighting factors of the loss function $w^\sigma$, $w^{\mathcal{D}\geq 0}$, number of training sequences $N^\text{seq}$ of length $N^\text{ts}$ and loss values $\mathcal{L}^\sigma$, $\mathcal{L}^{\mathcal{D}\geq0}$ after training for the given number of iterations. Validation load cases can be found in the figures given in Val.}
	\label{tab:ResRNNpsiParams}
	\centering
	\begin{small}
	\begin{tabular}{c c c c c c c c c c c c c}
		\toprule
		Mat. & $N^\text{c}$ & $N^\xi$  & $N_2^{\text{FF}^\xi}$ & $N_2^{\text{FF}^\psi}$ & ${\mathcal{A}^l_2}^{\text{FF}^\xi}$ & ${\mathcal{A}^l_2}^{\text{FF}^\psi}$ & $w^\sigma$ / $w^{\mathcal{D}\geq 0}$ & $N^\text{seq}$ & $N^\text{ts}$ & Iter. & $\mathcal{L}^\sigma$ / $\mathcal{L}^{\mathcal{D}\geq 0}$ & Val. \\
		\midrule
		V1 & 6 & 1 & 10 & 10 & $\tanh$ & $\tanh$ & 1/5 & 100 & 100 & 5079 & $\num[exponent-product=\ensuremath{\cdot},round-mode=figures,round-precision=1]{2e-3}$/$\num[exponent-product=\ensuremath{\cdot},round-mode=figures,round-precision=1]{4e-7}$ & \reff{fig:RNNpsi01} \\
		V2 & 6 & 2 & 10 & 10 & $\tanh$ & $\tanh$ & 1/5 & 100 & 100 & 5244 & $\num[exponent-product=\ensuremath{\cdot},round-mode=figures,round-precision=1]{8e-3}$/$\num[exponent-product=\ensuremath{\cdot},round-mode=figures,round-precision=1]{3e-5}$ & \reff{fig:RNNpsi02} \\
		P1 & 10 & 1 & 10 & 10 & $\relu$ & $\tanh$ & 1/5 & 100 & 100 & 5329 & $\num[exponent-product=\ensuremath{\cdot},round-mode=figures,round-precision=1]{7e-3}$/$\num[exponent-product=\ensuremath{\cdot},round-mode=figures,round-precision=1]{1e-4}$ & \reff{fig:RNNpsi03} \\
		P2 & 12 & 3 & 15 & 20 & $\relu$ & $\tanh$ & 1/5 & 100 & 100 & 5140 & $\num[exponent-product=\ensuremath{\cdot},round-mode=figures,round-precision=1]{1e-3}$/$\num[exponent-product=\ensuremath{\cdot},round-mode=figures,round-precision=1]{5e-6}$ & \reff{fig:RNNpsi04} \\
		\bottomrule
	\end{tabular}
	\end{small}
\end{table}
{
	\begin{figure}[h]
		\graphicspath{{images/Results/}}
		\centering
		\includegraphics{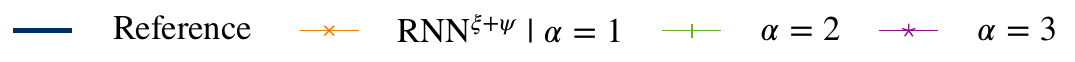}\\
		\setlength{\figw}{4cm}
		\setlength{\figh}{4cm}
		\subfloat[\label{fig:RNNpsi01}]{\includegraphics{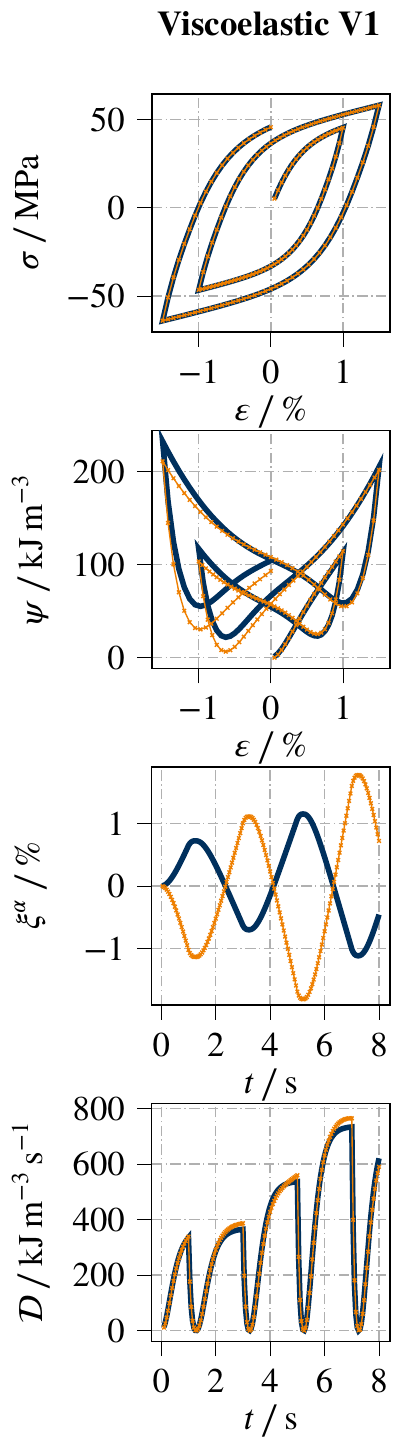}}
		\hfil
		\subfloat[\label{fig:RNNpsi02}]{\includegraphics{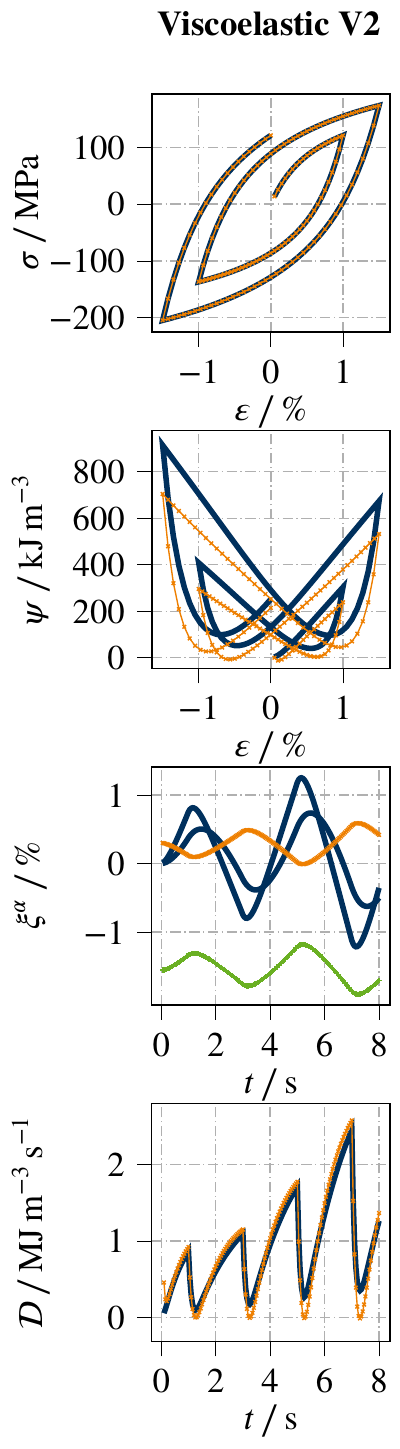}}
		\hfil
		\subfloat[\label{fig:RNNpsi03}]{\includegraphics{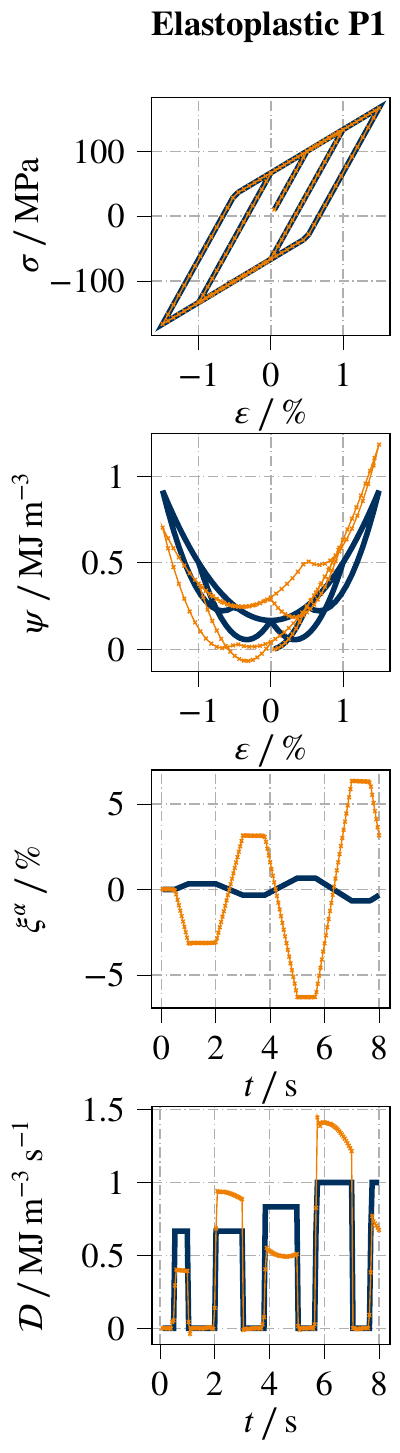}}
		\hfil
		\subfloat[\label{fig:RNNpsi04}]{\includegraphics{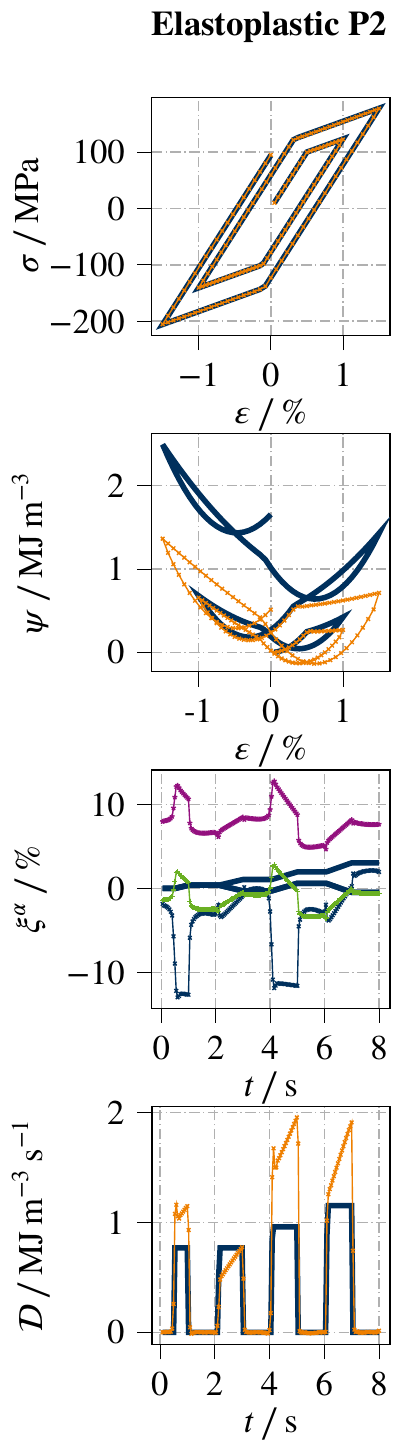}}
		\caption{Prediction of stress, free energy, internal variables and dissipation rate for the four test materials using RNN$^{\xi+\psi}$.}
		\label{fig:ResRNNpsi}
	\end{figure}
}

\clearpage

\subsubsection{Neural networks enforcing physics in a strong form}\label{subsec:NN_strong}

\paragraph{FNN$^{\psi+\phi}$}

Finally, the analysis of approaches belonging to the class of \emph{NNs enforcing physics in a strong form} is done by using the reference data generated with V1, V2, P1, and P2. The first model is the FNN$^{\psi+\phi}$ model according to Huang~et~al.\cite{Huang2022}. Here, the network hyperparameters and the information on the training are given in \reft{tab:ResFFBiot} and the results achieved for the considered validation path are given in \reff{fig:ResFFBiot}.

One can see, that the stress predictions are very precise for both viscoelastic materials, V1 and V2. 
In addition to the stress-strain plots, diagrams showing internal variables $\xi^\alpha$, free energy $\psi$, internal forces $\tau^\alpha$, dissipation rate $\mathcal D$, and dissipation potential $\phi$ are added.
From that one can see that all of these quantities are almost identical to the reference model, although these values do not explicitly appear in the loss, cf. Eq.~\eqref{eq:loss_FNN_psi_phi}. Only for V1 there is a visible discrepancy in the dissipation potential. Furthermore, in contrast to the former two models, the important condition \mbox{$\mathcal{D}\geq 0$} is now guaranteed for all admissible load paths by construction of the network architecture. However, FNN$^{\psi+\phi}$ requires the provision of information about the internal variables for the training data when no adapted training is used\cite{Asad2022a}.

Regarding rate independent materials, the FNN$^{\psi+\phi}$ model is able to rebuild the response of the elastoplastic material P1. Interestingly, however, the internal forces $\tau^\alpha$ do not match with the reference. This is due to the fact that $\xi^1=\varepsilon^\text{pl}$ and $\xi^2=\alpha$ coincide, i.e. $\alpha=\varepsilon^\text{pl}$, cf. Tab.~\ref{tab:analyt_models}. Therefore, there are any number of ways to divide the internal forces, all of which lead to the same, good result, even if the relation between the internal variables is not incorporated into the network, c.f. Remark \ref{remark:constraints}. Furthermore, regarding the plots of $\mathcal D$ and $\phi$, one can see that the the original non-continuous curve shapes at the zero line are approximated by smooth ones. This process is often called a regularization of the rate independent model, i.e., an approximation by a rate dependent one\cite{Nagler2022}. Due to the choice of the softplus activation function this results all by itself within the training process. Regarding the results for P2, a poor prediction becomes apparent in Fig.~\ref{fig:ResFFBiot} as soon as the plastic regime is reached. The stress response is, similar to FNN$^\sigma$, approximated with only kinematic hardening instead of kinematic and isotropic hardening. According to the authors, this is due to the fact that there is no function $\phi$, for which \refe{eq:Biot} can be applied without explicitly considering the relation between the rates of the plastic strain $\dot{\varepsilon}^{\text{pl}}$ and the isotropic hardening variable $\dot{\hat{\alpha}}$, c.f. Remark \ref{remark:constraints}.  If this relation, i.e., $\dot{\hat{\alpha}} = |\dot{\varepsilon}^{\text{pl}}|$, is incorporated into the model architecture, the model is able to make accurate predictions. The authors have already verified this. 

\begin{table}[h]
	\centering
	\caption{FNN$^{\psi+\phi}$ architectures: network hyperparameters with neurons in the hidden layer of the feedforward networks $N_2^{{\phi}^{\text{con}}}$, $N_2^{{\phi}^{\text{+}}}$, $N_2^{\psi}$ with activations ${\mathcal{A}^l_2}^{{\phi}^{\text{con}}}$, ${\mathcal{A}^l_2}^{{\phi}^{\text{+}}}$, ${\mathcal{A}^l_2}^{\psi}$ as well as weighting factors of the loss function $w^\sigma$, $w^{\text{Biot}}$, number of training data $N^\text{ds}$ and number of iterations resulting in the final loss values $\mathcal L^\sigma$, $\mathcal{L}^{\text{Biot}}$. The figures in Val. show the validation cases.}
	\label{tab:ResFFBiot}
	\centering
	\begin{small}
	\begin{tabular}{c c c c c c c c c c c c c c}
		\toprule
		Mat. & $N_2^{{\phi}^{\text{con}}}$ & $N_2^{{\phi}^{\text{+}}}$ & $N_2^{\psi}$ & ${\mathcal{A}^l_2}^{{\phi}^{\text{con}}}$ & ${\mathcal{A}^l_2}^{{\phi}^{\text{+}}}$ & ${\mathcal{A}^l_2}^{\psi}$ & $w^\sigma$ / $w^{\text{Biot}}$ & $N^\text{ds}$ & Val. & $\mathcal{L}^\sigma$ / $\mathcal{L}^{\text{Biot}}$ & Fig. \\
		\midrule
		V1 & 20 & 20 & 15 & $\softplus$ & $\softplus$ & $\tanh$ & 1/0.3 & 1000 & 1251 & $\num[exponent-product=\ensuremath{\cdot},round-mode=figures,round-precision=1]{6e-4}$/$\num[exponent-product=\ensuremath{\cdot},round-mode=figures,round-precision=1]{1e-2}$ & \reff{fig:FFBiot01} \\
		V2 & 20 & 20 & 15 & $\softplus$ & $\softplus$ & $\tanh$ & 1/0.3 & 1000 & 995 & $\num[exponent-product=\ensuremath{\cdot},round-mode=figures,round-precision=1]{6e-5}$/$\num[exponent-product=\ensuremath{\cdot},round-mode=figures,round-precision=1]{2e-4}$ & \reff{fig:FFBiot02} \\
		P1 & 20 & 20 & 15 & $\softplus$ & $\softplus$ & $\tanh$ & 1/0.3 & 1000 & 1047 & $\num[exponent-product=\ensuremath{\cdot},round-mode=figures,round-precision=1]{2e-4}$/$\num[exponent-product=\ensuremath{\cdot},round-mode=figures,round-precision=1]{2e-2}$ & \reff{fig:FFBiot03} \\
		P2 & 20 & 20 & 15 & $\softplus$ & $\softplus$ & $\tanh$ & 1/0.3 & 1000 & 832 & $\num[exponent-product=\ensuremath{\cdot},round-mode=figures,round-precision=1]{4e-4}$/$\num[exponent-product=\ensuremath{\cdot},round-mode=figures,round-precision=1]{6e-2}$ & \reff{fig:FFBiot04} \\
		\bottomrule
	\end{tabular}
	\end{small}
\end{table}
{
	\begin{figure}[h]
		\graphicspath{{images/Results/}}
		\centering
		\includegraphics{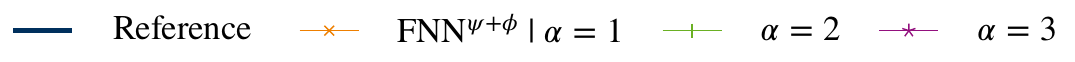}\\
		\setlength{\figw}{4cm}
		\setlength{\figh}{4cm}
		\subfloat[\label{fig:FFBiot01}]{\includegraphics{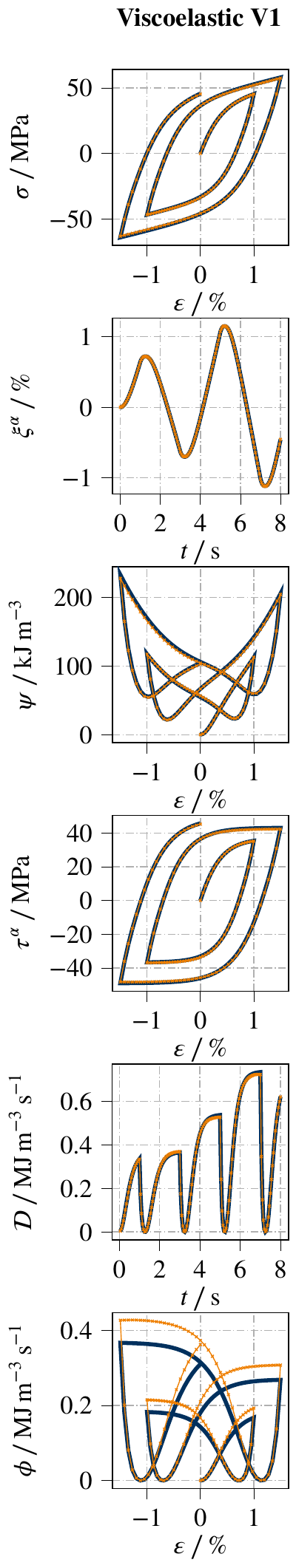}}
		\hfil
		\subfloat[\label{fig:FFBiot02}]{\includegraphics{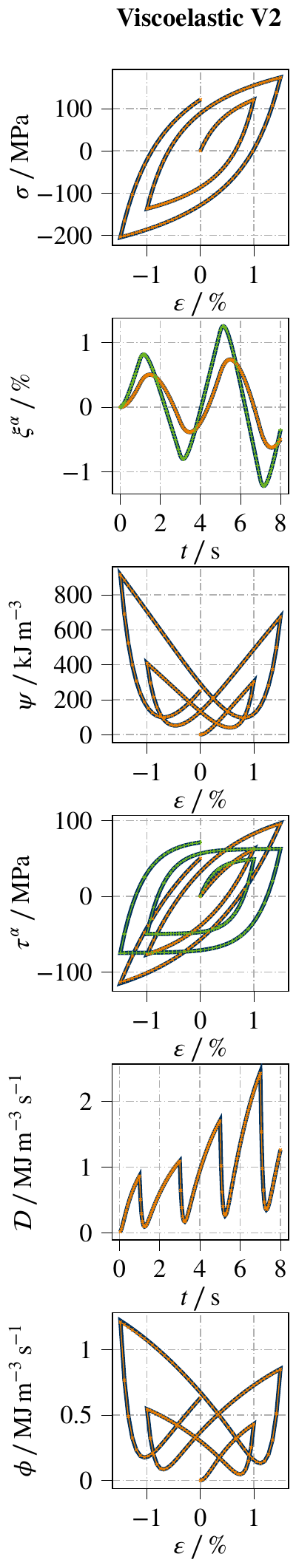}}
		\hfil
		\subfloat[\label{fig:FFBiot03}]{\includegraphics{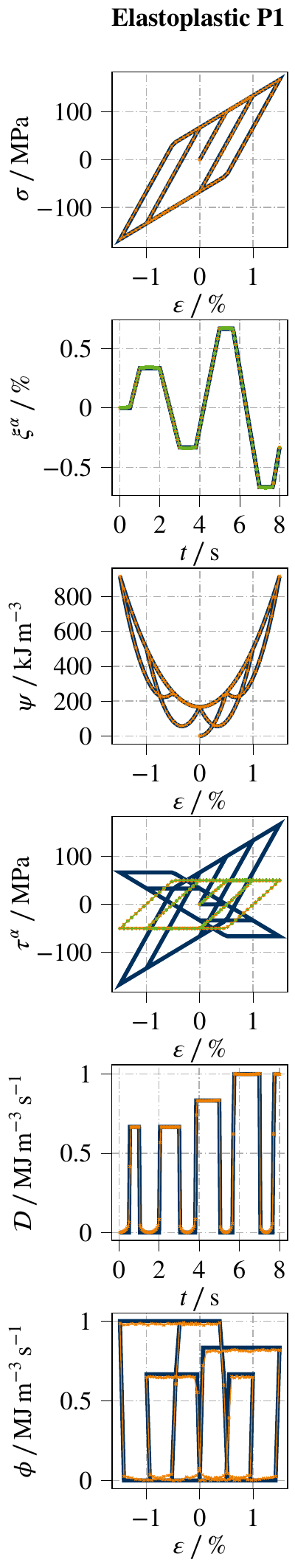}}
		\hfil
		\subfloat[\label{fig:FFBiot04}]{\includegraphics{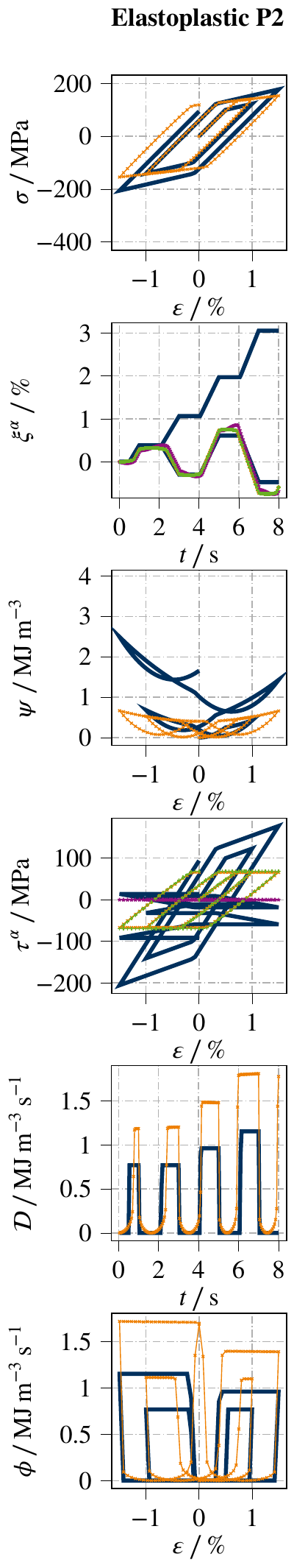}}
		\hfil
		\caption{{Prediction of stress, internal variables, free energy, internal forces, dissipation rate and dissipation potential for the four test materials using FNN$^{\psi+\phi}$.}}
		\label{fig:ResFFBiot}
	\end{figure}
}

\paragraph{FNN$^{\psi+\phi^*}$}
Instead of using the dissipation potential $\phi$, the model FNN$^{\psi+\phi^*}$ according to As'ad~and~Farhat\cite{Asad2022} makes use of the dual dissipation potential $\phi^*$. Here, the network hyperparameters and the information on the training are given in \reft{tab:ResFFBiotStar}. With that, the results given in \reff{fig:ResFFBiotStar} could be achieved for the considered validation path.

Similar to FNN$^{\psi+\phi*}$, the predictions for stress, free energy, dissipation rate, etc. are very precise for the viscoelastic materials V1 and V2. Again, only for V1 there is a visible discrepancy in the dual dissipation potential. FNN$^{\psi+\phi^*}$ requires the provision of information about the internal variables for the training data when no adapted training is used.

Regarding the prediction of FNN$^{\psi+\phi^*}$ for the elastoplastic materials P1 and P2, a rather poor result shows up in Fig.~\ref{fig:ResFFBiotStar} even for kinematic hardening. However, this is not surprising, since the dual dissipation potentials $\phi^*$ of the reference models P1 and P2 have a shape that cannot be reasonably represented with the selected activation functions, cf. Tab.~\ref{tab:analyt_models}, and restrictions to the networks weights.

\begin{table}[h]
	\centering
	\caption{FNN$^{\psi+\phi^*}$ architectures: network hyperparameters with neurons in the hidden layer of the feedforward networks $N_2^{{\phi^*}^{\text{con}}}$, $N_2^{{\phi^*}^{\text{+}}}$, $N_2^{\psi}$ with activations ${\mathcal{A}^l_2}^{{\phi^*}^{\text{con}}}$, ${\mathcal{A}^l_2}^{{\phi^*}^{\text{+}}}$, ${\mathcal{A}^l_2}^{\psi}$ as well as weighting factors of the loss function $w^\sigma$, $w^{\dot{\xi}}$, number of training data $N^\text{ds}$ and number of iterations and final loss values $\mathcal L^\sigma$, $\mathcal{L}^{\dot{\xi}}$. The figures showing the validation load cases are linked in Val.}
	\label{tab:ResFFBiotStar}
	\centering
	\begin{small}
	\begin{tabular}{c c c c c c c c c c c c}
		\toprule
		Mat. & $N_2^{{\phi^*}^{\text{con}}}$ & $N_2^{{\phi^*}^{\text{+}}}$ & $N_2^{\psi}$ & ${\mathcal{A}^l_2}^{{\phi^*}^{\text{con}}}$ & ${\mathcal{A}^l_2}^{{\phi^*}^{\text{+}}}$ & ${\mathcal{A}^l_2}^{\psi}$ & $w^\sigma$ / $w^{\dot{\xi}}$ & $N^\text{ds}$ & Iter. & $\mathcal{L}^\sigma$ / $\mathcal{L}^{\dot{\xi}}$ & Val. \\
		\midrule
		V1 & 20 & 20 & 15 & $\softplus$ & $\softplus$ & $\tanh$ & 1/1 & 1000 & 2429 & $\num[exponent-product=\ensuremath{\cdot},round-mode=figures,round-precision=1]{1e-4}$/$\num[exponent-product=\ensuremath{\cdot},round-mode=figures,round-precision=1]{1e-4}$ & \reff{fig:FFBiotStar01} \\
		V2 & 20 & 20 & 15 & $\softplus$ & $\softplus$ & $\tanh$ & 1/1 & 1000 & 859 & $\num[exponent-product=\ensuremath{\cdot},round-mode=figures,round-precision=1]{1e-4}$/$\num[exponent-product=\ensuremath{\cdot},round-mode=figures,round-precision=1]{1e-5}$ & \reff{fig:FFBiotStar02} \\
		P1 & 20 & 20 & 15 & $\softplus$ & $\softplus$ & $\tanh$ & 1/1 & 1000 & 525 & $\num[exponent-product=\ensuremath{\cdot},round-mode=figures,round-precision=1]{8e-5}$/$\num[exponent-product=\ensuremath{\cdot},round-mode=figures,round-precision=1]{8e-3}$ & \reff{fig:FFBiotStar03} \\
		P2 & 20 & 20 & 15 & $\softplus$ & $\softplus$ & $\tanh$ & 1/1 & 1000 & 673 & $\num[exponent-product=\ensuremath{\cdot},round-mode=figures,round-precision=1]{1e-5}$/$\num[exponent-product=\ensuremath{\cdot},round-mode=figures,round-precision=1]{9e-3}$ & \reff{fig:FFBiotStar04} \\
		\bottomrule
	\end{tabular}
	\end{small}
\end{table}
{
	\begin{figure}[h]
		\graphicspath{{images/Results/}}
		\centering
		\includegraphics{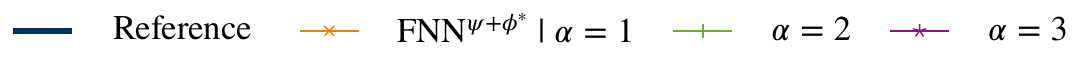}\\
		\setlength{\figw}{4cm}
		\setlength{\figh}{4cm}
		\subfloat[\label{fig:FFBiotStar01}]{\includegraphics{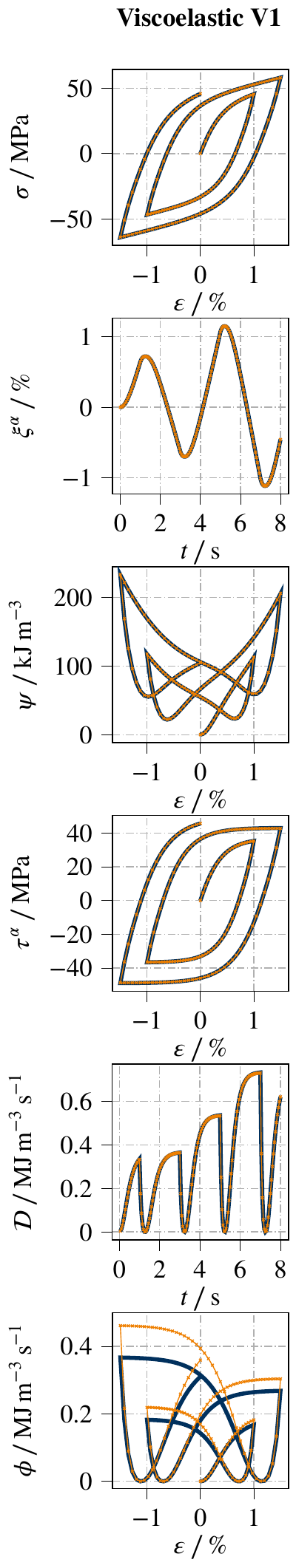}}
		\hfil
		\subfloat[\label{fig:FFBiotStar02}]{\includegraphics{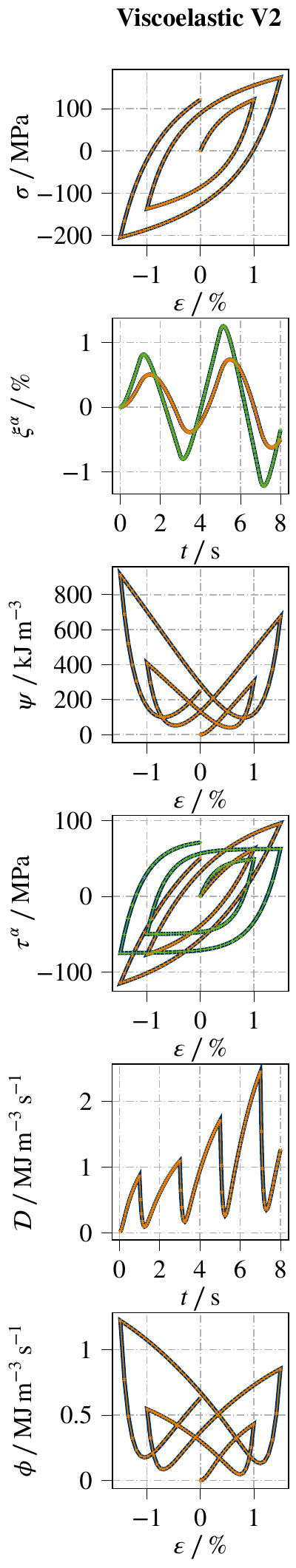}}
		\hfil
		\subfloat[\label{fig:FFBiotStar03}]{\includegraphics{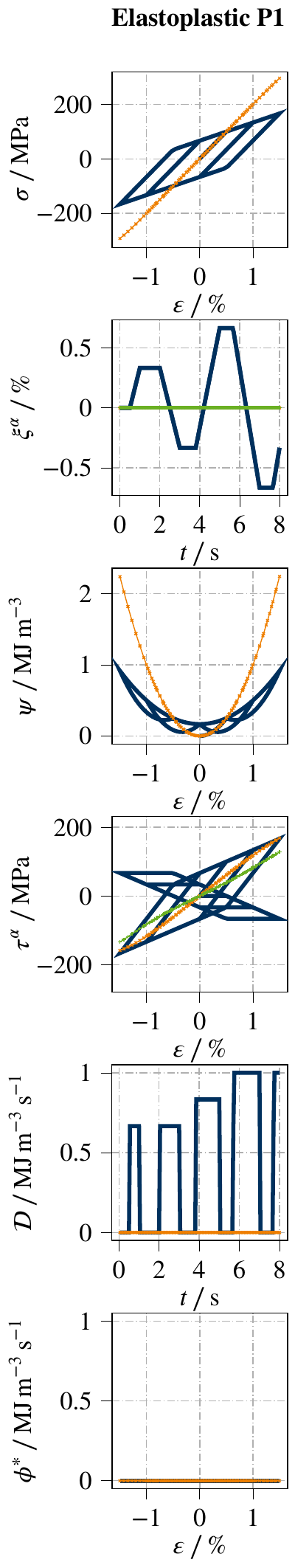}}
		\hfil
		\subfloat[\label{fig:FFBiotStar04}]{\includegraphics{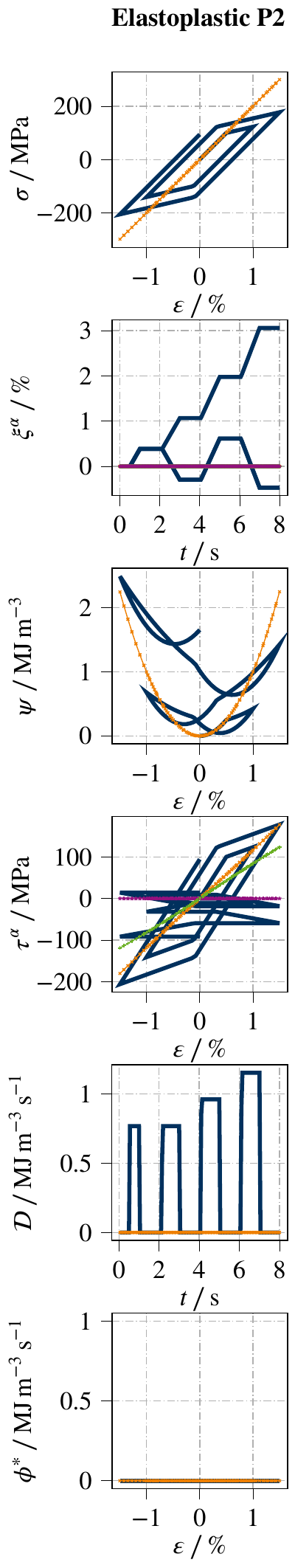}}
		\hfil
		\caption{Prediction of stress, internal variables, free energy, internal forces, dissipation rate and dual dissipation potential for the four test materials using FNN$^{\psi+\phi^*}$.}
		\label{fig:ResFFBiotStar}
	\end{figure}
}

\paragraph{FNN$^{\psi+\phi+\xi}$}
This architecture models the free energy and dissipation potential without requiring knowledge about the internal variables during training. The architecture is restricted to a single internal variable, i.e., a single FNN for the internal variable with scalar output. Using the network parameters in \reft{tab:ResFF3}, the results in \reff{fig:ResFF3} could be achieved. These results show, that the architecture is able to learn reasonable representations of the potentials for V1, V2 and P1, but not P2. All of the predictions are less precise compared to FNN$^{\psi+\phi}$. Similar to RNN$^{\xi+\psi}$, the internal variables are not learned exactly as in the reference model. However, normalization would show that the courses are similar. Because other potentials are learned if $\xi(t)$ differs from the reference model, this leads to equivalent results. 

Interestingly, the architecture finds a set of potentials for V2, that allows for surprisingly accurate stress predictions, although only one internal variable is used. The prediction for P1 shows a rather strong rate dependency. This can be attributed to a second regularization mechanism: besides the regularization of the dissipation potential, the sudden increase of $\dot{\varepsilon}^\text{pl}$ when leaving the elastic region cannot be modeled accurately by the FNN $\xi^\text{NN}(t)$. P2 in turn cannot be modeled at all, since at least two internal variables are necessary to describe the path dependency and, which is more important, due to the reasons discussed for FNN$^{\psi+\phi}$ above. The architecture FNN$^{\psi+\phi+\xi}$ is thus more suitable for viscoelastic materials.

\begin{table}[h]
	\centering
	\caption{FNN$^{\psi+\phi+\xi}$ architectures: network hyperparameters with neurons in the hidden layer of the feedforward networks $N_2^{{\phi}^{\text{con}}}$, $N_2^{{\phi}^{\text{+}}}$, $N_2^{\psi}$, $N_2^{\xi}$ with activations ${\mathcal{A}^l_2}^{{\phi}^{\text{con}}}$, ${\mathcal{A}^l_2}^{{\phi}^{\text{+}}}$, ${\mathcal{A}^l_2}^{\psi}$, ${\mathcal{A}^l_2}^{\xi}$ as well as weighting factors of the loss function $w^\sigma$, $w^{\text{Biot}}$, $w^{|\xi|}$, number of training data $N^\text{ds}$ and number of iterations and final loss values $\mathcal L^\sigma$, $\mathcal{L}^{\text{Biot}}$. The validation load cases are linked in Val.}
	\label{tab:ResFF3}
	\centering
	\begin{small}
	\begin{tabular}{c c c c c c c c c c c c c c}
		\toprule
		Mat. & $N_2^{{\phi}^{\text{con}}}$ & $N_2^{{\phi}^{\text{+}}}$ & $N_2^{\psi}$ & $N_2^{\xi}$ & ${\mathcal{A}^l_2}^{{\phi}^{\text{con}}}$ & ${\mathcal{A}^l_2}^{{\phi}^{\text{+}}}$ & ${\mathcal{A}^l_2}^{\psi}$ & ${\mathcal{A}^l_2}^{\xi}$ & $w^\sigma$/$w^{\text{Biot}}$/$w^{|\xi|}$ & $N^\text{ds}$ & Iter. & $\mathcal{L}^\sigma$ / $\mathcal{L}^{\dot{\xi}}$ & Val. \\
		\midrule
		V1 & 20 & 20 & 15 & 40 & $\softplus$ & $\softplus$ & $\tanh$ & $\tanh$ & 10/1/1 & 900 & 920 & $\num[exponent-product=\ensuremath{\cdot},round-mode=figures,round-precision=1]{5e-2}$/$\num[exponent-product=\ensuremath{\cdot},round-mode=figures,round-precision=1]{2e-1}$ & \ref{fig:FF301} \\
		V2 & 20 & 20 & 15 & 40 & $\softplus$ & $\softplus$ & $\tanh$ & $\tanh$ & 10/1/1 & 900 & 1161 & $\num[exponent-product=\ensuremath{\cdot},round-mode=figures,round-precision=1]{4e-2}$/$\num[exponent-product=\ensuremath{\cdot},round-mode=figures,round-precision=1]{3e-1}$ & \ref{fig:FF302} \\
		P1 & 20 & 20 & 15 & 75 & $\softplus$ & $\softplus$ & $\tanh$ & $\tanh$ & 10/1/1 & 900 & 1588 & $\num[exponent-product=\ensuremath{\cdot},round-mode=figures,round-precision=1]{4e-2}$/$\num[exponent-product=\ensuremath{\cdot},round-mode=figures,round-precision=1]{1e-1}$ & \ref{fig:FF303} \\
		P2 & 20 & 20 & 15 & 75 & $\softplus$ & $\softplus$ & $\tanh$ & $\tanh$ & 10/1/1 & 900 & 1163 & $\num[exponent-product=\ensuremath{\cdot},round-mode=figures,round-precision=1]{3e-2}$/$\num[exponent-product=\ensuremath{\cdot},round-mode=figures,round-precision=1]{2e-1}$ & \ref{fig:FF304} \\
		\bottomrule
	\end{tabular}
	\end{small}
\end{table}

{
	\begin{figure}[h]
		\graphicspath{{images/Results/}}
		\centering
		\includegraphics{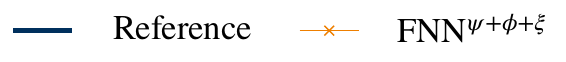}\\
		\setlength{\figw}{4cm}
		\setlength{\figh}{4cm}
		\subfloat[\label{fig:FF301}]{\includegraphics{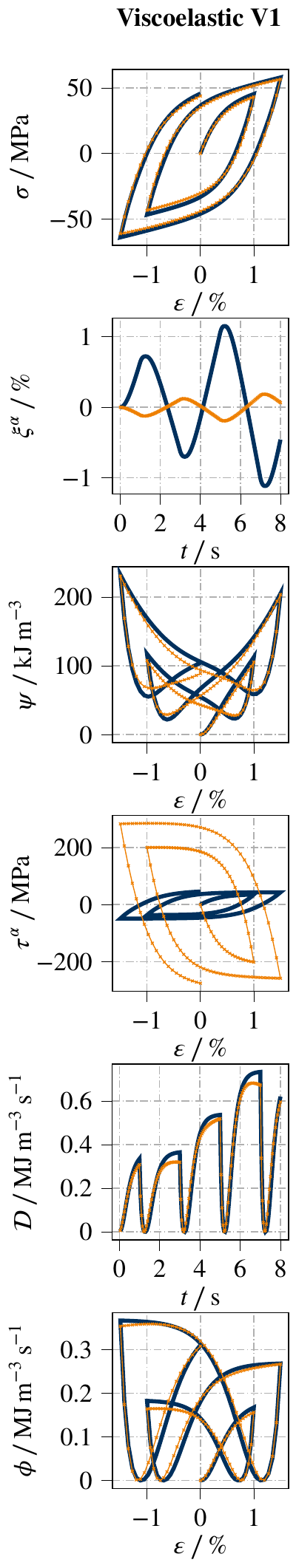}}
		\hfil
		\subfloat[\label{fig:FF302}]{\includegraphics{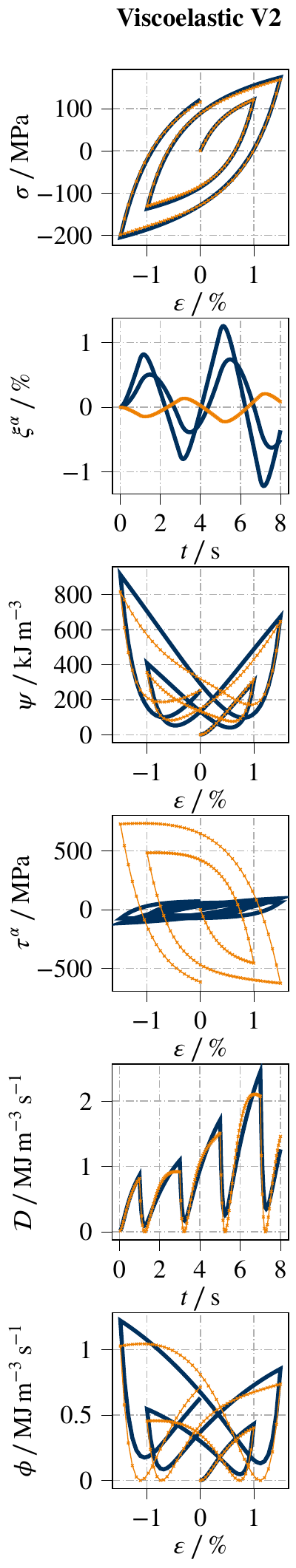}}
		\hfil
		\subfloat[\label{fig:FF303}]{\includegraphics{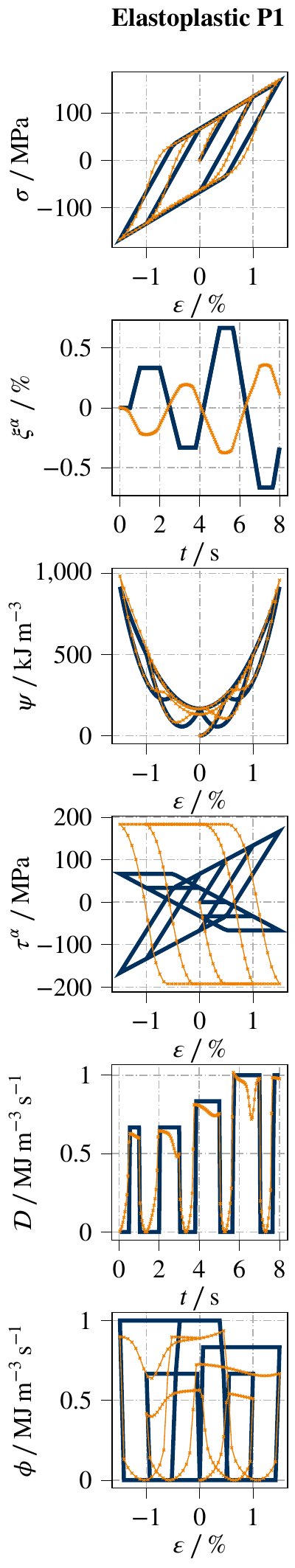}}
		\hfil
		\subfloat[\label{fig:FF304}]{\includegraphics{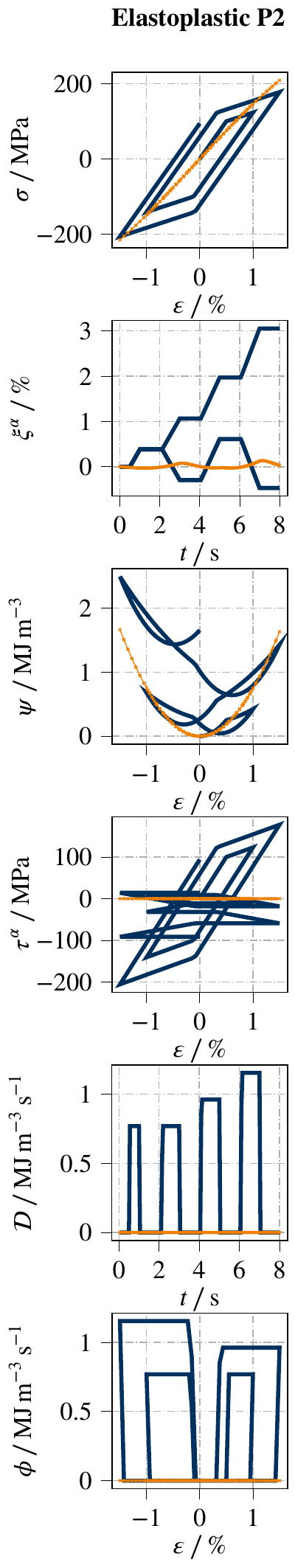}}
		\hfil
		\caption{Prediction of stress, internal variables, free energy, internal forces, dissipation rate and dissipation potential for the four test materials using the advanced training method FNN$^{\psi+\phi+\xi}$.}
		\label{fig:ResFF3}
	\end{figure}
}

\clearpage

\subsection{Validation path with extrapolation}\label{subsec:ResExtrapolation}
After investigating which NN-based approaches can reproduce which material behavior well, i.e. V1, V2, P1 and P2, the extrapolation capability of these approaches is now compared. This is done using the path given in \reff{fig:Extrapol}. Since all NN-based approaches have shown that they can reproduce the viscoelastic material V1 well, only this will be considered in the following.
The exact same models that were validated with the strain path without extrapolation are now used to carry out the extrapolation study.
The results of this final study are given in Fig.~\ref{fig:ExtraPol}. 

\subsubsection{Black box NNs}
To start with, the \emph{extrapolation} capabilities of the first \emph{black box model} FNN$^\sigma$ are evaluated. As can be seen, the model is easily able to extrapolate to strains of up to $\varepsilon=3\,\%$ which is outside the training range $\varepsilon^\text{train}\in[-2,2]\,\%$.
After reentering the training range without notable inaccuracies, the model fails to extrapolate into ranges of higher strain rates and exhibits large errors. However, after a short relaxation, the model is able to produce accurate predictions for strain increments outside the training range and is even able to yield reasonable values for the stress up to $\SI{4}{\percent}$.
Thus, the FNN$^\sigma$ model is surprisingly good at extrapolating, except for increased strain rates $\dot \varepsilon$. However, it should be noted that there is no possibility to make statements about the thermodynamics of the model.

The second \emph{black box model}, RNN$^\sigma$, initially shows a similarly good prediction quality, but also fails to predict stresses for larger strain rates and deviates for time increments of $\Delta t = \SI{0.2}{\second}$. It should be noted that the extrapolation behavior of RNNs often differs significantly for different training configurations. Even strong oscillations could be observed in some cases. Thus, compared to the FNN$^\sigma$ model, the RNN$^\sigma$ is worse at extrapolating. However, it can be used for a broader class of material behavior compared to FNN$^\sigma$, see Figs.~\ref{fig:ResFFsig} and \ref{fig:ResRNNsig}. The missing possibility to make statements about the thermodynamics remains.

\subsubsection{NNs enforcing physics in a weak form}
Now, the \emph{extrapolation} behavior of \emph{NNs enforcing physics in a weak form} is considered. As shown in Fig.~\ref{fig:ExtraPol}, the model FNN$^{\xi+\psi}$ agrees very well with the reference V1 up to maximum strains of $2.5\,\%$, but slightly deviates above. After reentering the training range, precise predictions are made again. Increased strain rates do lead to errors as well, but these are substantially reduced compared to the previous models. The model is able to cope with increased time increments and again precisely predicts the stress up to $2.5\,\%$. This also applies to the free energy $\psi$ and the dissipation rate, for which $\mathcal D\ge 0$ applies up to that point. Thus, the model does not violate the second law of thermodynamics up to this point of the loading path. However, when the strain is further increased to $\varepsilon>3\,\%$, significant errors in $\sigma$ occur. Furthermore, one can see that the model predicts negative values for $\mathcal D$.
Thus, FNN$^{\xi+\psi}$ is very good at extrapolation for the most part of the loading path. However, from a certain level of strains, unphysical predictions can be seen. This is due to the fact that the fulfillment of the second law of thermodynamics is enforced only by a penalty term in the loss and is not fulfilled a priori.

The RNN$^{\xi+\psi}$ model shows similarly good results, but is more precise in strains up to $3\,\%$.  As with the FNN$^{\xi+\psi}$, completely unphysical predictions with $\mathcal D<0$ occur for strains greater than $3\,\%$.
Thus, RNN$^{\xi+\psi}$ provides acceptable results when extrapolating up to $\varepsilon=3\,\%$. For this, however, no inner variables are necessary for the training here.
Finally, as it is also the case for FNN$^{\xi+\psi}$, unphysical predictions may occur. 

\subsubsection{NNs enforcing physics in a strong form}
Lastly, the \emph{extrapolation} behavior of the three \emph{NNs enforcing physics in a strong form}, FNN$^{\psi+\phi}$, FNN$^{\psi+\phi^*}$, and FNN$^{\psi+\phi}$, is analyzed. As shown in Fig.~\ref{fig:ExtraPol}, highly accurate predictions can be achieved with FNN$^{\psi+\phi}$ and FNN$^{\psi+\phi^*}$ for $\sigma$, $\psi$, as well as $\mathcal{D}$. This applies to the entire load path, except for the last piece of FNN$^{\psi+\phi^*}$ with $\varepsilon>4\,\%$. FNN$^{\psi+\phi}$ is even capable to produce very precise results for strains of $\varepsilon=6\,\%$. The extrapolation using FNN$^{\psi+\phi+\xi}$ is not as accurate as the two former architectures, but still good considering the mediocre interpolation results. Summarizing, neither highly increased strains or strain rates nor time increments outside of the training range lead to considerable deviations from the expected material response. Particularly noteworthy here is that in any case $\mathcal D\ge0$ is ensured. Thus, the predictions are always in accordance with the second law given by the CDI \eqref{eq:CDI}.
All in all, this model class is best suited for extrapolation which is due to the strong physical background inserted here.

{
	\begin{figure}[h]
		\graphicspath{{images/Results/}}
		\centering
		\includegraphics{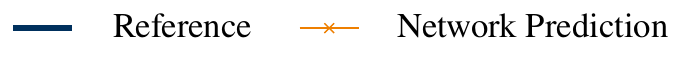}\\
		\setlength{\figw}{4cm}
		\setlength{\figh}{4cm}
		\subfloat[\label{fig:ExtraPolStress}]{\includegraphics{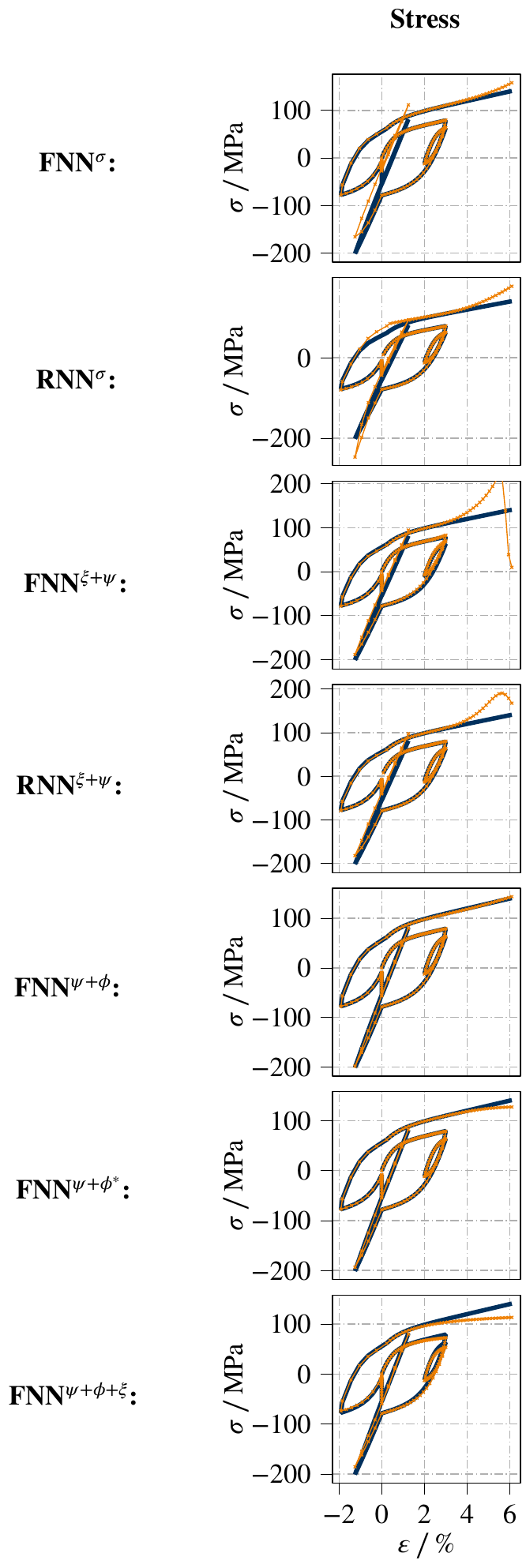}}
		\hfil
		\subfloat[\label{fig:ExtraPolEnergy}]{\includegraphics{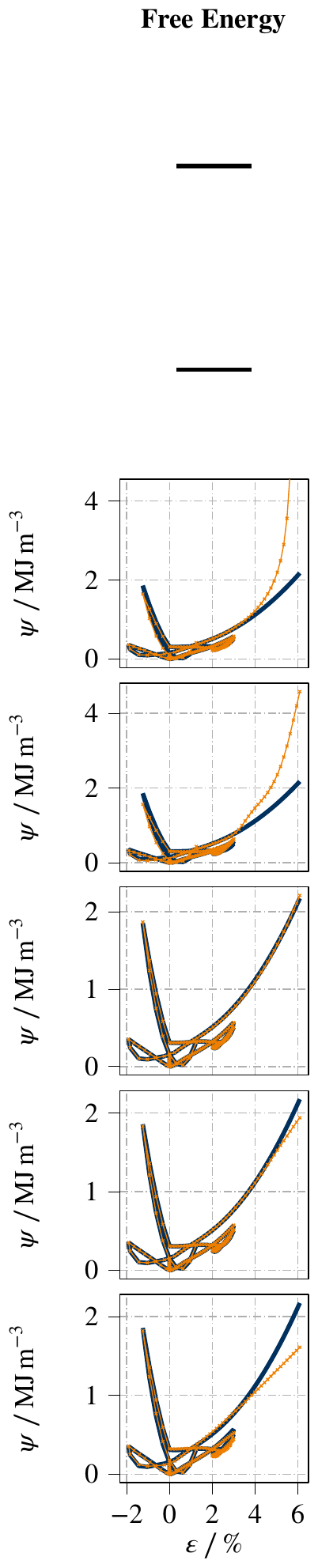}}
		\hfil
		\subfloat[\label{fig:ExtraPolDissipation}]{\includegraphics{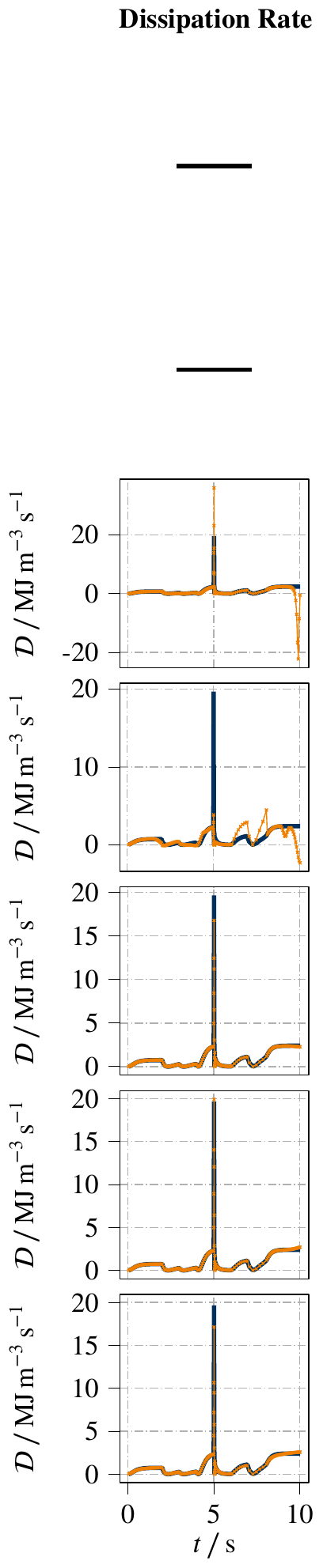}}
		\hfil
		\caption{Predictions of the six considered NNs tested for an extrapolation path with $\varepsilon$, $\dot \varepsilon$, and $\Delta t$ not included in the training data set: \textbf{(a)} stress $\sigma$, \textbf{(b)} free energy $\psi$, and \textbf{(c)} dissipation rate $\mathcal D$. The viscoelastic model V1 serves as a reference and has been used for generation of training data.}
		\label{fig:ExtraPol}
	\end{figure}
}

\clearpage

\section{Conclusions}\label{sec:Concl}

In this work, a classification of the variety of NN-based approaches to modeling inelastic constitutive behavior with particular attention to the thermodynamic framework as well as a unified formulation of these approaches is provided. To this end, a division of NN-based approaches into \emph{black box NNs, NNs enforcing physics in a weak form} and \emph{NNs enforcing physics in a strong form} is made and an application to both 1D elastoplastic and viscoelastic data is made.  

After a compact literature review, a short overview on continuum based constitutive modeling including standard viscoelastic and elastoplastic models and a condensed repetition to the basis of FNNs and RNNs is given. 
Based on this, a total of seven NN-based approaches are presented with a detailed description on training and application. It is shown in which way the second law of thermodynamics is taken into account in the respective model and what data are necessary for training.
Thereafter, the generation of training data and the application of the seven NN-based approaches to these data are shown. It can be seen that all considered models are able to represent viscoelasticity, whereas elastoplasticity cannot be represented by all approaches. Furthermore, the models' extrapolation capabilities are analyzed.

In summary, the results of this work show that NN-based models are promising for the description of complex inelastic behavior and prove to be very flexible. They have the potential to replace the time-consuming task of classical constitutive model formulation and calibration piecemeal and to enable automated workflows. 
\emph{Black box models}, however, do not allow any conclusions to be drawn as to whether the processes described by the respective model are embedded in a meaningful thermodynamic framework and are therefore not recommended for use. 
Furthermore, when applied to unknown load paths outside the training domain, the poor extrapolation capability of NNs can lead to large errors within stress predictions.
\emph{NNs enforcing physics in a weak form} on the other hand, have a higher content of physics included into the model. Only the class of \emph{NNs enforcing physics in a strong form} can really ensure that no violation of thermodynamics occurs. Moreover, they also have shown to be characterized by the best extrapolation behavior of the considered models.

Regarding the follow-up of the study presented here, several extensions are planned in the future. For instance, an extension to the general 3D case at finite strains has to be made. Thus, a variety of further physical principles and conditions have to be included in the comparison for this, e.g., principles as objectivity or material symmetry.

\section*{CRediT authorship contribution statement}

\textbf{Max Rosenkranz:} Conceptualization, Formal analysis, Investigation, Methodology, Visualization, Software, Validation, Writing - original draft, Writing - review and editing. 
\textbf{Karl A. Kalina:} Conceptualization, Formal analysis, Methodology, Writing - original draft, Writing - review and editing. 
\textbf{J\"org Brummund:} Formal analysis, Methodology, Writing - review and editing. 
\textbf{Markus K\"astner:} Funding acquisition, Resources, Writing - review and editing.

\bibliographystyle{unsrtnat}
\bibliography{ReferencesFinal}  

\begin{thebibliography}{61}
\providecommand{\natexlab}[1]{#1}
\providecommand{\url}[1]{\texttt{#1}}
\expandafter\ifx\csname urlstyle\endcsname\relax
  \providecommand{\doi}[1]{doi: #1}\else
  \providecommand{\doi}{doi: \begingroup \urlstyle{rm}\Url}\fi

\bibitem[Bock et~al.(2019)Bock, Aydin, Cyron, Huber, Kalidindi, and
  Klusemann]{Bock2019}
Frederic~E. Bock, Roland~C. Aydin, Christian~J. Cyron, Norbert Huber, Surya~R.
  Kalidindi, and Benjamin Klusemann.
\newblock A {{Review}} of the {{Application}} of {{Machine Learning}} and
  {{Data Mining Approaches}} in {{Continuum Materials Mechanics}}.
\newblock \emph{Frontiers in Materials}, 6:\penalty0 110, May 2019.
\newblock ISSN 2296-8016.
\newblock \doi{10.3389/fmats.2019.00110}.

\bibitem[Mont{\'a}ns et~al.(2019)Mont{\'a}ns, Chinesta, {G{\'o}mez-Bombarelli},
  and Kutz]{Montans2019}
Francisco~J. Mont{\'a}ns, Francisco Chinesta, Rafael {G{\'o}mez-Bombarelli},
  and J.~Nathan Kutz.
\newblock Data-driven modeling and learning in science and engineering.
\newblock \emph{Comptes Rendus M{\'e}canique}, 347\penalty0 (11):\penalty0
  845--855, November 2019.
\newblock ISSN 16310721.
\newblock \doi{10.1016/j.crme.2019.11.009}.

\bibitem[Ghaboussi et~al.(1991)Ghaboussi, Garrett, and Wu]{Ghaboussi1991}
J.~Ghaboussi, J.~H. Garrett, and X.~Wu.
\newblock Knowledge-{{Based Modeling}} of {{Material Behavior}} with {{Neural
  Networks}}.
\newblock \emph{Journal of Engineering Mechanics}, 117\penalty0 (1):\penalty0
  132--153, January 1991.
\newblock ISSN 0733-9399, 1943-7889.
\newblock \doi{10.1061/(ASCE)0733-9399(1991)117:1(132)}.

\bibitem[Raissi et~al.(2019)Raissi, Perdikaris, and Karniadakis]{Raissi2019}
M.~Raissi, P.~Perdikaris, and G.E. Karniadakis.
\newblock Physics-informed neural networks: {{A}} deep learning framework for
  solving forward and inverse problems involving nonlinear partial differential
  equations.
\newblock \emph{Journal of Computational Physics}, 378:\penalty0 686--707,
  February 2019.
\newblock ISSN 00219991.
\newblock \doi{10.1016/j.jcp.2018.10.045}.

\bibitem[Henkes et~al.(2022{\natexlab{a}})Henkes, Wessels, and
  Mahnken]{Henkes2022}
Alexander Henkes, Henning Wessels, and Rolf Mahnken.
\newblock Physics informed neural networks for continuum micromechanics.
\newblock \emph{Computer Methods in Applied Mechanics and Engineering},
  393:\penalty0 114790, April 2022{\natexlab{a}}.
\newblock ISSN 0045-7825.
\newblock \doi{10.1016/j.cma.2022.114790}.

\bibitem[As'ad et~al.(2022)As'ad, Avery, and Farhat]{Asad2022}
Faisal As'ad, Philip Avery, and Charbel Farhat.
\newblock A mechanics-informed artificial neural network approach in
  data-driven constitutive modeling.
\newblock \emph{International Journal for Numerical Methods in Engineering},
  123\penalty0 (12):\penalty0 2738--2759, 2022.
\newblock ISSN 1097-0207.
\newblock \doi{10.1002/nme.6957}.
\newblock \_eprint: https://onlinelibrary.wiley.com/doi/pdf/10.1002/nme.6957.

\bibitem[As'ad and Farhat(2022)]{Asad2022a}
Faisal As'ad and Charbel Farhat.
\newblock A {{Mechanics}}-{{Informed Neural Network Framework}} for
  {{Data}}-{{Driven Nonlinear Viscoelasticity}}.
\newblock 2022.
\newblock \doi{10.13140/RG.2.2.21694.36168}.

\bibitem[Klein et~al.(2022)Klein, Ortigosa, {Mart{\'i}nez-Frutos}, and
  Weeger]{Klein2022}
Dominik~K. Klein, Rogelio Ortigosa, Jes{\'u}s {Mart{\'i}nez-Frutos}, and Oliver
  Weeger.
\newblock Finite electro-elasticity with physics-augmented neural networks.
\newblock \emph{Computer Methods in Applied Mechanics and Engineering},
  400:\penalty0 115501, October 2022.
\newblock ISSN 0045-7825.
\newblock \doi{10.1016/j.cma.2022.115501}.

\bibitem[Linden et~al.(2023)Linden, Klein, Kalina, Brummund, Weeger, and
  K{\"a}stner]{Linden2023}
Lennart Linden, Dominik~K. Klein, Karl~A. Kalina, J{\"o}rg Brummund, Oliver
  Weeger, and Markus K{\"a}stner.
\newblock Neural networks meet hyperelasticity: {{A}} guide to enforcing
  physics.
\newblock February 2023.
\newblock \doi{https://doi.org/10.48550/arXiv.2302.02403}.

\bibitem[Kalina et~al.(2023)Kalina, Linden, Brummund, and
  K{\"a}stner]{Kalina2023}
Karl~A. Kalina, Lennart Linden, J{\"o}rg Brummund, and Markus K{\"a}stner.
\newblock {{FE}}{{{\textsuperscript{ANN}}}}: An efficient data-driven
  multiscale approach based on physics-constrained neural networks and
  automated data mining.
\newblock \emph{Computational Mechanics}, February 2023.
\newblock ISSN 1432-0924.
\newblock \doi{10.1007/s00466-022-02260-0}.

\bibitem[Masi et~al.(2021)Masi, Stefanou, Vannucci, and
  {Maffi-Berthier}]{Masi2021}
Filippo Masi, Ioannis Stefanou, Paolo Vannucci, and Victor {Maffi-Berthier}.
\newblock Thermodynamics-based {{Artificial Neural Networks}} for constitutive
  modeling.
\newblock \emph{Journal of the Mechanics and Physics of Solids}, 147:\penalty0
  104277, February 2021.
\newblock ISSN 0022-5096.
\newblock \doi{10.1016/j.jmps.2020.104277}.

\bibitem[Shen et~al.(2004)Shen, Chandrashekhara, Breig, and Oliver]{Shen2004}
Y.~Shen, K.~Chandrashekhara, W.~F. Breig, and L.~R. Oliver.
\newblock Neural {{Network Based Constitutive Model}} for {{Rubber Material}}.
\newblock \emph{Rubber Chemistry and Technology}, 77\penalty0 (2):\penalty0
  257--277, May 2004.
\newblock ISSN 1943-4804, 0035-9475.
\newblock \doi{10.5254/1.3547822}.

\bibitem[Liang and Chandrashekhara(2008)]{Liang2008}
G.~Liang and K.~Chandrashekhara.
\newblock Neural network based constitutive model for elastomeric foams.
\newblock \emph{Engineering Structures}, 30\penalty0 (7):\penalty0 2002--2011,
  July 2008.
\newblock ISSN 01410296.
\newblock \doi{10.1016/j.engstruct.2007.12.021}.

\bibitem[Linka et~al.(2021)Linka, Hillg{\"a}rtner, Abdolazizi, Aydin, Itskov,
  and Cyron]{Linka2021}
Kevin Linka, Markus Hillg{\"a}rtner, Kian~P. Abdolazizi, Roland~C. Aydin,
  Mikhail Itskov, and Christian~J. Cyron.
\newblock Constitutive artificial neural networks: {{A}} fast and general
  approach to predictive data-driven constitutive modeling by deep learning.
\newblock \emph{Journal of Computational Physics}, 429:\penalty0 110010, March
  2021.
\newblock ISSN 00219991.
\newblock \doi{10.1016/j.jcp.2020.110010}.

\bibitem[Linden et~al.(2021)Linden, Kalina, Brummund, Metsch, and
  K{\"a}stner]{Linden2021}
Lennart Linden, Karl~A. Kalina, J{\"o}rg Brummund, Philipp Metsch, and Markus
  K{\"a}stner.
\newblock Thermodynamically consistent constitutive modeling of isotropic
  hyperelasticity based on artificial neural networks.
\newblock \emph{PAMM}, 21\penalty0 (1), December 2021.
\newblock ISSN 1617-7061, 1617-7061.
\newblock \doi{10.1002/pamm.202100144}.

\bibitem[Klein et~al.(2021)Klein, Fern{\'a}ndez, Martin, Neff, and
  Weeger]{Klein2021}
Dominik~K. Klein, Mauricio Fern{\'a}ndez, Robert~J. Martin, Patrizio Neff, and
  Oliver Weeger.
\newblock Polyconvex anisotropic hyperelasticity with neural networks.
\newblock \emph{Journal of the Mechanics and Physics of Solids}, page 104703,
  November 2021.
\newblock ISSN 00225096.
\newblock \doi{10.1016/j.jmps.2021.104703}.

\bibitem[Fuhg et~al.(2022{\natexlab{a}})Fuhg, Bouklas, and Jones]{Fuhg2022b}
Jan~N. Fuhg, Nikolaos Bouklas, and Reese~E. Jones.
\newblock Learning hyperelastic anisotropy from data via a tensor basis neural
  network.
\newblock \emph{Journal of the Mechanics and Physics of Solids}, 168:\penalty0
  105022, November 2022{\natexlab{a}}.
\newblock ISSN 00225096.
\newblock \doi{10.1016/j.jmps.2022.105022}.

\bibitem[Tac et~al.(2022)Tac, Sahli~Costabal, and Tepole]{Tac2022a}
Vahidullah Tac, Francisco Sahli~Costabal, and Adrian~B. Tepole.
\newblock Data-driven tissue mechanics with polyconvex neural ordinary
  differential equations.
\newblock \emph{Computer Methods in Applied Mechanics and Engineering},
  398:\penalty0 115248, August 2022.
\newblock ISSN 0045-7825.
\newblock \doi{10.1016/j.cma.2022.115248}.

\bibitem[Vlassis et~al.(2020)Vlassis, Ma, and Sun]{Vlassis2020}
Nikolaos~N. Vlassis, Ran Ma, and WaiChing Sun.
\newblock Geometric deep learning for computational mechanics {{Part I}}:
  Anisotropic hyperelasticity.
\newblock \emph{Computer Methods in Applied Mechanics and Engineering},
  371:\penalty0 113299, November 2020.
\newblock ISSN 0045-7825.
\newblock \doi{10.1016/j.cma.2020.113299}.

\bibitem[Vlassis et~al.(2022)Vlassis, Zhao, Ma, Sewell, and Sun]{Vlassis2022a}
Nikolaos~N. Vlassis, Puhan Zhao, Ran Ma, Tommy Sewell, and WaiChing Sun.
\newblock Molecular dynamics inferred transfer learning models for
  finite-strain hyperelasticity of monoclinic crystals: {{Sobolev}} training
  and validations against physical constraints.
\newblock \emph{International Journal for Numerical Methods in Engineering},
  123\penalty0 (17):\penalty0 3922--3949, 2022.
\newblock ISSN 1097-0207.
\newblock \doi{10.1002/nme.6992}.
\newblock \_eprint: https://onlinelibrary.wiley.com/doi/pdf/10.1002/nme.6992.

\bibitem[Kalina et~al.(2021)Kalina, Linden, Brummund, Metsch, and
  K{\"a}stner]{Kalina2021}
Karl~A. Kalina, Lennart Linden, J{\"o}rg Brummund, Philipp Metsch, and Markus
  K{\"a}stner.
\newblock Automated constitutive modeling of isotropic hyperelasticity based on
  artificial neural networks.
\newblock \emph{Computational Mechanics}, October 2021.
\newblock ISSN 0178-7675, 1432-0924.
\newblock \doi{10.1007/s00466-021-02090-6}.

\bibitem[Furukawa and Yagawa(1998)]{Furukawa1998}
Tomonari Furukawa and Genki Yagawa.
\newblock Implicit constitutive modelling for viscoplasticity using neural
  networks.
\newblock \emph{International Journal for Numerical Methods in Engineering},
  43\penalty0 (2):\penalty0 195--219, 1998.
\newblock ISSN 1097-0207.
\newblock
  \doi{10.1002/(SICI)1097-0207(19980930)43:2<195::AID-NME418>3.0.CO;2-6}.
\newblock \_eprint:
  https://onlinelibrary.wiley.com/doi/pdf/10.1002/\%28SICI\%291097-0207\%2819980930\%2943\%3A2\%3C195\%3A\%3AAID-NME418\%3E3.0.CO\%3B2-6.

\bibitem[Ghaboussi and Sidarta(1998)]{Ghaboussi1998a}
J.~Ghaboussi and D.E. Sidarta.
\newblock New nested adaptive neural networks ({{NANN}}) for constitutive
  modeling.
\newblock \emph{Computers and Geotechnics}, 22\penalty0 (1):\penalty0 29--52,
  January 1998.
\newblock ISSN 0266352X.
\newblock \doi{10.1016/S0266-352X(97)00034-7}.

\bibitem[Hashash et~al.(2004)Hashash, Jung, and Ghaboussi]{Hashash2004}
Y.~M.~A. Hashash, S.~Jung, and J.~Ghaboussi.
\newblock Numerical implementation of a neural network based material model in
  finite element analysis: {{NEURAL NETWORK BASED MATERIAL MODEL}}.
\newblock \emph{International Journal for Numerical Methods in Engineering},
  59\penalty0 (7):\penalty0 989--1005, February 2004.
\newblock ISSN 00295981.
\newblock \doi{10.1002/nme.905}.

\bibitem[{Al-Haik} et~al.(2006){Al-Haik}, Hussaini, and
  Garmestani]{Al-Haik2006}
M.S. {Al-Haik}, M.Y. Hussaini, and H.~Garmestani.
\newblock Prediction of nonlinear viscoelastic behavior of polymeric composites
  using an artificial neural network.
\newblock \emph{International Journal of Plasticity}, 22\penalty0 (7):\penalty0
  1367--1392, July 2006.
\newblock ISSN 07496419.
\newblock \doi{10.1016/j.ijplas.2005.09.002}.

\bibitem[Jung and Ghaboussi(2006)]{Jung2006}
Sungmoon Jung and Jamshid Ghaboussi.
\newblock Neural network constitutive model for rate-dependent materials.
\newblock \emph{Computers \& Structures}, 84\penalty0 (15-16):\penalty0
  955--963, June 2006.
\newblock ISSN 00457949.
\newblock \doi{10.1016/j.compstruc.2006.02.015}.

\bibitem[Hochreiter and Schmidhuber(1997)]{Hochreiter1997}
Sepp Hochreiter and J{\"u}rgen Schmidhuber.
\newblock Long {{Short}}-{{Term Memory}}.
\newblock \emph{Neural Computation}, 9\penalty0 (8):\penalty0 1735--1780,
  November 1997.
\newblock ISSN 0899-7667.
\newblock \doi{10.1162/neco.1997.9.8.1735}.
\newblock Conference Name: Neural Computation.

\bibitem[Ghavamian and Simone(2019)]{Ghavamian2019}
F.~Ghavamian and A.~Simone.
\newblock Accelerating multiscale finite element simulations of
  history-dependent materials using a recurrent neural network.
\newblock \emph{Computer Methods in Applied Mechanics and Engineering},
  357:\penalty0 112594, December 2019.
\newblock ISSN 00457825.
\newblock \doi{10.1016/j.cma.2019.112594}.

\bibitem[Wu et~al.(2020)Wu, Nguyen, Kilingar, and Noels]{Wu2020}
Ling Wu, Van~Dung Nguyen, Nanda~Gopala Kilingar, and Ludovic Noels.
\newblock A recurrent neural network-accelerated multi-scale model for
  elasto-plastic heterogeneous materials subjected to random cyclic and
  non-proportional loading paths.
\newblock \emph{Computer Methods in Applied Mechanics and Engineering},
  369:\penalty0 113234, September 2020.
\newblock ISSN 00457825.
\newblock \doi{10.1016/j.cma.2020.113234}.

\bibitem[Li and Zhuang(2020)]{Li2020a}
Bin Li and Xiaoying Zhuang.
\newblock Multiscale computation on feedforward neural network and recurrent
  neural network.
\newblock \emph{Frontiers of Structural and Civil Engineering}, 14\penalty0
  (6):\penalty0 1285--1298, December 2020.
\newblock ISSN 2095-2430, 2095-2449.
\newblock \doi{10.1007/s11709-020-0691-7}.

\bibitem[Fuchs et~al.(2021)Fuchs, Heider, Wang, Sun, and Kaliske]{Fuchs2021}
Alexander Fuchs, Yousef Heider, Kun Wang, WaiChing Sun, and Michael Kaliske.
\newblock {{DNN2}}: {{A}} hyper-parameter reinforcement learning game for
  self-design of neural network based elasto-plastic constitutive descriptions.
\newblock \emph{Computers \& Structures}, 249:\penalty0 106505, June 2021.
\newblock ISSN 00457949.
\newblock \doi{10.1016/j.compstruc.2021.106505}.

\bibitem[Henkes et~al.(2022{\natexlab{b}})Henkes, Eshraghian, and
  Wessels]{Henkes2022a}
Alexander Henkes, Jason~K. Eshraghian, and Henning Wessels.
\newblock Spiking neural networks for nonlinear regression.
\newblock October 2022{\natexlab{b}}.
\newblock \doi{10.48550/arXiv.2210.03515}.

\bibitem[Bonatti and Mohr(2022)]{Bonatti2022}
Colin Bonatti and Dirk Mohr.
\newblock On the importance of self-consistency in recurrent neural network
  models representing elasto-plastic solids.
\newblock \emph{Journal of the Mechanics and Physics of Solids}, 158:\penalty0
  104697, January 2022.
\newblock ISSN 00225096.
\newblock \doi{10.1016/j.jmps.2021.104697}.

\bibitem[Bonatti et~al.(2022)Bonatti, Berisha, and Mohr]{Bonatti2022a}
Colin Bonatti, Bekim Berisha, and Dirk Mohr.
\newblock From {{CP}}-{{FFT}} to {{CP}}-{{RNN}}: {{Recurrent}} neural network
  surrogate model of crystal plasticity.
\newblock \emph{International Journal of Plasticity}, 158:\penalty0 103430,
  November 2022.
\newblock ISSN 0749-6419.
\newblock \doi{10.1016/j.ijplas.2022.103430}.

\bibitem[Heider et~al.(2020)Heider, Wang, and Sun]{Heider2020}
Yousef Heider, Kun Wang, and WaiChing Sun.
\newblock {{SO}}(3)-invariance of informed-graph-based deep neural network for
  anisotropic elastoplastic materials.
\newblock \emph{Computer Methods in Applied Mechanics and Engineering},
  363:\penalty0 112875, May 2020.
\newblock ISSN 00457825.
\newblock \doi{10.1016/j.cma.2020.112875}.

\bibitem[Rocha et~al.(2023)Rocha, Kerfriden, and {van der Meer}]{Rocha2023}
I.~B. C.~M. Rocha, P.~Kerfriden, and F.~P. {van der Meer}.
\newblock Machine learning of evolving physics-based material models for
  multiscale solid mechanics.
\newblock January 2023.
\newblock \doi{10.48550/arXiv.2301.13547}.

\bibitem[Liu et~al.(2019)Liu, Wu, and Koishi]{Liu2019a}
Zeliang Liu, C.T. Wu, and M.~Koishi.
\newblock A deep material network for multiscale topology learning and
  accelerated nonlinear modeling of heterogeneous materials.
\newblock \emph{Computer Methods in Applied Mechanics and Engineering},
  345:\penalty0 1138--1168, March 2019.
\newblock ISSN 00457825.
\newblock \doi{10.1016/j.cma.2018.09.020}.

\bibitem[Liu and Wu(2019)]{Liu2019}
Zeliang Liu and C.T. Wu.
\newblock Exploring the {{3D}} architectures of deep material network in
  data-driven multiscale mechanics.
\newblock \emph{Journal of the Mechanics and Physics of Solids}, 127:\penalty0
  20--46, June 2019.
\newblock ISSN 00225096.
\newblock \doi{10.1016/j.jmps.2019.03.004}.

\bibitem[Gajek et~al.(2020)Gajek, Schneider, and B{\"o}hlke]{Gajek2020}
Sebastian Gajek, Matti Schneider, and Thomas B{\"o}hlke.
\newblock On the micromechanics of deep material networks.
\newblock \emph{Journal of the Mechanics and Physics of Solids}, 142:\penalty0
  103984, September 2020.
\newblock ISSN 00225096.
\newblock \doi{10.1016/j.jmps.2020.103984}.

\bibitem[Gajek et~al.(2022)Gajek, Schneider, and B{\"o}hlke]{Gajek2022}
Sebastian Gajek, Matti Schneider, and Thomas B{\"o}hlke.
\newblock An {{FE}}-{{DMN}} method for the multiscale analysis of
  thermomechanical composites.
\newblock \emph{Computational Mechanics}, February 2022.
\newblock ISSN 1432-0924.
\newblock \doi{10.1007/s00466-021-02131-0}.

\bibitem[Settgast et~al.(2020)Settgast, H{\"u}tter, Kuna, and
  Abendroth]{Settgast2020}
Christoph Settgast, Geralf H{\"u}tter, Meinhard Kuna, and Martin Abendroth.
\newblock A hybrid approach to simulate the homogenized irreversible
  elastic-plastic deformations and damage of foams by neural networks.
\newblock \emph{International Journal of Plasticity}, 126:\penalty0 102624,
  March 2020.
\newblock ISSN 07496419.
\newblock \doi{10.1016/j.ijplas.2019.11.003}.

\bibitem[Malik et~al.(2021)Malik, Abendroth, H{\"u}tter, and Kiefer]{Malik2021}
Alexander Malik, Martin Abendroth, Geralf H{\"u}tter, and Bjoern Kiefer.
\newblock A {{Hybrid Approach Employing Neural Networks}} to {{Simulate}} the
  {{Elasto}}-{{Plastic Deformation Behavior}} of {{3D}}-{{Foam Structures}}.
\newblock \emph{Advanced Engineering Materials}, n/a\penalty0 (n/a):\penalty0
  2100641, 2021.
\newblock ISSN 1527-2648.
\newblock \doi{10.1002/adem.202100641}.
\newblock \_eprint:
  https://onlinelibrary.wiley.com/doi/pdf/10.1002/adem.202100641.

\bibitem[Vlassis and Sun(2021{\natexlab{a}})]{Vlassis2021}
Nikolaos~N. Vlassis and WaiChing Sun.
\newblock Sobolev training of thermodynamic-informed neural networks for
  interpretable elasto-plasticity models with level set hardening.
\newblock \emph{Computer Methods in Applied Mechanics and Engineering},
  377:\penalty0 113695, April 2021{\natexlab{a}}.
\newblock ISSN 00457825.
\newblock \doi{10.1016/j.cma.2021.113695}.

\bibitem[Vlassis and Sun(2021{\natexlab{b}})]{Vlassis2021b}
Nikolaos~Napoleon Vlassis and Waiching Sun.
\newblock Component-based machine learning paradigm for discovering
  rate-dependent and pressure-sensitive level-set plasticity models.
\newblock \emph{Journal of Applied Mechanics}, pages 1--13, October
  2021{\natexlab{b}}.
\newblock ISSN 0021-8936, 1528-9036.
\newblock \doi{10.1115/1.4052684}.

\bibitem[Fuhg et~al.(2022{\natexlab{b}})Fuhg, Hamel, Johnson, Jones, and
  Bouklas]{Fuhg2022d}
Jan~N. Fuhg, Craig~M. Hamel, Kyle Johnson, Reese Jones, and Nikolaos Bouklas.
\newblock Modular machine learning-based elastoplasticity: Generalization in
  the context of limited data.
\newblock October 2022{\natexlab{b}}.
\newblock \doi{10.48550/arXiv.2210.08343}.

\bibitem[Zopf and Kaliske(2017)]{Zopf2017}
C.~Zopf and M.~Kaliske.
\newblock Numerical characterisation of uncured elastomers by a neural network
  based approach.
\newblock \emph{Computers \& Structures}, 182:\penalty0 504--525, April 2017.
\newblock ISSN 00457949.
\newblock \doi{10.1016/j.compstruc.2016.12.012}.

\bibitem[Masi and Stefanou(2022)]{Masi2022}
Filippo Masi and Ioannis Stefanou.
\newblock Multiscale modeling of inelastic materials with
  {{Thermodynamics}}-based {{Artificial Neural Networks}} ({{TANN}}).
\newblock \emph{Computer Methods in Applied Mechanics and Engineering},
  398:\penalty0 115190, August 2022.
\newblock ISSN 0045-7825.
\newblock \doi{10.1016/j.cma.2022.115190}.

\bibitem[He and Chen(2022)]{He2022}
Xiaolong He and Jiun-Shyan Chen.
\newblock Thermodynamically consistent machine-learned internal state variable
  approach for data-driven modeling of path-dependent materials.
\newblock \emph{Computer Methods in Applied Mechanics and Engineering}, page
  115348, July 2022.
\newblock ISSN 00457825.
\newblock \doi{10.1016/j.cma.2022.115348}.

\bibitem[Huang et~al.(2022)Huang, He, Chem, and Reina]{Huang2022}
Shenglin Huang, Zequn He, Bryan Chem, and Celia Reina.
\newblock Variational {{Onsager Neural Networks}} ({{VONNs}}): {{A}}
  thermodynamics-based variational learning strategy for non-equilibrium
  {{PDEs}}.
\newblock \emph{Journal of the Mechanics and Physics of Solids}, 163:\penalty0
  104856, June 2022.
\newblock ISSN 0022-5096.
\newblock \doi{10.1016/j.jmps.2022.104856}.

\bibitem[Miehe(2002)]{Miehe2002}
Christian Miehe.
\newblock Strain-driven homogenization of inelastic microstructures and
  composites based on an incremental variational formulation.
\newblock \emph{International Journal for Numerical Methods in Engineering},
  55\penalty0 (11):\penalty0 1285--1322, December 2002.
\newblock ISSN 0029-5981, 1097-0207.
\newblock \doi{10.1002/nme.515}.

\bibitem[Amos et~al.(2017)Amos, Xu, and Kolter]{Amos2017}
Brandon Amos, Lei Xu, and J.~Zico Kolter.
\newblock Input {{Convex Neural Networks}}.
\newblock In \emph{Proceedings of the 34th {{International Conference}} on
  {{Machine Learning}}}, pages 146--155. {PMLR}, July 2017.

\bibitem[Tac et~al.(2023)Tac, Rausch, Sahli~Costabal, and
  Buganza~Tepole]{Tac2023}
Vahidullah Tac, Manuel~K. Rausch, Francisco Sahli~Costabal, and Adrian
  Buganza~Tepole.
\newblock Data-{{Driven Anisotropic Finite Viscoelasticity Using Neural
  Ordinary Differential Equations}}.
\newblock \emph{SSRN Electronic Journal}, 2023.
\newblock ISSN 1556-5068.
\newblock \doi{10.2139/ssrn.4332501}.

\bibitem[Haupt(2000)]{Haupt2000}
Peter Haupt.
\newblock \emph{Continuum {{Mechanics}} and {{Theory}} of {{Materials}}}.
\newblock {Springer Berlin Heidelberg}, {Berlin, Heidelberg}, 2000.
\newblock ISBN 978-3-662-04109-3.
\newblock OCLC: 851363702.

\bibitem[Miehe et~al.(2002)Miehe, Schotte, and Lambrecht]{Miehe2002a}
C~Miehe, J~Schotte, and M~Lambrecht.
\newblock Homogenization of inelastic solid materials at finite strains based
  on incremental minimization principles. {{Application}} to the texture
  analysis of polycrystals.
\newblock \emph{J. Mech. Phys. Solids}, page~45, 2002.

\bibitem[Miehe et~al.(2011)Miehe, Kiefer, and Rosato]{Miehe2011a}
Christian Miehe, Bj{\"o}rn Kiefer, and Daniele Rosato.
\newblock An incremental variational formulation of dissipative
  magnetostriction at the macroscopic continuum level.
\newblock \emph{International Journal of Solids and Structures}, 48\penalty0
  (13):\penalty0 1846--1866, June 2011.
\newblock ISSN 00207683.
\newblock \doi{10.1016/j.ijsolstr.2011.02.011}.

\bibitem[Simo and Hughes(2000)]{Simo2000}
Juan~Carlos Simo and Thomas J.~R. Hughes.
\newblock \emph{Computational Inelasticity}.
\newblock Number~7 in Interdisciplinary Applied Mathematics {{Mechanics}} and
  Materials. {Springer}, {New York, NY}, corr. 2. print edition, 2000.
\newblock ISBN 978-0-387-97520-7 978-1-4757-7169-5.
\newblock OCLC: 254534973.

\bibitem[Kruse et~al.(2016)Kruse, Borgelt, Braune, Mostaghim, and
  Steinbrecher]{Kruse2016}
Rudolf Kruse, Christian Borgelt, Christian Braune, Sanaz Mostaghim, and
  Matthias Steinbrecher.
\newblock \emph{Computational {{Intelligence}}}.
\newblock Texts in {{Computer Science}}. {Springer London}, {London}, 2016.
\newblock ISBN 978-1-4471-7294-9 978-1-4471-7296-3.
\newblock \doi{10.1007/978-1-4471-7296-3}.

\bibitem[Kollmannsberger et~al.(2021)Kollmannsberger, D'Angella, Jokeit, and
  Herrmann]{Kollmannsberger2021}
Stefan Kollmannsberger, Davide D'Angella, Moritz Jokeit, and Leon Herrmann.
\newblock \emph{Deep {{Learning}} in {{Computational Mechanics}}: {{An
  Introductory Course}}}, volume 977 of \emph{Studies in {{Computational
  Intelligence}}}.
\newblock {Springer International Publishing}, {Cham}, 2021.
\newblock ISBN 978-3-030-76586-6 978-3-030-76587-3.
\newblock \doi{10.1007/978-3-030-76587-3}.

\bibitem[Nagler et~al.(2022)Nagler, Pechstein, and Humer]{Nagler2022}
Michaela Nagler, Astrid Pechstein, and Alexander Humer.
\newblock A mixed finite element formulation for elastoplasticity.
\newblock \emph{International Journal for Numerical Methods in Engineering},
  123\penalty0 (21):\penalty0 5346--5368, 2022.
\newblock ISSN 1097-0207.
\newblock \doi{10.1002/nme.7070}.
\newblock \_eprint: https://onlinelibrary.wiley.com/doi/pdf/10.1002/nme.7070.

\bibitem[Ladev{\`e}ze et~al.(2019)Ladev{\`e}ze, N{\'e}ron, and
  Gerbaud]{Ladeveze2019}
Pierre Ladev{\`e}ze, David N{\'e}ron, and Paul-William Gerbaud.
\newblock Data-driven computation for history-dependent materials.
\newblock \emph{Comptes Rendus M{\'e}canique}, 347\penalty0 (11):\penalty0
  831--844, November 2019.
\newblock ISSN 16310721.
\newblock \doi{10.1016/j.crme.2019.11.008}.

\bibitem[Gerbaud et~al.(2022)Gerbaud, N{\'e}ron, and Ladev{\`e}ze]{Gerbaud2022}
Paul-William Gerbaud, David N{\'e}ron, and Pierre Ladev{\`e}ze.
\newblock Data-driven elasto-(visco)-plasticity involving hidden state
  variables.
\newblock \emph{Computer Methods in Applied Mechanics and Engineering},
  402:\penalty0 115394, December 2022.
\newblock ISSN 00457825.
\newblock \doi{10.1016/j.cma.2022.115394}.

\end{thebibliography}

\end{document}